\documentclass[a4paper,11pt]{article}
\pdfoutput=1 

\usepackage{jheppub} 
\usepackage{comment}
\usepackage[bottom]{footmisc}
\usepackage{amssymb}
\usepackage{extarrows}
\usepackage{amsmath}
\usepackage{amsthm}
\usepackage[dvipsnames]{xcolor}
\usepackage{epsfig}
\usepackage{dcolumn}
\usepackage{tikz}
\usetikzlibrary{shapes.geometric, arrows}
\usepackage{upgreek}
\usepackage{setspace}
\usepackage{array,multirow,bigdelim,arydshln}
\usepackage{appendix}
\usepackage{colortbl}
\usepackage{xparse}
\usepackage{stmaryrd}
\usepackage[T1]{fontenc} 
\usepackage{mathtools}
\usepackage{physics} 
\usepackage{adjustbox}
\usepackage{multirow}
\usepackage{graphicx} 
\usepackage{float} 
\graphicspath{{./images/}}
\usepackage[nottoc]{tocbibind}
\usepackage{hyperref}
\usepackage[utf8]{inputenc}
\hypersetup{
	colorlinks,
	urlcolor=Maroon,
	linkcolor=Maroon,
	citecolor=Maroon
	}
    \usepackage{CJKutf8}

\NewDocumentCommand{\binomial}{omm}
 {%
  \genfrac(){0pt}{}{#2}{#3}%
  \IfValueT{#1}{_{\!#1}}%
 }
\NewDocumentCommand{\eulerian}{omm}
 {%
  \genfrac<>{0pt}{}{#2}{#3}%
  \IfValueT{#1}{_{\!#1}}%
 }

\usepackage{latexsym}
\usepackage{tikz}
\usepackage{tabularray}

\usetikzlibrary{calc} 
\allowdisplaybreaks

\newcommand{\trphi}{\mathrm{tr}\,\phi^3}
\newcommand{\beps}{\boldsymbol{\epsilon}}
\newcommand{\bk}{\boldsymbol{k}}
\tikzset{
  polygon edge/.style={line width=0.95pt},
  triangulation/.style={line width=0.95pt},
  vertex dot/.style={circle,fill=black,inner sep=1.6pt},
  boundary label/.style={font=\small},
  channel label/.style={font=\small,fill=white,inner sep=1pt}
}
\theoremstyle{plain}

\theoremstyle{definition}

\newcommand{\ii}{\mathrm{i}}

\newcommand{\Disc}{\operatorname{Disc}}
\newcommand{\YM}{\mathrm{YM}}
\newcommand{\vect}[1]{\mathbf{#1}}

\newcommand{\xx}{\mathbf{x}}
\newcommand{\calD}{\mathcal{D}}

\newcommand{\dS}{\mathrm{dS}}
\newcommand{\AdS}{\mathrm{AdS}}

\def\bea#1\eea{\begin{eqnarray}#1\end{eqnarray}}
\def\be#1\ee{\begin{equation}#1\end{equation}}
\def\ba#1\ea{\begin{align}#1\end{align}}

\usepackage{amsmath}
\allowdisplaybreaks
\usepackage{multicol}
\usepackage{bbm}
\usepackage{enumerate}

\usepackage{amsthm}
\usepackage{mathrsfs}
\usepackage{upgreek}
\usepackage{amssymb}

\usepackage{bm}
\usepackage{setspace}
\usepackage{array,multirow,arydshln}
\usepackage{bigdelim}
\usepackage{scalerel}
\usepackage{diagbox}
\usepackage{tabularx}

\usetikzlibrary{shapes.geometric,arrows,arrows.meta,decorations.pathmorphing,decorations.markings,patterns}

\usetikzlibrary{decorations.pathmorphing}

\tikzset{
  gluon/.style={
    decorate,
    decoration={snake, amplitude=1.8pt, segment length=7pt},
    line width=0.8pt
  },
  scalar/.style={
    dashed,
    line width=0.9pt
  },
  bluefactor/.style={
    color=blue!70!black,
    line width=1.2pt,
    line cap=round,
    line join=round
  }
}

\newcommand{\uk}[1]{k_{\underline{#1}}}

\newcommand{\FourPointExchange}{%
\begin{tikzpicture}[x=1cm,y=1cm,every node/.style={font=\small}]
  \coordinate (L) at (-1.2,0);
  \coordinate (R) at (1.2,0);

  \draw[gluon] (-2.8,-1.4) -- (L);
  \draw[gluon] (-2.8, 1.4) -- (L);
  \draw[gluon] (L) -- (R);
  \draw[gluon] (R) -- ( 2.8, 1.4);
  \draw[gluon] (R) -- ( 2.8,-1.4);

  \node[above] at (0,0.18) {$\uk{12}$};

  \node[left] at (-3.00,-1.40) {$1$};
  \node[left] at (-3.00, 1.40) {$2$};
  \node[right] at (3.00, 1.40) {$3$};
  \node[right] at (3.00,-1.40) {$4$};
\end{tikzpicture}%
}

\newcommand{\FourPointDashedExchange}{%
\begin{tikzpicture}[x=1cm,y=1cm,every node/.style={font=\small}]
  \coordinate (L) at (-1.2,0);
  \coordinate (R) at (1.2,0);

  \draw[gluon] (-2.8,-1.4) -- (L);
  \draw[gluon] (-2.8, 1.4) -- (L);
  \draw[scalar] (L) -- (R);
  \draw[gluon] (R) -- ( 2.8, 1.4);
  \draw[gluon] (R) -- ( 2.8,-1.4);

  \node[above] at (0,0.18) {$\uk{12}$};

  \node[left] at (-3.00,-1.40) {$1$};
  \node[left] at (-3.00, 1.40) {$2$};
  \node[right] at (3.00, 1.40) {$3$};
  \node[right] at (3.00,-1.40) {$4$};
\end{tikzpicture}%
}

\newcommand{\FourPointContact}{%
\begin{tikzpicture}[x=1cm,y=1cm,every node/.style={font=\small}]
  \coordinate (C) at (0,0);

  \draw[gluon] (-2.8,-1.4) -- (C);
  \draw[gluon] (-2.8, 1.4) -- (C);
  \draw[gluon] (C) -- ( 2.8, 1.4);
  \draw[gluon] (C) -- ( 2.8,-1.4);

  \node[left] at (-3.00,-1.40) {$1$};
  \node[left] at (-3.00, 1.40) {$2$};
  \node[right] at (3.00, 1.40) {$3$};
  \node[right] at (3.00,-1.40) {$4$};
\end{tikzpicture}%
}

\newcommand{\LocalCollapseDiagram}[4]{%
\begin{tikzpicture}[baseline=-0.6ex,x=0.85cm,y=0.85cm,every node/.style={font=\scriptsize}]
  \coordinate (O) at (0,0);
  \coordinate (A) at (-1.35,-0.95);
  \coordinate (B) at (-1.35, 0.95);
  \coordinate (C) at ( 1.35, 0.95);
  \coordinate (D) at ( 1.35,-0.95);

  \draw[line width=0.85pt] (O) -- (A);
  \draw[line width=0.85pt] (O) -- (B);
  \draw[line width=0.85pt] (O) -- (C);
  \draw[line width=0.85pt] (O) -- (D);

  \fill (O) circle (1.5pt);

  \node[draw,circle,fill=white,fill opacity=1,text opacity=1,inner sep=1.0pt,minimum size=12pt] at (A) {$A$};
  \node[draw,circle,fill=white,fill opacity=1,text opacity=1,inner sep=1.0pt,minimum size=12pt] at (B) {$B$};
  \node[draw,circle,fill=white,fill opacity=1,text opacity=1,inner sep=1.0pt,minimum size=12pt] at (C) {$C$};
  \node[draw,circle,fill=white,fill opacity=1,text opacity=1,inner sep=1.0pt,minimum size=12pt] at (D) {$D$};

  \ifnum#1=1
    \draw[line width=0.8pt] (-0.62,-0.18) -- (-0.32,-0.52);
  \fi
  \ifnum#2=1
    \draw[line width=0.8pt] (-0.62,0.18) -- (-0.32,0.52);
  \fi
  \ifnum#3=1
    \draw[line width=0.8pt] (0.32,0.52) -- (0.62,0.18);
  \fi
  \ifnum#4=1
    \draw[line width=0.8pt] (0.32,-0.52) -- (0.62,-0.18);
  \fi
\end{tikzpicture}%
}

\newcommand{\LocalOperatorExpansionSketch}{%
\[
\resizebox{0.95\textwidth}{!}{$
\begin{aligned}
  \psi_{(A,B)\,||\,(C,D)}^{\mathrm{sc}}
  =
  \frac{1}{16\,q^2}\Big[&
  (k_A-k_B)(k_C-k_D)\,\LocalCollapseDiagram{0}{0}{0}{0}\\[0.4em]
  &-(k_A-k_B)\Big(
    \LocalCollapseDiagram{0}{0}{1}{0}
    -
    \LocalCollapseDiagram{0}{0}{0}{1}
  \Big)\\[0.4em]
  &-(k_C-k_D)\Big(
    \LocalCollapseDiagram{1}{0}{0}{0}
    -
    \LocalCollapseDiagram{0}{1}{0}{0}
  \Big)\\[0.4em]
  &+\Big(
    \LocalCollapseDiagram{1}{0}{1}{0}
    -
    \LocalCollapseDiagram{1}{0}{0}{1}
    -
    \LocalCollapseDiagram{0}{1}{1}{0}
    +
    \LocalCollapseDiagram{0}{1}{0}{1}
  \Big)
  \Big].
\end{aligned}
$}
\]
}

\newcommand{\LocalOperatorDoubleSketch}{%
\begin{tikzpicture}[x=1cm,y=1cm,every node/.style={font=\small}]
  \coordinate (L) at (-1.45,0);
  \coordinate (R) at ( 1.45,0);
  \coordinate (A) at (-3.15,-1.65);
  \coordinate (B) at (-3.10, 1.65);
  \coordinate (C) at ( 3.10, 1.65);
  \coordinate (D) at ( 3.15,-1.65);

  \draw[line width=0.9pt] (L) -- (A);
  \draw[line width=0.9pt] (L) -- (B);
  \draw[scalar] (L) -- (R);
  \draw[line width=0.9pt] (R) -- (C);
  \draw[line width=0.9pt] (R) -- (D);

  \fill (L) circle (1.6pt);
  \fill (R) circle (1.6pt);

  \node[draw,circle,fill=white,fill opacity=1,text opacity=1,inner sep=1.5pt,minimum size=18pt] at (A) {$A$};
  \node[draw,circle,fill=white,fill opacity=1,text opacity=1,inner sep=1.5pt,minimum size=18pt] at (B) {$B$};
  \node[draw,circle,fill=white,fill opacity=1,text opacity=1,inner sep=1.5pt,minimum size=18pt] at (C) {$C$};
  \node[draw,circle,fill=white,fill opacity=1,text opacity=1,inner sep=1.5pt,minimum size=18pt] at (D) {$D$};

  \node[above] at (0,0.16) {$q$};
\end{tikzpicture}%
}

\newcommand{\FivePointDoubleExchange}[2]{%
\begin{tikzpicture}[x=1cm,y=1cm,every node/.style={font=\small}]
  \coordinate (L) at (-2.2,0);
  \coordinate (M) at (0,0);
  \coordinate (R) at (2.2,0);

  \draw[gluon] (-3.8,-1.3) -- (L);
  \draw[gluon] (-3.8, 1.3) -- (L);
  \draw[#1] (L) -- (M);
  \draw[gluon] (M) -- (0,2.2);
  \draw[#2] (M) -- (R);
  \draw[gluon] (R) -- ( 3.8, 1.3);
  \draw[gluon] (R) -- ( 3.8,-1.3);

  \node[above] at (-1.10,0.18) {$\uk{12}$};
  \node[above] at ( 1.10,0.18) {$\uk{45}$};

  \node[left] at (-4.00,-1.30) {$1$};
  \node[left] at (-4.00, 1.30) {$2$};
  \node[above] at (0,2.35) {$3$};
  \node[right] at (4.00, 1.30) {$4$};
  \node[right] at (4.00,-1.30) {$5$};
\end{tikzpicture}%
}

\newcommand{\FivePointSingleExchange}{%
\begin{tikzpicture}[x=1cm,y=1cm,every node/.style={font=\small}]
  \coordinate (L) at (-2.1,0);
  \coordinate (C) at (1.0,0);

  \draw[gluon] (-3.7,-1.3) -- (L);
  \draw[gluon] (-3.7, 1.3) -- (L);
  \draw[gluon] (L) -- (C);
  \draw[gluon] (C) -- (0.8,2.2);
  \draw[gluon] (C) -- (3.6, 1.2);
  \draw[gluon] (C) -- (3.6,-1.2);

  \node[above] at (-0.55,0.18) {$\uk{12}$};

  \node[left] at (-3.90,-1.30) {$1$};
  \node[left] at (-3.90, 1.30) {$2$};
  \node[above] at (0.8,2.35) {$3$};
  \node[right] at (3.80, 1.20) {$4$};
  \node[right] at (3.80,-1.20) {$5$};
\end{tikzpicture}%
}

\newcommand{\SixPointChain}[3]{%
\begin{tikzpicture}[x=1cm,y=1cm,every node/.style={font=\small}]
  \coordinate (A) at (-3.0,0);
  \coordinate (B) at (-1.0,0);
  \coordinate (C) at ( 1.0,0);
  \coordinate (D) at ( 3.0,0);

  \draw[gluon] (A) -- (-4.5, 1.2);
  \draw[gluon] (A) -- (-4.5,-1.2);
  \draw[#1]    (A) -- (B);
  \draw[gluon] (B) -- (-1.0, 2.1);
  \draw[#2]    (B) -- (C);
  \draw[gluon] (C) -- ( 1.0, 2.1);
  \draw[#3]    (C) -- (D);
  \draw[gluon] (D) -- ( 4.5, 1.2);
  \draw[gluon] (D) -- ( 4.5,-1.2);

  \node[above] at (-2.0,0.16) {$\uk{12}$};
  \node[above] at ( 0.0,0.16) {$\uk{123}$};
  \node[above] at ( 2.0,0.16) {$\uk{56}$};

  \node[left]  at (-4.65, 1.2) {$2$};
  \node[left]  at (-4.65,-1.2) {$1$};
  \node[above] at (-1.0, 2.25) {$3$};
  \node[above] at ( 1.0, 2.25) {$4$};
  \node[right] at ( 4.65, 1.2) {$5$};
  \node[right] at ( 4.65,-1.2) {$6$};
\end{tikzpicture}%
}

\newcommand{\SixPointStar}[3]{%
\begin{tikzpicture}[x=1cm,y=1cm,every node/.style={font=\small}]
  \coordinate (O) at (0,0);
  \coordinate (L) at (-2.2,0);
  \coordinate (T) at (0,2.2);
  \coordinate (R) at (2.2,0);

  \draw[#1] (O) -- (L);
  \draw[#2] (O) -- (T);
  \draw[#3] (O) -- (R);

  \draw[gluon] (L) -- (-3.8, 1.1);
  \draw[gluon] (L) -- (-3.8,-1.1);
  \draw[gluon] (T) -- (-0.9, 3.6);
  \draw[gluon] (T) -- ( 0.9, 3.6);
  \draw[gluon] (R) -- ( 3.8, 1.1);
  \draw[gluon] (R) -- ( 3.8,-1.1);

  \node[above] at (-1.1,0.14) {$\uk{12}$};
  \node[right] at (0.12,1.15) {$\uk{34}$};
  \node[above] at ( 1.1,0.14) {$\uk{56}$};

  \node[left]  at (-3.95, 1.1) {$2$};
  \node[left]  at (-3.95,-1.1) {$1$};
  \node[above] at (-0.9, 3.78) {$3$};
  \node[above] at ( 0.9, 3.78) {$4$};
  \node[right] at ( 3.95, 1.1) {$5$};
  \node[right] at ( 3.95,-1.1) {$6$};
\end{tikzpicture}%
}

\newcommand{\SixPointStarFigureSixteen}{%
\begin{tikzpicture}[x=1cm,y=1cm,every node/.style={font=\small}]
  \coordinate (O) at (0,0);
  \coordinate (L) at (-2.2,0);
  \coordinate (T) at (0,2.2);
  \coordinate (R) at (2.2,0);

  \draw[gluon]  (O) -- (L);
  \draw[scalar] (O) -- (T);
  \draw[gluon]  (O) -- (R);

  \draw[gluon] (L) -- (-3.8, 1.1);
  \draw[gluon] (L) -- (-3.8,-1.1);
  \draw[gluon] (T) -- (-0.9, 3.6);
  \draw[gluon] (T) -- ( 0.9, 3.6);
  \draw[gluon] (R) -- ( 3.8, 1.1);
  \draw[gluon] (R) -- ( 3.8,-1.1);

  \node[above] at (-1.1,0.14) {$\uk{56}$};
  \node[left]  at (-0.12,1.15) {$\uk{12}$};
  \node[above] at ( 1.1,0.14) {$\uk{34}$};

  \node[left]  at (-3.95, 1.1) {$5$};
  \node[left]  at (-3.95,-1.1) {$6$};
  \node[above] at (-0.9, 3.78) {$1$};
  \node[above] at ( 0.9, 3.78) {$2$};
  \node[right] at ( 3.95, 1.1) {$3$};
  \node[right] at ( 3.95,-1.1) {$4$};
\end{tikzpicture}%
}

\newcommand{\SixPointQuarticEnd}[2]{%
\begin{tikzpicture}[x=1cm,y=1cm,every node/.style={font=\small}]
  \coordinate (A) at (-2.8,0);
  \coordinate (B) at (-0.6,0);
  \coordinate (Q) at (2.0,0);

  \draw[gluon] (A) -- (-4.3, 1.1);
  \draw[gluon] (A) -- (-4.3,-1.1);
  \draw[#1]    (A) -- (B);
  \draw[gluon] (B) -- (-0.6, 2.1);
  \draw[#2]    (B) -- (Q);
  \draw[gluon] (Q) -- ( 1.1, 2.6);
  \draw[gluon] (Q) -- ( 3.8, 1.1);
  \draw[gluon] (Q) -- ( 3.8,-1.1);

  \node[above] at (-1.7,0.14) {$\uk{12}$};
  \node[above] at ( 0.7,0.14) {$\uk{123}$};

  \node[left]  at (-4.45, 1.1) {$2$};
  \node[left]  at (-4.45,-1.1) {$1$};
  \node[above] at (-0.6, 2.25) {$3$};
  \node[above] at ( 1.1, 2.78) {$4$};
  \node[right] at ( 3.95, 1.1) {$5$};
  \node[right] at ( 3.95,-1.1) {$6$};
\end{tikzpicture}%
}

\newcommand{\SixPointQuarticMid}[2]{%
\begin{tikzpicture}[x=1cm,y=1cm,every node/.style={font=\small}]
  \coordinate (L) at (-2.4,0);
  \coordinate (Q) at (0,0);
  \coordinate (R) at (2.4,0);

  \draw[gluon] (L) -- (-3.9, 1.1);
  \draw[gluon] (L) -- (-3.9,-1.1);
  \draw[#1]    (L) -- (Q);
  \draw[gluon] (Q) -- (-0.9, 2.5);
  \draw[gluon] (Q) -- ( 0.9, 2.5);
  \draw[#2]    (Q) -- (R);
  \draw[gluon] (R) -- ( 3.9, 1.1);
  \draw[gluon] (R) -- ( 3.9,-1.1);

  \node[above] at (-1.2,0.14) {$\uk{12}$};
  \node[above] at ( 1.2,0.14) {$\uk{56}$};

  \node[left]  at (-4.05, 1.1) {$1$};
  \node[left]  at (-4.05,-1.1) {$2$};
  \node[above] at (-0.9, 2.68) {$3$};
  \node[above] at ( 0.9, 2.68) {$4$};
  \node[right] at ( 4.05, 1.1) {$5$};
  \node[right] at ( 4.05,-1.1) {$6$};
\end{tikzpicture}%
}

\newcommand{\SixPointQuarticPair}[1]{%
\begin{tikzpicture}[x=1cm,y=1cm,every node/.style={font=\small}]
  \coordinate (L) at (-2.2,0);
  \coordinate (R) at ( 2.2,0);

  \draw[gluon] (L) -- (-3.6, 1.6);
  \draw[gluon] (L) -- (-4.0, 0.0);
  \draw[gluon] (L) -- (-3.6,-1.6);
  \draw[#1]    (L) -- (R);
  \draw[gluon] (R) -- ( 3.6, 1.6);
  \draw[gluon] (R) -- ( 4.0, 0.0);
  \draw[gluon] (R) -- ( 3.6,-1.6);

  \node[above] at (0,0.14) {$\uk{123}$};

  \node[left]  at (-3.75, 1.6) {$1$};
  \node[left]  at (-4.15, 0.0) {$2$};
  \node[left]  at (-3.75,-1.6) {$3$};
  \node[right] at ( 3.75, 1.6) {$4$};
  \node[right] at ( 4.15, 0.0) {$5$};
  \node[right] at ( 3.75,-1.6) {$6$};
\end{tikzpicture}%
}

\newcommand{\DiagramFigure}[3]{%
\begin{figure}[H]
  \centering
  #1
  \caption{#2}
  \label{#3}
\end{figure}
}

\newcommand{\CompactDiagramCell}[2]{%
\begin{minipage}[t]{0.31\textwidth}
  \centering
  \resizebox{\linewidth}{!}{#1}

  \vspace{0.25em}
  {\footnotesize #2}
\end{minipage}
}

\newcommand{\CompactDiagramCellWide}[2]{%
\begin{minipage}[t]{0.48\textwidth}
  \centering
  \resizebox{0.92\linewidth}{!}{#1}

  \vspace{0.25em}
  {\footnotesize #2}
\end{minipage}
}

\usepackage[percent]{overpic}
\usepackage{multirow} 
\usepackage{slashed}

\title{From Cosmological Cuts to Yang--Mills Wavefunctions in de Sitter Space}

\author[1]{Song He (何颂)}
\emailAdd{songhe@itp.ac.cn}

\author[2]{Jiajie Mei (梅嘉杰)}
\emailAdd{j.mei@uva.nl}

\author[1]{Yuyu Mo (莫裕宇)}
\emailAdd{moyuyu@itp.ac.cn}

\affiliation[1]{New Cornerstone Laboratory, Institute of Theoretical Physics, Chinese Academy of Sciences, Beijing 100190, China}
\affiliation[2]{Institute of Physics, University of Amsterdam, Amsterdam, 1098 XH, The Netherlands}

\abstract{
We study tree-level Yang--Mills wavefunctions in four-dimensional de Sitter
space using their discontinuities.  Cosmological cuts factorize gluon
discontinuities into lower-point wavefunctions glued by cut propagators and
transverse projectors.  For ray-like trees and one-loop \(n\)-gons, the maximal
cuts take a particularly simple form: a scalar \(\phi^3\) discontinuity dressed
by an ordered Yang--Mills numerator built from local gluing maps.

We then use these cuts as reconstruction data for the four-, five-, and
six-gluon wavefunctions in momentum space.  The result separates into a
cut-detectable part obtained from lower-point gluing and a cut-invisible
completion fixed by current conservation and the flat-space limit.  Through six
points, the terms without longitudinal propagators follow the pole structure of
color-ordered scalar \(\phi^3+\phi^4\) wavefunctions, dressed by local
Yang--Mills numerators.  Longitudinal propagators collapse part of this scalar
structure into contact-type contributions, with the first internal-line
corrections appearing at six points.  The reconstructed expressions agree with
direct momentum-space Feynman-rule computations and give concrete low-point
data for an all-\(n\) organization of spinning de Sitter wavefunctions.}

\keywords{de Sitter space, cosmological correlators, scattering amplitudes,
Yang--Mills theory}

\begin{document}
\begin{CJK*}{UTF8}{gbsn}
\maketitle
\end{CJK*}
\addtocontents{toc}{\protect\setcounter{tocdepth}{2}}

\numberwithin{equation}{section}

		\tikzset{
		particles/.style={dashed, postaction={decorate},
			decoration={markings,mark=at position .5 with {\arrow[scale=1.5]{>}}
		}}
	}
	\tikzset{
		particle/.style={draw=black, postaction={decorate},
			decoration={markings,mark=at position .5 with {\arrow[scale=1.1]{>}}
		}}
	}
	\def  \layersep {.6cm}
 
\section{Introduction}


Cosmological unitarity gives direct access to the singularity structure of the
Bunch--Davies wavefunction and related observables.  A central result is the
cosmological optical theorem, which relates wavefunction coefficients to their
Hermitian-analytic images
\cite{Goodhew:2020hob,Melville:2021lst,Baumann:2021fxj,Goodhew:2021oqg}.
For exchange diagrams, these relations expose the same physical information
that one would like to use recursively: a singularity in an internal energy
factorizes the observable into lower-point data.

This factorization naturally supports reconstruction approaches to cosmological
observables.
Locality and unitarity restrict the allowed total- and partial-energy
singularities, and in favorable cases these singularities reconstruct rational
wavefunction or correlator data from lower-point or flat-space input
\cite{Baumann:2020dch,Pajer:2020wxk,Jazayeri:2021fvk}.  Discontinuities also
enter directly in recent single-cut, Schwinger--Keldysh, dispersive,
physical-cut-basis, and dressing-rule approaches to cosmological correlators
\cite{Tong:2021wai,Ema:2024hkj,Das:2025qsh,Colipi-Marchant:2025oin,
Chowdhury:2026upp,Das:2026vfv,Ansari:2026xkm,De:2024zic,Liu:2024xyi}.
In this paper we use such singularity data constructively: the cuts are not
only checks on an answer, but input for reconstructing the wavefunction.

For spinning fields, the general cutting rules are known, but explicit
high-point implementations are still scarce.  General bosonic formulas imply
tree-level factorization for exchanges of arbitrary integer spin
\cite{Melville:2021lst,Goodhew:2021oqg}; the practical problem is to turn this
factorization into usable reconstruction rules.  Yang--Mills theory is a
useful test case because nontrivial tensor structures already appear at low
multiplicity.  Related bootstrap constructions,
factorization analyses, AdS
recursion relations, soft theorems, and amplitude-inspired representations in
(A)dS provide complementary ways of organizing the same class of
observables
{ \cite{Armstrong:2022mfr,Chowdhury:2024wwe,Mei:2025jko,He:2024olr,
Chen:2025foq,Mei:2023jkb,Gomez:2026yno,
Raju:2010by,Raju:2011mp,Raju:2012zr,Raju:2012zs}}.

There is also a useful analogy with flat-space amplitudes.  In flat-space
scattering, sufficiently sharp cuts isolate rigid data: generalized cuts and
leading singularities constrain loop integrands, while on-shell diagrams
organize such residues in terms of cells of the positive Grassmannian
\cite{Bern:1994zx,Britto:2004nc,Cachazo:2008hp,
Arkani-Hamed:2010zjl,
Arkani-Hamed:2010pyv,
Arkani-Hamed:2012zlh}.  This viewpoint is intertwined with positive-geometric
structures such as the amplituhedron, positive geometries and canonical forms,
kinematic associahedra and scattering forms, and more recent surface-based
organizations of amplitudes
\cite{Arkani-Hamed:2013jha,
Arkani-Hamed:2017tmz,
Arkani-Hamed:2017mur,
Arkani-Hamed:2023swr,
Arkani-Hamed:2023jry,
Arkani-Hamed:2024nhp,
Arkani-Hamed:2024tzl,
Cao:2025lzv}.  Cosmological observables have their own geometric and
combinatorial structures, including cosmological polytopes, scattering facets,
Steinmann-type compatibility constraints, cosmohedra, and correlator polytopes
\cite{Arkani-Hamed:2017fdk,
Arkani-Hamed:2018bjr,
Benincasa:2020aoj,
Benincasa:2024leu,
Arkani-Hamed:2024jbp,
Figueiredo:2025daa}.  The analogy is not literal, because cosmological
wavefunctions have a different analytic structure from flat-space amplitudes
\cite{Salcedo:2022aal,
Lee:2023kno,
Albayrak:2023hie}.  Still, it suggests a concrete question: how much of a
spinning cosmological wavefunction is already fixed by the most restrictive
discontinuities?

The aim of this paper is to make the discontinuity-based reconstruction of
tree-level Yang--Mills de Sitter wavefunctions explicit in momentum space.
Closely related ideas have been used to bootstrap four- and five-point gluon
and graviton wavefunctions in Mellin-momentum space
\cite{Mei:2024abu,Mei:2024sqz}.  Here we formulate the gluing rule for gluon
discontinuities, turn it into a reconstruction procedure, and work out the
four-, five-, and six-point wavefunction coefficients in momentum space.
The maximal discontinuities play the role of cosmological leading
singularities, while lower-codimension cuts supply the remaining gluing data.
The explicit answers separate into a cut-detectable part and a smaller
cut-invisible completion fixed by current conservation, and the flat-space limit.
We check the reconstruction against direct Yang--Mills momentum-space Feynman
rules and use the low-point results to identify the first ingredients of an
all-\(n\) organization.  Section~\ref{gluon_reconstruction} gives the
reconstruction strategy; the four-, five-, and six-point cases are then treated
in Sections~\ref{subsec:four-point-reconstruction}--\ref{subsec:six-point-reconstruction},
before the all-\(n\) pattern is summarized in
Section~\ref{subsec:all-n-structure}.

\textbf{Notation and conventions}

Throughout the paper we write
\[
  k_a:=|\bk_a|
\]
for the external energies.  When a subset of external momenta is separated by
an internal channel \(I\), we denote the corresponding exchanged spatial
momentum and energy by
\[
  \bk_I := \sum_{a\in I}\bk_a,
  \qquad
  k_I := |\bk_I| .
\]
For ordered planar channels it is useful to use the shorthand
\[
  x_{ij}:=\left|\bk_i+\bk_{i+1}+\cdots+\bk_{j-1}\right|=\uk{i\,i+1\cdots j-1},
  \qquad i<j,
\]
so that, for example, \(x_{13}=|\bk_1+\bk_2|\).  We also denote the
\(n\)-point total energy by
\[
  E_n:=\sum_{a=1}^{n}k_a=k_1+k_2+\cdots+k_n .
\]

We work in \(\dS_4\), with metric
\begin{equation}
  ds_{\dS}^2
  =
  \frac{\ell_{\dS}^2}{\eta^2}
  \left(
    -d\eta^2+\delta_{ij}dx^i dx^j
  \right).
  \label{eq:ds-metric}
\end{equation}
Here \(\eta\in(-\infty,\eta_\ast)\) is conformal time, and
\(\eta_\ast\to0^-\) is the late-time cutoff at which the wavefunction is
evaluated.  We set \(H=1/\ell_{\dS}=1\) in the following.

In practice, it is convenient to compute the corresponding diagrams in
Euclidean \(\AdS_4\),
\[
  ds_{\mathrm{EAdS}}^2
  =
  \frac{\ell_{\AdS}^2}{z^2}
  \left(
    dz^2+\delta_{ij}dx^i dx^j
  \right),
\]
and then analytically continue
\[
  z=-i\eta,\qquad
  z_\ast=-i\eta_\ast,\qquad
  \ell_{\AdS}^2=-\ell_{\dS}^2 .
\]
Equivalently, one may choose the branch
\(\ell_{\AdS}=-i\ell_{\dS}\).  With these continuations the Euclidean
\(\AdS\) metric maps to the Lorentzian \(\dS\) metric above
\cite{Maldacena:2002vr,Anninos:2014lwa,Bzowski:2023nef}.
\section{Review of wavefunction coefficients and cutting rules}
\label{section2}

This section fixes the conventions used in the reconstruction.  We first
recall the late-time Bunch--Davies wavefunction coefficients, then review the
energy-sign-flip discontinuity for tree-level rational wavefunctions.  We
finally specialize the cutting rule to gluons, where the sum over exchanged
physical states is implemented by a transverse projector.

\subsection{Wavefunction coefficients}

Let \(\Phi(\eta,\xx)\) denote a bulk field in \(\dS_4\), and let
\(\varphi(\xx)\) be its late-time boundary value,
\begin{equation}
  \varphi(\xx)=\Phi(\eta_\ast,\xx),
  \qquad
  \eta_\ast\to 0^- .
  \label{eq:boundary-field}
\end{equation}
We work with the late-time wavefunction \(\Psi[\varphi]\), viewed as a
functional of the boundary configuration \(\varphi(\xx)\):
\begin{equation}
  \Psi[\varphi]
  =
  \langle \varphi,\eta_\ast|\,
  U\!\left(\eta_\ast,-\infty\right)
  |\Omega_{\rm BD}\rangle ,
  \label{eq:wavefunction-operator}
\end{equation}
where \(|\varphi,\eta_\ast\rangle\) is the field eigenstate at the late-time
slice, \(|\Omega_{\rm BD}\rangle\) is the Bunch--Davies initial state, and
\(U\) is the time-evolution operator from the deformed past contour to
\(\eta_\ast\).  In the path-integral representation this becomes
\begin{equation}
  \Psi[\varphi]
  =
  \int_{\Phi(\eta\to-\infty,\xx)=0}
       ^{\Phi(\eta_\ast,\xx)=\varphi(\xx)}
  \calD\Phi\,
  \exp\!\left(\ii S[\Phi]\right).
  \label{eq:wavefunction-path-integral}
\end{equation}
Here the lower endpoint of the time contour is deformed as
\(-\infty \to -\infty(1-\ii\epsilon)\).  This is the standard
prescription implementing the Bunch--Davies initial condition in the
asymptotic past
\cite{Maldacena:2002vr,Weinberg:2005vy,Goodhew:2020hob,Melville:2021lst,Salcedo:2022aal}.

From now on, we work in momentum space and allow for a collection of boundary
fields \(\varphi_\alpha(\bk)\), where the label \(\alpha\) denotes the field
species and, when appropriate, any discrete external-state labels such as
color or helicity.  Since the wavefunction is a functional of these boundary
fields, it admits the perturbative functional Taylor expansion
\begin{align}
  \Psi[\varphi_\alpha]
  =
  \exp \Bigg[
  -\sum_{n=2}^{\infty}
  \frac{1}{n!}
  \sum_{\alpha_1,\ldots,\alpha_n}
  \int
  \prod_{a=1}^{n}
  \frac{\dd^3\bk_a}{(2\pi)^3}\,
  (2\pi)^3
  \delta^{(3)}
  \!\left(\sum_{a=1}^{n}\bk_a\right)
  \psi_n^{\alpha_1\cdots\alpha_n}(\bk_1,\ldots,\bk_n)
  \prod_{a=1}^{n}\varphi_{\alpha_a}(\bk_a)
  \Bigg].
  \label{eq:wavefunction-momentum}
\end{align}
Here \(\varphi_\alpha(\bk)\) is the Fourier transform of the boundary field,
and \(\psi_n^{\alpha_1\cdots\alpha_n}\) is the
momentum-conserving-delta-function-stripped \(n\)-point coefficient, evaluated
on \(\sum_{a=1}^n \bk_a=0\).  These coefficients are constrained by the
conformal symmetry of the late-time boundary and can be computed by analytic
continuation from AdS Witten diagrams
\cite{Maldacena:2011nz,Baumann:2020dch,Anninos:2014lwa,Bzowski:2023nef,
Maldacena:2002vr,McFadden:2009fg,Sleight:2021plv}.  For gluons, one also sums
over color and helicity states.

The corresponding color-dressed coefficient can be decomposed in a trace
basis as
\begin{equation}
  \psi_n^{a_1\cdots a_n;\,h_1\cdots h_n}
  (\bk_1,\ldots,\bk_n)
  =
  \sum_{\sigma\in S_n/\mathbb Z_n}
  \Tr\!\left(
    T^{a_{\sigma(1)}}\cdots T^{a_{\sigma(n)}}
  \right)
  \psi_n\!\left[
    \sigma(1)^{h_{\sigma(1)}},\ldots,
    \sigma(n)^{h_{\sigma(n)}}
  \right].
  \label{eq:gluon-trace-basis}
\end{equation}
Here \(\psi_n[\sigma(1)^{h_{\sigma(1)}},\ldots,
\sigma(n)^{h_{\sigma(n)}}]\) is the color-ordered gluon wavefunction
coefficient.  It no longer carries explicit color indices, but
depends on the ordering, helicities, momenta, and energies.

In this work we focus on the color-ordered, polarization-contracted,
transverse part of the gluon wavefunction coefficient.  The helicity of each
external gluon is carried by its polarization vector, and we write
\(\beps_a\equiv\beps_a^{h_a}(\bk_a)\) while leaving the helicity labels
implicit throughout:
\begin{equation}
  \psi_n^{\YM}
  \bigl(
    \{\beps_a,\bk_a,k_a\}_{a=1}^{n}
  \bigr)
  :=
  \beps_1^{i_1}\cdots \beps_n^{i_n}\,
  \psi^{\YM}_{n;i_1\cdots i_n}
  \bigl(
    \bk_1,\ldots,\bk_n
  \bigr),
  \label{eq:ym-contracted-wavefunction}
\end{equation}
The external polarizations satisfy
\[
  \beps_a\cdot\bk_a=0,
  \qquad a=1,\ldots,n .
\]
This transverse, polarization-contracted object is the basic quantity whose
discontinuities we study and whose full form we later reconstruct.  The
longitudinal components of the full Yang--Mills wavefunction coefficient are
fixed by the transverse Ward identities 
\cite{Bzowski:2013sza,Bzowski:2023nef,Armstrong:2020woi}.

\subsection{Cutting rules and discontinuities}
The cosmological optical theorem constrains Hermitian-analytic
discontinuities of wavefunction coefficients.  The associated cutting rules
relate these discontinuities, defined by subtracting the corresponding
Hermitian-analytic image, to products of lower-point data
\cite{Pajer:2020wxk,Goodhew:2020hob,Melville:2021lst,
Jazayeri:2021fvk,Goodhew:2021oqg}.

We will use the same discontinuities as reconstruction data.  The basic
mechanism is already visible at the level of a single internal
bulk-to-bulk propagator: its discontinuity separates an exchange diagram into
lower-point wavefunction or correlator data.  Conversely, such factorized data
can be used as input for dispersive or cut-based reconstruction of full
wavefunctions and correlators
\cite{Meltzer:2021zin,Baumann:2021fxj,Werth:2024mjg,
Liu:2024xyi,Das:2025qsh,Colipi-Marchant:2025oin,Das:2026vfv}.  In this paper
we focus on wavefunction coefficients.

It is convenient to record the spectral representation here:
\begin{equation}
G_{\Delta}(k,z_1,z_2)
=
-i\int_{0}^{\infty}\frac{dp}{2\pi i}\,
\frac{p^{d+1-2\Delta}}{k^2+p^2}
\bigl(\phi_{\Delta}(z_1,ip)-\phi_{\Delta}(z_1,-ip)\bigr)
\bigl(\phi_{\Delta}(z_2,ip)-\phi_{\Delta}(z_2,-ip)\bigr),
\label{eq:btb-spectral-representation}
\end{equation}
where
$\phi_{\Delta}(z,k_I)$ is the corresponding bulk-to-boundary mode.

At tree level, the non-analytic dependence on an internal exchange energy
\(k_I\) comes from the corresponding bulk-to-bulk propagator.  The
factorization across the channel \(I\) is therefore the discontinuity of this
Green function.  Using the spectral representation gives\footnote{This follows
from the standard
distributional identity
\[
  \frac{1}{k^2+p^2+i0}
  -
  \frac{1}{k^2+p^2-i0}
  =
  -2\pi i\,\delta(k^2+p^2),
\]
which localizes the spectral integral in \eqref{eq:btb-spectral-representation}.
Here \(k^2\) should be understood as a complex variable.}
\begin{equation}
\Disc_{k_I^2} G_{\Delta}(z_1,z_2;k_I)
 =
  -P_{k_I}\,
  \bigl[
    \phi_{\Delta}(z_1,k_I)-\phi_{\Delta}(z_1,-k_I)
  \bigr]
  \bigl[
    \phi_{\Delta}(z_2,k_I)-\phi_{\Delta}(z_2,-k_I)
  \bigr] .
  \label{eq:btb-propagator-disc}
\end{equation}
Here \(P_{k_I}\) denotes the power spectrum,
$
  P_{k_I}
  =
  \frac{1}{2}k_I^{d-2\Delta}.
$
Thus the discontinuity of the bulk-to-bulk propagator factorizes into two
bulk-to-boundary differences times the power spectrum.

Diagrammatically, cutting the internal line in channel \(I\) separates the full
wavefunction into left and right sub-wavefunction coefficients:
\begin{equation}
  \Disc_{k_I^2}\psi_n
  =
 - (\psi_L(k_I)-\psi_L(-k_I))\;
  P_I\;
( \psi_R(k_I)-\psi_R(-k_I) ).
  \label{eq:cutting-rule}
\end{equation}
For the rational functions considered here, the discontinuity of the full
wavefunction is given by the difference\footnote{This comes from the fact that
\(k_I=\sqrt{k_I^2}\) and
\(\Disc_{k_I^2}f(k_I^2)=f({k_I^2})-f({e^{2 i \pi}k_I^2})\).}
\begin{equation}
\Disc_{k_I^2}F(k_I)=F(k_I)-F(-k_I).
\label{signflip}
\end{equation}
In the rest of the text, when no confusion can arise,
we write $\Disc_{k_I^2}F(k_I):=\Disc_{k_I}F(k_I)$.

It is also useful to define iterated discontinuities.  Let
$S=\{I_1,\ldots,I_m\}$
be a set of mutually compatible channels.  We define
\begin{equation}
  \Disc_S
  :=
  \prod_{I\in S}\Disc_{k_I}.
  \label{eq:iterated-disc-review}
\end{equation}
Compatibility means that the chosen channels can be cut simultaneously and
nontrivially on the same tree graph.  Repeated use of
\eqref{eq:cutting-rule} then decomposes the
original wavefunction coefficient into lower-point building blocks glued along
all channels in \(S\).  In particular, maximal compatible cuts, where every
internal channel of a given tree topology is cut, localize the answer as much
as possible and will play the role of cosmological analogues of leading
singularities in the discussion below.

\subsubsection{Flat-space massless trace $\phi^3$ maximal discontinuities}
\label{Flatspace_scalar_maximal_disc}
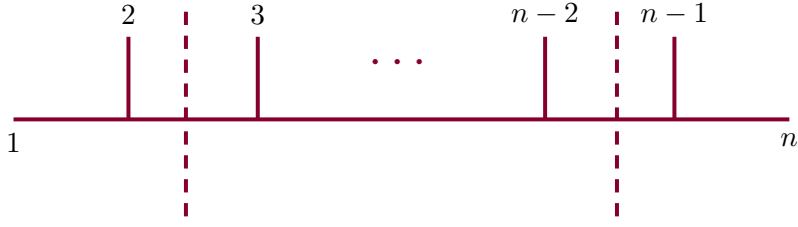
\begin{figure}[H]
\centering
\begin{tikzpicture}[x=1cm,y=1cm,scale=0.95]
  \def\laddercolor{purple!70!black}
  \draw[\laddercolor,line width=1.5pt]
    (-5.4,0) -- (5.4,0);
  \node[below] at (-5.4,-0.05) {$1$};
  \node[below] at (5.4,-0.05) {$n$};
  \foreach \x/\lab in {-3.8/2,-2.0/3,2.0/{n-2},3.8/{n-1}} {
    \draw[\laddercolor,line width=1.5pt]
      (\x,0) -- ++(0,1.15);
    \node[above] at (\x,1.18) {$\lab$};
  }
  \foreach \x in {-3.0,3.0} {
    \draw[\laddercolor,line width=1.5pt,dashed,dash pattern=on 5pt off 4pt]
      (\x,-1.35) -- ++(0,2.9);
  }
  \node[\laddercolor,scale=1.7] at (0,0.78) {$\cdots$};
\end{tikzpicture}
\caption{Maximal discontinuity on a half ladder.}
\label{fig:half-ladder-max-disc}
\end{figure}

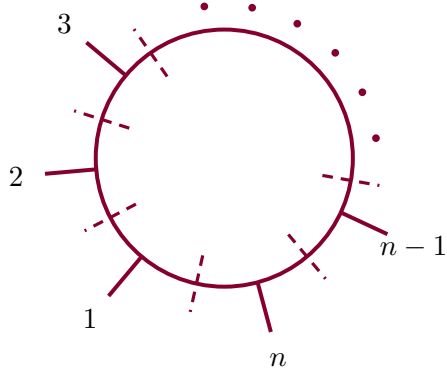
\begin{figure}[H]
\centering
\begin{tikzpicture}[x=1cm,y=1cm,scale=0.85]
  \def\ngoncolor{purple!70!black}
  \draw[\ngoncolor,line width=1.5pt] (0,0) circle (2.0);
  \foreach \ang in {-152.5,162.5,125,-10,-50,-102.5} {
    \draw[\ngoncolor,line width=1.2pt,dashed,dash pattern=on 4pt off 3pt]
      ({1.55*cos(\ang)},{1.55*sin(\ang)})
      -- ({2.45*cos(\ang)},{2.45*sin(\ang)});
  }
  \foreach \ang/\lab in {230/1,185/2,140/3,-25/{n-1},-75/n} {
    \draw[\ngoncolor,line width=1.5pt]
      ({2*cos(\ang)},{2*sin(\ang)}) -- ({2.8*cos(\ang)},{2.8*sin(\ang)});
    \node at ({3.25*cos(\ang)},{3.25*sin(\ang)}) {$\lab$};
  }
  \foreach \ang in {97.5,79.5,61.5,43.5,25.5,7.5} {
    \fill[\ngoncolor] ({2.38*cos(\ang)},{2.38*sin(\ang)}) circle (1.7pt);
  }
\end{tikzpicture}
\caption{Maximal cut of one-loop \(n\)-gon.}
\label{fig:one-loop-ngon-max-cut}
\end{figure}

As a simple toy model, consider the flat-space massless trace \(\phi^3\)
wavefunction.
We start with the scalar cubic three-point building block
\begin{equation}
 \psi_{3}^{\trphi}(k_1,k_2,k_3)
  =
  C_3(k_1,k_2,k_3)
  :=
  \frac{1}{k_1+k_2+k_3},
  \label{eq:scalar-cubic-block}
\end{equation}
where \(k_1,k_2,k_3\) are energies.  If one of these arguments is a cut
energy, the discontinuity is implemented by the sign flip defined in
\eqref{signflip}.  For example,
\begin{align}
  \Disc_{k_1} C_3(k_1,k_2,k_3)
  &=
  C_3(k_1,k_2,k_3)-C_3(-k_1,k_2,k_3)
  \nonumber\\
  &=
  -\frac{2k_1}{(k_1+k_2+k_3)(-k_1+k_2+k_3)} .
  \label{eq:single-disc-cubic-block}
\end{align}
Similarly, the double discontinuity in two energy slots is
\begin{align}
 & \Disc_{k_1}\Disc_{k_3} C_3(k_1,k_2,k_3)\\
  =&
  C_3(k_1,k_2,k_3)
  -
  C_3(k_1,k_2,-k_3)
  -
  C_3(-k_1,k_2,k_3)
  +
  C_3(-k_1,k_2,-k_3)
  \nonumber\\
  =&
  \frac{8k_1k_2k_3}
  {\bigl((k_1+k_2)^2-k_3^2\bigr)\bigl((k_2-k_1)^2-k_3^2\bigr)} .
  \label{eq:double-disc-cubic-block}
\end{align}
These two elementary expressions are the only scalar ingredients needed for
the maximal cuts considered below.  The tree half-ladder assignment is
illustrated in Fig.~\ref{fig:half-ladder-max-disc}: endpoint vertices carry one
cut channel and therefore one discontinuity, while internal vertices carry two
cut channels and therefore a double discontinuity.  In a one-loop
\(n\)-gon, every cubic vertex is adjacent to two cut loop lines and therefore
also carries a double discontinuity, as illustrated in
Fig.~\ref{fig:one-loop-ngon-max-cut}.

\paragraph{Tree-level half-ladder.}

Consider the color ordering \((1,2,\ldots,n)\).  In the notation introduced
in the introduction, the internal channels of the half-ladder are
\[
  x_{13},x_{14},\ldots,x_{1,n-1}.
\]
The maximal discontinuity for this
topology is therefore
\begin{equation}
  \Disc_{\rm max}
  :=
  \prod_{m=2}^{n-2}
  \Disc_{x_{1,m+1}} .
  \label{eq:half-ladder-max-disc-operator}
\end{equation}

Since each internal bulk-to-bulk propagator is cut according to
\eqref{eq:cutting-rule}, applying the maximal discontinuity operator
\eqref{eq:half-ladder-max-disc-operator} makes the scalar cubic half-ladder
factorize vertex by vertex:
\begin{align}
  &\Disc_{\rm max}
  \psi^{\phi^3}_{n,\mathrm{HL}}
  =
  \left(
    \prod_{m=2}^{n-2}
    \frac{1}{2x_{1,m+1}}
  \right)
  \Big[
    \Disc_{x_{13}}
    C_3(k_1,k_2,x_{13})
  \Big]
  \nonumber\\
  &\times
  \left(
    \prod_{m=3}^{n-2}
    \Disc_{x_{1m}}\Disc_{x_{1,m+1}}
    C_3(x_{1m},k_m,x_{1,m+1})
  \right)
  \Big[
    \Disc_{x_{1,n-1}}
    C_3(x_{1,n-1},k_{n-1},k_n)
  \Big],
  \label{eq:scalar-half-ladder-max-disc}
\end{align}
where we have used the flat-space massless power spectrum \(P_I=1/(2k_I)\).
The product over \(m=3,\ldots,n-2\) is understood to be absent when the range is
empty.  The first and last brackets correspond to the two endpoint vertices of
the half-ladder, while the middle product corresponds to the internal cubic
vertices.  Using \eqref{eq:single-disc-cubic-block} and
\eqref{eq:double-disc-cubic-block}, this expression becomes a local rational
function of the external energies and the half-ladder channel variables
\(x_{13},\ldots,x_{1,n-1}\).

\paragraph{One-loop \(n\)-gon.}

For the one-loop \(n\)-gon we need loop-line energies, which are not among the
planar tree-channel variables \(x_{ij}\).  Let \(\boldsymbol\ell\) be the loop
spatial momentum and define
\begin{equation}
  q_0(\boldsymbol\ell):=|\boldsymbol\ell|,
  \qquad
  q_a(\boldsymbol\ell)
  :=
  \left|
    \boldsymbol\ell+\bk_1+\cdots+\bk_a
  \right|,
  \qquad
  a=1,\ldots,n .
  \label{eq:loop-line-energies}
\end{equation}
Momentum conservation gives \(q_n(\boldsymbol\ell)=q_0(\boldsymbol\ell)\).
Below we suppress the explicit \(\boldsymbol\ell\)-dependence and simply write
\(q_a\).  The \(a\)-th cubic vertex is adjacent to the two cut loop energies
\(q_{a-1}\) and \(q_a\), with the cyclic identification \(q_n\equiv q_0\).

The scalar maximal-discontinuity integrand is then
\begin{align}
  \Disc_{\rm max}
  \mathcal I^{\phi^3}_{n,n\text{-}\mathrm{gon}}
  (\boldsymbol\ell)
  =
  \left(
    \prod_{a=0}^{n-1}
    \frac{1}{2q_a}
  \right)
  \left(
    \prod_{a=1}^{n}
    \Disc_{q_{a-1}}\Disc_{q_a}
    C_3(q_{a-1},k_a,q_a)
  \right),
  \qquad
  q_n\equiv q_0 .
  \label{eq:scalar-ngon-max-disc-integrand}
\end{align}
The full scalar one-loop maximal discontinuity is obtained by integrating over
the loop spatial momentum,
\begin{equation}
  \Disc_{\rm max}
  \psi^{\phi^3}_{n,n\text{-}\mathrm{gon}}
  =
  \int
  \frac{\dd^3\boldsymbol\ell}{(2\pi)^3}\,
  \Disc_{\rm max}
  \mathcal I^{\phi^3}_{n,n\text{-}\mathrm{gon}}
  (\boldsymbol\ell).
  \label{eq:scalar-ngon-max-disc}
\end{equation}
These scalar expressions will serve as the energy factors that are dressed by
Yang--Mills tensor numerators in the ray-like tree and one-loop \(n\)-gon
examples below.

\subsection{Discontinuities of gluon wavefunction coefficients}
\label{subsec:gluon-discontinuities}

For the reader's convenience, we record the Yang--Mills bulk-to-boundary and
bulk-to-bulk propagators here.  The bulk-to-boundary propagator is
\begin{equation}
\label{eq:ym-btb-propagator}
\phi^i(z,k)\equiv\beps^i \phi(z,k)=\beps^i e^{-zk},
\end{equation}
while the transverse bulk-to-bulk propagator is written in spectral form as
\begin{equation}
\label{eq:ym-transverse-btb-spectral}
\begin{aligned}
G^\perp_{ij}(z_1,z_2;k_I)
&=
-\frac{1}{4}\Pi_{ij}^{(\bk_I)}
\int_{-\infty}^{\infty}\frac{dp}{\pi}\,
\frac{(\phi(z_1,ip)-\phi(z_1,-ip))(\phi(z_2,ip)-\phi(z_2,-ip))}
{k_I^2+p^2},
\end{aligned}
\end{equation}
where
$
  \Pi_{ij}^{(\bk_I)}
  =
  \delta_{ij}
  -\frac{\bk_I^i \bk_I^j}{k_I^2}.
$
The longitudinal part is analytic in \(k_I\) and does not contribute to the
discontinuity considered in this subsection.
The scalar part appearing in
\eqref{eq:ym-btb-propagator} and
\eqref{eq:ym-transverse-btb-spectral} is the flat-space massless scalar
bulk-to-boundary and bulk-to-bulk propagator \cite{Arkani-Hamed:2017fdk}.  This
follows from the conformally coupled nature of Yang--Mills theory in (A)dS.

For Yang--Mills wavefunction coefficients, the scalar discontinuity is dressed
by the tensor factor in the propagator.  The transverse projector
\(\Pi^{ij}(\bk_I)\) in \eqref{eq:ym-transverse-btb-spectral} is analytic in
\(k_I\), so it factors out of the energy discontinuity.  It is also the
helicity sum over the two cut gluon polarizations,
$
  \sum_h
  \epsilon_i^{(h)}(\bk_I)\,
  \epsilon_j^{(h)\,*}(\bk_I)
  =
  \Pi_{ij}(\bk_I).
$
The right-hand side of \eqref{eq:btb-propagator-disc} is therefore replaced by
energy-flipped differences of gluon bulk-to-boundary propagators\footnote{The polarization vector \(\epsilon^{(h)}(\bk)\) does not change
under the energy sign flip \(k\to -k\).  The polarization depends on
the spatial momentum \(\bk\), not on the sign choice of the energy.  See
\cite{Goodhew:2021oqg}.}
(cf. \eqref{eq:ym-btb-propagator}). 

The single-cut rule becomes
\begin{align}
  \Disc_{k_I}\psi_n^{\YM}
  =
  P^{\YM}(k_I)\,
  \Bigg[
    \frac{\partial}{\partial \epsilon_I^i}
    \Disc_{k_I}
    \psi_L^{\YM}
    \bigl(
      \epsilon_I,-\bk_I,k_I
    \bigr)
  \Bigg]
  \Pi^{ij}(\bk_I)
  \Bigg[
    \frac{\partial}{\partial \tilde\epsilon_I^j}
    \Disc_{k_I}
    \psi_R^{\YM}
    \bigl(
      \tilde\epsilon_I,\bk_I,k_I
    \bigr)
  \Bigg] .
  \label{eq:ym-single-cut}
\end{align}
Here \(P^{\YM}(k_I)=1/(2k_I)\) is the Yang--Mills power spectrum on the cut
line, and \(\Pi^{ij}(\bk_I)\) performs the helicity sum.  The discontinuity
\(\Disc_{k_I}\) acts only on the scalar energy dependence, not on the tensor
structure.
Differentiating with respect to the auxiliary polarizations,
$
  \frac{\partial}{\partial \epsilon_I^i},
\; 
  \frac{\partial}{\partial \tilde\epsilon_I^j},
$
strips off the polarization of the lower-point wavefunction coefficient and
exposes the vector index to be glued by the projector.  Thus the discontinuity
in channel \(I\) factorizes into two lower-point discontinuities connected by
the physical polarization sum of the exchanged gluon.

For a compatible set of channels \(S\), the iterated Yang--Mills cut
is obtained by applying \eqref{eq:ym-single-cut} to every channel in \(S\).
Let \(\mathcal C(\mathcal T/S)\) denote the set of connected components
obtained from the tree topology \(\mathcal T\) after cutting all edges in
\(S\).  For a component \(C\in\mathcal C(\mathcal T/S)\), we define
\[
  \mathrm{Inc}(C)
  :=
  \{\, e\in S \;|\; e \text{ is incident on } C \,\}.
\]
Thus \(\mathrm{Inc}(C)\) is the set of cut lines attached to the lower-point
wavefunction coefficient associated with \(C\).  For each incident cut edge
\(e\in \mathrm{Inc}(C)\), we introduce an auxiliary polarization
\(\epsilon_{e,C}\) on the corresponding cut leg.  Differentiating with respect
to these auxiliary polarizations opens the vector indices that will be glued
to the neighboring components.

With this notation, repeated use of the single-cut rule gives the
multi-cut formula
\begin{align}
  \Disc_S \psi_n^{\YM}
  =
  \left[
    \prod_{e\in S}
    P^{\YM}(k_e)
  \right]
  \left[
    \prod_{e=(C,C')\in S}
    \Pi^{\,i_{e,C}i_{e,C'}}(\bk_e)
  \right]
  \prod_{C\in\mathcal C(\mathcal T/S)}
  \left[
    \left(
      \prod_{e\in \mathrm{Inc}(C)}
      \frac{\partial}{\partial \epsilon_{e,C}^{\,i_{e,C}}}
    \right)
    \Disc_{\mathrm{Inc}(C)}
    \psi_C^{\YM}
  \right] .
  \label{eq:ym-compatible-cut}
\end{align}
Here \(e=(C,C')\) means that the cut edge \(e\) connects the two components
\(C\) and \(C'\).  The indices \(i_{e,C}\) and \(i_{e,C'}\) are the open vector
indices exposed on the two sides of the cut, and they are contracted by the
transverse projector \(\Pi^{\,i_{e,C}i_{e,C'}}(\bk_e)\).  The notation
\(\Disc_{\mathrm{Inc}(C)}\) means the product of discontinuities in the cut
energies carried by the edges incident on \(C\),
\[
  \Disc_{\mathrm{Inc}(C)}
  :=
  \prod_{e\in\mathrm{Inc}(C)}\Disc_{k_e}.
\]

This formula separates the scalar and spinning data.  The cut propagators and
discontinuities determine the energy dependence; the tensor structure is built
from lower-point Yang--Mills tensors and transverse projectors.  We now turn to
maximal compatible cuts, where every internal channel of the chosen topology is
cut.

\subsubsection{Gluon maximal discontinuities}

\label{subsec:raylike-maxdisc}

Among compatible cuts, maximal cuts are the most restrictive: every internal
channel of a chosen topology is cut, leaving a local gluing problem.  For the
ray-like tree (half-ladder) and its one-loop \(n\)-gon analogue, this gives
compact closed-form expressions while retaining nontrivial tensor structure.

We begin with the three-point wavefunction coefficient,
\begin{equation}
\psi_3(\beps_1,\beps_2,\beps_3;\bk_1,\bk_2,\bk_3;k_1,k_2,k_3)
=
V_{123}\,C_3(k_1,k_2,k_3).
\label{eq:ym-three-point-wavefunction}
\end{equation}
Here the tensor structure is carried by \(V_{123}\), while the scalar part is
encoded in the same cubic block \(C_3\) used in the scalar example
in Section \ref{Flatspace_scalar_maximal_disc}.  Explicitly,
\begin{equation*}
V_{123}
=
\beps_1^i \beps_2^j \beps_3^k V_{ijk}(\bk_1,\bk_2,\bk_3)
=
(\beps_1\!\cdot\!\bk_2)(\beps_2\!\cdot\!\beps_3)
+(\beps_1\!\cdot\!\beps_3)(\beps_2\!\cdot\!\bk_3)
+(\beps_1\!\cdot\!\beps_2)(\beps_3\!\cdot\!\bk_1).
\end{equation*}
Here \(V_{ijk}\) is the three-gluon vertex
\begin{equation}
V_{ijk}(\vect p,\vect q,\vect r)
=
\frac{1}{2}\left(\delta_{ij}(\vect p-\vect q)_k
+\delta_{jk}(\vect q-\vect r)_i
+\delta_{ki}(\vect r-\vect p)_j\right).
\end{equation}

It follows from \eqref{eq:ym-compatible-cut} that the tree-level half-ladder
maximal discontinuity is given by
\begin{equation}
\Disc_{\max}\psi^{\rm YM}_{n,\mathrm{HL}}(1,\dots,n)
=
V^{12 i_1}\,
\Pi_{i_1 i_2}\,
V^{i_2 3 i_3}\,
\Pi_{i_3 i_4}\,
\cdots\,
\Pi_{i_{n-2} i_{n-1}}\,
V^{i_{n-1}\, n-1\, n}
 \Disc_{\rm max}
  \psi^{\phi^3}_{n,\mathrm{HL}}
\, .
\end{equation}
Here
$
  \Disc_{\rm max}
  \psi^{\phi^3}_{n,\mathrm{HL}}$
is given in \eqref{eq:scalar-half-ladder-max-disc}.  Thus, the maximal
discontinuity of the color-ordered
Yang--Mills wavefunction on a ray-like triangulation factorizes into a purely
scalar $\phi^3$ maximal discontinuity times a tensor numerator built by gluing
three-point Yang--Mills vertices with transverse projectors.

The same logic extends to the one-loop \(n\)-gon, where the open chain closes
into a cyclic trace:
\begin{equation}
\Disc_{\max}\psi^{\rm YM}_{n,n\text{-}\mathrm{gon}}
=
\int\!\frac{\dd^3\boldsymbol\ell}{(2\pi)^3}\,
V^{i_{n+1}\,1\,i_1}\,
\Pi_{i_1 i_2}\,
V^{i_2\,2\,i_3}\,
\Pi_{i_3 i_4}\,
\cdots\,
V^{i_n\,n\,i_{n+1}}
\Disc_{\max}\mathcal I^{\phi^3}_{n,n\text{-}\mathrm{gon}}(\boldsymbol\ell)\, .
\end{equation}
The scalar integrand appearing here is given in
\eqref{eq:scalar-ngon-max-disc-integrand}.

Thus, the ray-like tree and its one-loop \(n\)-gon analogue form a particularly
clean sector for testing amplitude-inspired organizations of singularity data.
At the same time, the cosmological character of the problem remains manifest:
the scalar factor is still a cosmological maximal discontinuity rather than an
ordinary flat-space leading singularity.

The rest of the section gives a compact recursive formula for the tensor part
of the tree and one-loop factorizations, summarized in
\eqref{eq:tree-factorization}, and
\eqref{eq:loop-factorization}.

\paragraph{Local gluing map.}
The tensor structure is encoded in a local gluing map built from the
three-gluon vertex.  Since every maximal-cut formula below is obtained by
iterating the same elementary step, it is useful to isolate that step once and
for all.  We use the all-outgoing convention
\begin{equation}
V_{ijk}(p,q,r)
=
\frac{1}{2}\left(\delta_{ij}(p-q)_k+\delta_{jk}(q-r)_i+\delta_{ki}(r-p)_j\right),
\qquad
p+q+r=0.
\label{eq:ym-vertex}
\end{equation}
Let $J_i$ be an open current carrying spatial momentum $P$ and satisfying
$P\!\cdot\!J=0$, and let $\beps_i$ be an external polarization carrying
spatial momentum $k$ with $k\!\cdot\!\beps=0$.  We define the single-step
gluing map by
\begin{equation}
\mathcal G_{P,k}[J,\beps]_i
:=
\Pi_{ij}(P+k)\,J_m\beps_n\,V_{mn}{}^{j}(P,k,-P-k).
\label{eq:gluing-map-def}
\end{equation}
A direct contraction gives the explicit form
\begin{equation}
\mathcal G_{P,k}[J,\beps]
=\frac{1}{2}
\Pi(P+k) \cdot \Big[
(J\!\cdot\!\beps)(P-k)
-2(\beps\!\cdot\!P)\,J
+2(J\!\cdot\!k)\,\beps
\Big].
\label{eq:gluing-map}
\end{equation}
In particular,
\begin{equation}
(P+k)\cdot \mathcal G_{P,k}[J,\beps]=0,
\label{eq:gluing-transverse}
\end{equation}
so the gluing map preserves transversality.  This is the basic local ingredient
behind all of the maximal-cut numerators below.

\paragraph{Tree-level half-ladder.}
Consider the color ordering $(1,2,\dots,n)$ and define the partial momenta
\begin{equation}
Q_m:=\bk_1+\cdots+\bk_m,
\qquad
q_m:=|Q_m|,
\qquad
m=1,\dots,n-1,
\label{eq:partial-momenta}
\end{equation}
so that $q_1=k_1$ and, by momentum conservation, $Q_{n-1}=-\bk_n$ and hence
$q_{n-1}=k_n$.

We now define the open current recursively,
\begin{equation}
J_1:=\beps_1,
\qquad
J_{1\cdots m}
:=
\mathcal G_{Q_{m-1},\,\bk_m}[J_{1\cdots m-1},\beps_m],
\qquad
m=2,\dots,n-1.
\label{eq:current-recursion}
\end{equation}
or, equivalently,
\begin{equation}
J_{1\cdots m}
=
\frac{1}{2}\Pi(Q_m)\Big[
(J_{1\cdots m-1}\!\cdot\!\beps_m)(Q_{m-1}-\bk_m)
-2(\beps_m\!\cdot\!Q_{m-1})\,J_{1\cdots m-1}
+2(J_{1\cdots m-1}\!\cdot\!\bk_m)\,\beps_m
\Big].
\label{eq:current-recursion-explicit}
\end{equation}
This makes manifest that each new leg is attached by a single local gluing
operation followed by a projection onto the transverse subspace.  In
particular, the full numerator is built by adjoining one external gluon at a
time.

The maximal discontinuity of the Yang--Mills half-ladder then factorizes as
\begin{equation}
\Disc_{\max}\psi^{\rm YM}_{n,\mathrm{HL}}(1,\dots,n)
=
\Disc_{\max}\psi^{\phi^3}_{n,\mathrm{HL}}(1,\dots,n)\,
\mathcal N^{\mathrm{HL}}_n(1,\dots,n),
\label{eq:tree-factorization}
\end{equation}
with tensor factor
\begin{equation}
\mathcal N^{\mathrm{HL}}_n(1,\dots,n)
=
\beps_n\cdot J_{1\cdots n-1},
\label{eq:tree-numerator}
\end{equation}
which is the natural open-chain analogue of a color-ordered amplitude numerator.

The proof follows directly from repeated use of the discontinuity formula
\eqref{eq:ym-single-cut}.  Once every internal line is cut, the time integrals
localize vertex by vertex and produce the scalar chain
\begin{equation}
(q_1,k_2,q_2),\ (q_2,k_3,q_3),\ \dots,\ (q_{n-2},k_{n-1},q_{n-1}),
\label{eq:tree-vertex-chain}
\end{equation}
which yields \eqref{eq:scalar-half-ladder-max-disc}.  The remaining index contractions
are exactly the iterated gluing maps
\eqref{eq:current-recursion}--\eqref{eq:current-recursion-explicit}, and the
final external polarization $\beps_n$ closes the chain, giving
\eqref{eq:tree-factorization} and \eqref{eq:tree-numerator}.

It is often convenient to rewrite the recursion as a transfer-matrix product.
Define
\begin{equation}
(\mathbb T_m)^i{}_{j}
=
\frac{1}{2}\Pi^i{}_{r}(Q_m)\Big[
(Q_{m-1}-\bk_m)^r\,\beps_{m\,j}
-2(\beps_m\!\cdot\!Q_{m-1})\,\delta^r{}_j
+2\beps_m^r\,(\bk_m)_j
\Big],
\qquad
m=2,\dots,n-1.
\label{eq:transfer-tree}
\end{equation}
Then
\begin{equation}
J_{1\cdots m}^i
=
(\mathbb T_m\mathbb T_{m-1}\cdots \mathbb T_2)^i{}_j\,\beps_1^j,
\label{eq:current-transfer}
\end{equation}
and hence
\begin{equation}
\mathcal N^{\mathrm{HL}}_n
=
\beps_n\cdot \mathbb T_{n-1}\mathbb T_{n-2}\cdots \mathbb T_2\,\beps_1.
\label{eq:tree-transfer-final}
\end{equation}
This transfer-matrix form is useful because it packages the entire ray-like
numerator into an ordered product of universal local factors.

\paragraph{One-loop $n$-gon.}
The same construction extends to the one-loop $n$-gon.  We define
\begin{equation}
K_a:=\bk_1+\cdots+\bk_a,
\qquad
K_0=0,
\qquad
K_n=0,
\label{eq:loop-partial}
\end{equation}
and loop-line momenta
\begin{equation}
L_a(\ell):=\ell+K_a,
\qquad
q_a:=|L_a(\ell)|,
\qquad
a=0,1,\dots,n,
\label{eq:loop-line-momenta}
\end{equation}
so that $L_n=L_0=\ell$ and $q_n=q_0=|\ell|$.

At the $a$-th vertex we define the local transfer matrix
\begin{equation}
(\mathbb T_a(\ell))^i{}_{j}
=
\frac{1}{2}\Pi^i{}_{r}(L_a)\Big[
(L_{a-1}-\bk_a)^r\,\beps_{a\,j}
-2(\beps_a\!\cdot\!L_{a-1})\,\delta^r{}_j
+2\beps_a^r\,(\bk_a)_j
\Big].
\label{eq:loop-transfer}
\end{equation}
Then the maximal discontinuity of the Yang--Mills one-loop $n$-gon factorizes
at the cut-integrand level as
\begin{equation}
\Disc_{\max}\psi^{\rm YM}_{n,\text{$n$-gon}}
=
\int\!\frac{d^3\ell}{(2\pi)^3}\,
\Disc_{\max}\psi^{\phi^3}_{n,\text{$n$-gon}}(\ell)\,
\Tr\!\big[\mathbb T_n(\ell)\mathbb T_{n-1}(\ell)\cdots \mathbb T_1(\ell)\big].
\label{eq:loop-factorization}
\end{equation}

The derivation parallels the tree case.  After cutting all \(n\) internal
lines, every propagator factorizes as in \eqref{eq:ym-single-cut}, the time
integrals localize into the product of scalar cubic factors
\eqref{eq:scalar-ngon-max-disc}, and the remaining tensor contractions are
local maps arranged around the loop.  Since the chain is now closed rather than
open, the final tensor structure is the cyclic trace in
\eqref{eq:loop-factorization} rather than the open current of
\eqref{eq:tree-transfer-final}.

Equations \eqref{eq:tree-factorization} and \eqref{eq:loop-factorization} show
that ray-like tree diagrams and one-loop $n$-gons are controlled by the same
universal ingredients: a scalar $\phi^3$ maximal discontinuity and an ordered
product of local Yang--Mills gluing maps dressed only by transverse projectors.
At tree level the result is an open-chain numerator, while at one loop it
closes into a cyclic trace.  This is the basic structural lesson carried into
the explicit reconstruction of full Yang--Mills wavefunctions in the
next section: maximal discontinuities isolate the part of the answer that is
closest to an amplitude-style numerator times a purely cosmological scalar
factor.

\section{Reconstructing Yang--Mills wavefunctions from discontinuities}
We now use the discontinuity formulas of Section~\ref{section2} to reconstruct tree-level
Yang--Mills wavefunctions.  The scalar warm-up isolates the spectral part of
the construction: each chord of a polygon is reconstructed by an integral over
discontinuities of lower-point blocks.  The Yang--Mills case then adds tensor
data by gluing each cut gluon line with a transverse projector.  This determines
the cut-detectable sector.  The remaining cut-invisible terms are fixed by
current conservation, spurious OPE-pole cancellation, and the flat-space
total-energy pole.

Consider a channel with internal energy \(k_I\).  If the wavefunction is
otherwise analytic in the complex \(k_I\)-plane and sufficiently well behaved
at infinity, its discontinuity determines the non-analytic part in this
channel up to terms analytic in \(k_I\).  We denote this part by
\(\psi_{n,I}^{\mathrm{NA}}\).  The corresponding Cauchy reconstruction is
\begin{equation}
\label{eq:na-dispersive-reconstruction}
\psi_{n,I}^{\mathrm{NA}}
=
-\frac{1}{2\pi i}
\int_0^\infty
\frac{d\kappa^2}{\kappa^2-k_I^2}\,
P_\kappa\,
\bigl[\psi_L(\kappa)-\psi_L(-\kappa)\bigr]\,
\bigl[\psi_R(\kappa)-\psi_R(-\kappa)\bigr].
\end{equation}
For the flat-space massless scalar wavefunction, and for the scalar
energy-dependent part of the Yang--Mills wavefunction, the cut-line power
spectrum is
$
  P_\kappa=\frac{1}{2\kappa}.
$
When the \(\kappa\)-integral is convergent at infinity, we deform the contour
from \(0\to\infty\) to \(0\to i\infty\), assuming that no pole in the complex
\(k_I^2\)-plane is crossed.  With \(\kappa=ip\), this gives
\begin{equation}
\label{eq:scalar-spectral-reconstruction}
\psi_{n,I}^{\mathrm{scalar}}
=
-i\int_0^\infty
\frac{dp}{2\pi i}\,
\frac{1}{k_I^2+p^2}\,
\bigl[\psi_L(ip)-\psi_L(-ip)\bigr]\,
\bigl[\psi_R(ip)-\psi_R(-ip)\bigr].
\end{equation}
For Yang--Mills, the same spectral reconstruction is dressed by the transverse
polarization sum on the cut gluon line\footnote{Equivalently,
\eqref{eq:ym-spectral-reconstruction} can be written as a sum over helicities of
the cut gluon
using
\(\sum_h \beps_I^{(h)i}\beps_I^{(-h)j}=\Pi^{ij}(\bk_I)\).}:
\begin{equation}
\label{eq:ym-spectral-reconstruction}
\begin{aligned}
\psi_{n,I}^{\mathrm{YM,reg}}
={}&
-i\int_0^\infty
\frac{dp}{2\pi i}\,
\frac{1}{k_I^2+p^2}\,
\left[
  \frac{\partial}{\partial \epsilon_I^i}\,
  \Disc_p\psi_L^{\mathrm{YM}}
  \bigl(\epsilon_I,-\bk_I,ip\bigr)
\right]
\Pi^{ij}(\bk_I)\\
&\times
\left[
  \frac{\partial}{\partial \tilde\epsilon_I^j}\,
  \Disc_p\psi_R^{\mathrm{YM}}
  \bigl(\tilde\epsilon_I,\bk_I,ip\bigr)
\right].
\end{aligned}
\end{equation}
Thus the scalar and Yang--Mills formulas both reconstruct the contribution of
the full bulk-to-bulk propagator in
\eqref{eq:btb-spectral-representation}.  We use
\eqref{eq:scalar-spectral-reconstruction} and
\eqref{eq:ym-spectral-reconstruction} below as the basic cut-reconstruction
formulas.

\subsection{Scalar warm-up: flat-space \texorpdfstring{\(\mathrm{Tr}\,\phi^3\)}{Tr phi3} reconstruction from discontinuities}
We first recall the scalar reconstruction in the simplest color-ordered
flat-space \(\trphi\) toy model.  The elementary three-point building block
\(\psi_{3}^{\operatorname{tr}\phi^3}=C_3(k_1,k_2,k_3)\) is the cubic scalar
block in \eqref{eq:scalar-cubic-block}.  Diagrammatically,
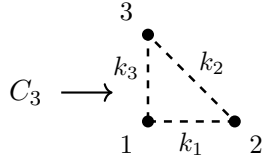
\[
\begin{tikzpicture}[baseline={(current bounding box.center)},scale=0.85]
  \node at (-1.9,0.45) {\(C_3\)};
  \draw[->,line width=0.8pt] (-1.35,0.45) -- (-0.55,0.45);
  \coordinate (A) at (0,0);
  \coordinate (B) at (1.35,0);
  \coordinate (C) at (0,1.35);
  \draw[polygon edge,dashed] (A) -- (B) -- (C) -- cycle;
  \node[vertex dot,label={[boundary label]-135:$1$}] at (A) {};
  \node[vertex dot,label={[boundary label]-45:$2$}] at (B) {};
  \node[vertex dot,label={[boundary label]135:$3$}] at (C) {};
  \node[boundary label] at ($(A)!0.5!(B)+(0,-0.35)$) {\(k_1\)};
  \node[boundary label] at ($(B)!0.5!(C)+(0.32,0.24)$) {\(k_2\)};
  \node[boundary label] at ($(C)!0.5!(A)+(-0.34,0.16)$) {\(k_3\)};
\end{tikzpicture}
\]

Whenever an edge of the triangle becomes an internal chord of a larger polygon,
we take the discontinuity in the corresponding internal energy.  For one
spectral slot this is denoted
\begin{equation}
  \label{eq:scalar-c3bar-toy}
  \bar C_3(k_1,k_2,ip)
  :=
  \Disc_p C_3(k_1,k_2,ip)
  =
  \frac{1}{k_1+k_2+ip}
  -\frac{1}{k_1+k_2-ip}.
\end{equation}
More generally, if several slots are replaced by spectral arguments, then
\(\bar C_3\) means the discontinuity in each of those slots.  For instance,
\begin{equation}
  \begin{aligned}
  \bar C_3(k_1,ip_2,ip_3)
  &:=
  \Disc_{p_2}\Disc_{p_3}C_3(k_1,ip_2,ip_3)
  \\
  &=
  C_3(k_1,ip_2,ip_3)-C_3(k_1,-ip_2,ip_3)
  -C_3(k_1,ip_2,-ip_3)+C_3(k_1,-ip_2,-ip_3).
  \end{aligned}
\end{equation}
The cases with more spectral arguments are defined analogously, e.g.
\[
  \bar C_3(ip_1,ip_2,ip_3)
  :=
  \Disc_{p_1}\Disc_{p_2}\Disc_{p_3}C_3(ip_1,ip_2,ip_3).
\]
This scalar notation is enough to reconstruct all cut contributions built from
three-point blocks.

At four points there are two planar triangulations,
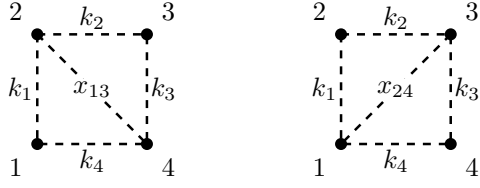
\[
\begin{tikzpicture}[baseline={(current bounding box.center)},scale=0.58]
  \coordinate (A) at (-1.25, 1.25);
  \coordinate (B) at ( 1.25, 1.25);
  \coordinate (C) at ( 1.25,-1.25);
  \coordinate (D) at (-1.25,-1.25);
  \draw[polygon edge,dashed] (A) -- (B) -- (C) -- (D) -- cycle;
  \draw[triangulation,dashed] (A) -- (C);
  \node[vertex dot,label={[boundary label]135:$2$}] at (A) {};
  \node[vertex dot,label={[boundary label]45:$3$}] at (B) {};
  \node[vertex dot,label={[boundary label]-45:$4$}] at (C) {};
  \node[vertex dot,label={[boundary label]-135:$1$}] at (D) {};
  \node[boundary label] at ($(D)!0.5!(A)+(-0.38,0)$) {\(k_1\)};
  \node[boundary label] at ($(A)!0.5!(B)+(0,0.38)$) {\(k_2\)};
  \node[boundary label] at ($(B)!0.5!(C)+(0.38,0)$) {\(k_3\)};
  \node[boundary label] at ($(C)!0.5!(D)+(0,-0.38)$) {\(k_4\)};
  \node[channel label] at ($(A)!0.5!(C)$) {\(x_{13}\)};
\end{tikzpicture}
\qquad\qquad
\begin{tikzpicture}[baseline={(current bounding box.center)},scale=0.58]
  \coordinate (A) at (-1.25, 1.25);
  \coordinate (B) at ( 1.25, 1.25);
  \coordinate (C) at ( 1.25,-1.25);
  \coordinate (D) at (-1.25,-1.25);
  \draw[polygon edge,dashed] (A) -- (B) -- (C) -- (D) -- cycle;
  \draw[triangulation,dashed] (B) -- (D);
  \node[vertex dot,label={[boundary label]135:$2$}] at (A) {};
  \node[vertex dot,label={[boundary label]45:$3$}] at (B) {};
  \node[vertex dot,label={[boundary label]-45:$4$}] at (C) {};
  \node[vertex dot,label={[boundary label]-135:$1$}] at (D) {};
  \node[boundary label] at ($(D)!0.5!(A)+(-0.38,0)$) {\(k_1\)};
  \node[boundary label] at ($(A)!0.5!(B)+(0,0.38)$) {\(k_2\)};
  \node[boundary label] at ($(B)!0.5!(C)+(0.38,0)$) {\(k_3\)};
  \node[boundary label] at ($(C)!0.5!(D)+(0,-0.38)$) {\(k_4\)};
  \node[channel label] at ($(B)!0.5!(D)$) {\(x_{24}\)};
\end{tikzpicture}
\]
with \(x_{ij}:=|\bk_i+\cdots+\bk_{j-1}|\).  The \(x_{13}\) channel is
reconstructed by the same spectral gluing rule that we will use below for
Yang--Mills, but without any tensor numerator:
\begin{equation}
  \label{eq:scalar-four-point-x13-integral}
  \psi^{\trphi}_{4,13}
  =
  -i\int_0^\infty
  \frac{dp_{13}}{2\pi i}\,
  \frac{
    \bar C_3(k_1,k_2,ip_{13})\,
    \bar C_3(ip_{13},k_3,k_4)}
  {x_{13}^2+p_{13}^2}.
\end{equation}
Closing the \(p_{13}\) contour in the upper half-plane gives
\begin{equation}
  \label{eq:scalar-four-point-x13-residue}
  \begin{aligned}
  \psi^{\trphi}_{4,13}
  &=
  -\frac{i}{2}
  \sum_{p_\star\in\{\,ix_{13},\,i(k_1+k_2),\,i(k_3+k_4)\,\}}
  \operatorname*{Res}_{p_{13}=p_\star}
  \left[
  \frac{
    \bar C_3(k_1,k_2,ip_{13})\,
    \bar C_3(ip_{13},k_3,k_4)}
  {x_{13}^2+p_{13}^2}
  \right]
  \\
  &=
  \frac{1}{
  (k_1+k_2+k_3+k_4)
  (k_1+k_2+x_{13})
  (k_3+k_4+x_{13})}.
  \end{aligned}
\end{equation}
The second triangulation is obtained by the cyclic shift
\((1,2,3,4)\mapsto(2,3,4,1)\).
Thus the cut reconstruction gives the planar scalar four-point coefficient
\begin{equation}
  \label{eq:scalar-four-point-reconstructed}
  \psi^{\trphi}_{4}
  =
  \psi^{\trphi}_{4,13}
  +\left.\psi^{\trphi}_{4,13}\right|_{2\leftrightarrow4}.
\end{equation}

The five-point reconstruction is completely analogous.  A useful representative
is the ray-like triangulation with internal edges \(x_{13}\) and \(x_{14}\):
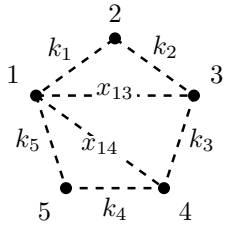
\[
\begin{tikzpicture}[baseline={(current bounding box.center)},scale=0.72]
  \coordinate (A) at ( 0.00, 1.50);
  \coordinate (B) at ( 1.45, 0.45);
  \coordinate (C) at ( 0.90,-1.25);
  \coordinate (D) at (-0.90,-1.25);
  \coordinate (E) at (-1.45, 0.45);
  \draw[polygon edge,dashed] (A) -- (B) -- (C) -- (D) -- (E) -- cycle;
  \draw[triangulation,dashed] (E) -- (B);
  \draw[triangulation,dashed] (E) -- (C);
  \node[vertex dot,label={[boundary label]90:$2$}] at (A) {};
  \node[vertex dot,label={[boundary label]20:$3$}] at (B) {};
  \node[vertex dot,label={[boundary label]-45:$4$}] at (C) {};
  \node[vertex dot,label={[boundary label]-135:$5$}] at (D) {};
  \node[vertex dot,label={[boundary label]160:$1$}] at (E) {};
  \node[boundary label] at ($(E)!0.5!(A)+(-0.28,0.34)$) {\(k_1\)};
  \node[boundary label] at ($(A)!0.5!(B)+(0.20,0.38)$) {\(k_2\)};
  \node[boundary label] at ($(B)!0.5!(C)+(0.42,0.03)$) {\(k_3\)};
  \node[boundary label] at ($(C)!0.5!(D)+(0,-0.38)$) {\(k_4\)};
  \node[boundary label] at ($(D)!0.5!(E)+(-0.42,0.03)$) {\(k_5\)};
  \node[channel label] at ($(E)!0.5!(B)+(0,0.08)$) {\(x_{13}\)};
  \node[channel label] at ($(E)!0.5!(C)+(0,-0.08)$) {\(x_{14}\)};
\end{tikzpicture}
\]
For the middle three-point block, \(\bar C_3(ip_{13},k_3,ip_{14})\)
denotes the double discontinuity in the two internal energies.  The
corresponding double spectral gluing is
\begin{equation}
  \label{eq:scalar-five-point-raylike-integral}
  \psi^{\trphi}_{5,13,14}
  =
  (-i)^2\int_0^\infty\frac{dp_{13}}{2\pi i}
  \int_0^\infty\frac{dp_{14}}{2\pi i}\,
  \frac{\bar C_3(k_1,k_2,ip_{13})\,
  \bar C_3(ip_{13},k_3,ip_{14})
  \bar C_3(ip_{14},k_4,k_5)}
  {(x_{13}^2+p_{13}^2)(x_{14}^2+p_{14}^2)}.
\end{equation}
Equivalently, closing the two \(p\)-contours in the upper half-plane gives the
ray-like scalar block
\begin{equation}
  \label{eq:scalar-five-point-raylike-result}
  \begin{aligned}
  \psi^{\trphi}_{5,13,14}
  =
  &\frac{1}{
  E_5\,(k_1+k_2+x_{13})\,
  (k_3+k_4+k_5+x_{13})\,
  (k_4+k_5+x_{14})\,
  (k_3+x_{13}+x_{14})}
  \\
  &+
  \frac{1}{
  E_5\,(k_1+k_2+x_{13})\,
  (k_1+k_2+k_3+x_{14})\,
  (k_4+k_5+x_{14})\,
  (k_3+x_{13}+x_{14})},
  \end{aligned}
\end{equation}
Here \(E_5:=\sum_{a=1}^5 k_a\) is the total energy.
The full planar scalar five-point coefficient is obtained by the cyclic sum
\begin{equation}
  \label{eq:scalar-five-point-reconstructed}
  \psi^{\trphi}_5
  =
  \psi^{\trphi}_{5,13,14}
  +\mathrm{cyc}_5 .
\end{equation}

At six points, the first ray-like triangulation is the direct continuation of
the same pattern, with internal edges \(x_{13}\), \(x_{14}\), and \(x_{15}\):
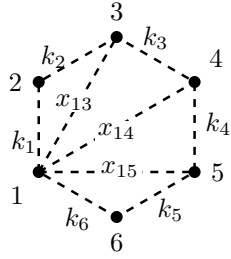
\[
\begin{tikzpicture}[baseline={(current bounding box.center)},scale=0.70]
  \coordinate (A) at (210:1.70);
  \coordinate (B) at (150:1.70);
  \coordinate (C) at ( 90:1.70);
  \coordinate (D) at ( 30:1.70);
  \coordinate (E) at (-30:1.70);
  \coordinate (F) at (-90:1.70);
  \draw[polygon edge,dashed] (A) -- (B) -- (C) -- (D) -- (E) -- (F) -- cycle;
  \draw[triangulation,dashed] (A) -- (C);
  \draw[triangulation,dashed] (A) -- (D);
  \draw[triangulation,dashed] (A) -- (E);
  \node[vertex dot,label={[boundary label]below left:$1$}] at (A) {};
  \node[vertex dot,label={[boundary label]left:$2$}] at (B) {};
  \node[vertex dot,label={[boundary label]above:$3$}] at (C) {};
  \node[vertex dot,label={[boundary label]above right:$4$}] at (D) {};
  \node[vertex dot,label={[boundary label]right:$5$}] at (E) {};
  \node[vertex dot,label={[boundary label]below:$6$}] at (F) {};
  \node[boundary label] at ($(A)!0.5!(B)+(-0.28,-0.28)$) {\(k_1\)};
  \node[boundary label] at ($(B)!0.5!(C)+(-0.45,0.06)$) {\(k_2\)};
  \node[boundary label] at ($(C)!0.5!(D)+(0,0.42)$) {\(k_3\)};
  \node[boundary label] at ($(D)!0.5!(E)+(0.45,0.06)$) {\(k_4\)};
  \node[boundary label] at ($(E)!0.5!(F)+(0.28,-0.28)$) {\(k_5\)};
  \node[boundary label] at ($(F)!0.5!(A)+(0,-0.45)$) {\(k_6\)};
  \node[channel label] at ($(A)!0.5!(C)+(-0.06,0.05)$) {\(x_{13}\)};
  \node[channel label] at ($(A)!0.5!(D)+(0,-0.08)$) {\(x_{14}\)};
  \node[channel label] at ($(A)!0.5!(E)+(0.06,0.05)$) {\(x_{15}\)};
\end{tikzpicture}
\]
The scalar gluing form is therefore
\begin{equation}
  \label{eq:scalar-six-point-raylike-integral}
  \begin{aligned}
  \psi^{\trphi}_{6,13,14,15}
  =
  i\int_0^\infty
  \frac{dp_{13}}{2\pi i}
  \frac{dp_{14}}{2\pi i}
  \frac{dp_{15}}{2\pi i}\,
  \frac{
  \bar C_3(k_1,k_2,ip_{13})\,
  \bar C_3(ip_{13},k_3,ip_{14})\,
  \bar C_3(ip_{14},k_4,ip_{15})\,
  \bar C_3(ip_{15},k_5,k_6)}
  {(x_{13}^2+p_{13}^2)
   (x_{14}^2+p_{14}^2)
   (x_{15}^2+p_{15}^2)} .
  \end{aligned}
\end{equation}

Using the shorthand \(k_{a\cdots b}:=k_a+\cdots+k_b\), we obtain
\begin{equation}
  \label{eq:scalar-six-point-raylike-result}
  \begin{aligned}
  \psi^{\trphi}_{6,13,14,15}
  &=
  \frac{1}{
  E_6\,
  (k_{12}+x_{13})\,
  (k_3+x_{13}+x_{14})\,
  (k_{56}+x_{15})\,
  (k_4+x_{14}+x_{15})}
  \\[-0.2em]
  &\quad\times
  \Bigg[
  \frac{1}{(k_{3456}+x_{13})(k_{456}+x_{14})}
  +\frac{1}{(k_{123}+x_{14})(k_{456}+x_{14})}
  \\[-0.2em]
  &\quad
  +\frac{1}{(k_{123}+x_{14})(k_{1234}+x_{15})}
  +\frac{1}{(k_{3456}+x_{13})(k_{34}+x_{13}+x_{15})}
  \\[-0.2em]
  &\quad
  +\frac{1}{(k_{1234}+x_{15})(k_{34}+x_{13}+x_{15})}
  \Bigg] .
  \end{aligned}
\end{equation}

The first non-ray-like triple cut is the snowflake triangulation with channels
\(x_{13}\), \(x_{35}\), and \(x_{15}\):
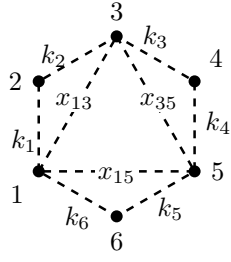
\[
\begin{tikzpicture}[baseline={(current bounding box.center)},scale=0.70]
  \coordinate (A) at (210:1.70);
  \coordinate (B) at (150:1.70);
  \coordinate (C) at ( 90:1.70);
  \coordinate (D) at ( 30:1.70);
  \coordinate (E) at (-30:1.70);
  \coordinate (F) at (-90:1.70);
  \draw[polygon edge,dashed] (A) -- (B) -- (C) -- (D) -- (E) -- (F) -- cycle;
  \draw[triangulation,dashed] (A) -- (C);
  \draw[triangulation,dashed] (C) -- (E);
  \draw[triangulation,dashed] (E) -- (A);
  \node[vertex dot,label={[boundary label]below left:$1$}] at (A) {};
  \node[vertex dot,label={[boundary label]left:$2$}] at (B) {};
  \node[vertex dot,label={[boundary label]above:$3$}] at (C) {};
  \node[vertex dot,label={[boundary label]above right:$4$}] at (D) {};
  \node[vertex dot,label={[boundary label]right:$5$}] at (E) {};
  \node[vertex dot,label={[boundary label]below:$6$}] at (F) {};
  \node[boundary label] at ($(A)!0.5!(B)+(-0.28,-0.28)$) {\(k_1\)};
  \node[boundary label] at ($(B)!0.5!(C)+(-0.45,0.06)$) {\(k_2\)};
  \node[boundary label] at ($(C)!0.5!(D)+(0,0.42)$) {\(k_3\)};
  \node[boundary label] at ($(D)!0.5!(E)+(0.45,0.06)$) {\(k_4\)};
  \node[boundary label] at ($(E)!0.5!(F)+(0.28,-0.28)$) {\(k_5\)};
  \node[boundary label] at ($(F)!0.5!(A)+(0,-0.45)$) {\(k_6\)};
  \node[channel label] at ($(A)!0.5!(C)+(-0.06,0.05)$) {\(x_{13}\)};
  \node[channel label] at ($(C)!0.5!(E)+(0.06,0.05)$) {\(x_{35}\)};
  \node[channel label] at ($(E)!0.5!(A)+(0,-0.08)$) {\(x_{15}\)};
\end{tikzpicture}
\]
Its scalar gluing expression is
\begin{equation}
  \label{eq:scalar-six-point-snowflake-integral}
  \begin{aligned}
  \psi^{\trphi}_{6,13,35,15}
  =
  i\int_0^\infty
  \frac{dp_{13}}{2\pi i}
  \frac{dp_{35}}{2\pi i}
  \frac{dp_{15}}{2\pi i}\,
  \frac{
  \bar C_3(k_1,k_2,ip_{13})\,
  \bar C_3(ip_{13},ip_{35},ip_{15})\,
  \bar C_3(ip_{35},k_3,k_4)\,
  \bar C_3(ip_{15},k_5,k_6)}
  {(x_{13}^2+p_{13}^2)
   (x_{35}^2+p_{35}^2)
   (x_{15}^2+p_{15}^2)} .
  \end{aligned}
\end{equation}
This gives
\begin{equation}
  \label{eq:scalar-six-point-snowflake-result}
  \begin{aligned}
  \psi^{\trphi}_{6,13,35,15}
  &=
  \frac{1}{
  E_6\,
  (k_{12}+x_{13})\,
  (k_{34}+x_{35})\,
  (k_{56}+x_{15})\,
  (x_{13}+x_{35}+x_{15})}
  \\[-0.2em]
  &\quad\times
  \Bigg[
  \frac{1}{
  (k_{34}+k_{56}+x_{13})\,
  (k_{34}+x_{13}+x_{15})}
  +
  \frac{1}{
  (k_{12}+k_{34}+x_{15})\,
  (k_{34}+x_{13}+x_{15})}
  \\[0.3em]
  &\qquad
  +
  \frac{1}{
  (k_{34}+k_{56}+x_{13})\,
  (k_{56}+x_{13}+x_{35})}
  +
  \frac{1}{
  (k_{12}+k_{56}+x_{35})\,
  (k_{56}+x_{13}+x_{35})}
  \\[0.3em]
  &\qquad
  +
  \frac{1}{
  (k_{12}+k_{34}+x_{15})\,
  (k_{12}+x_{15}+x_{35})}
  +
  \frac{1}{
  (k_{12}+k_{56}+x_{35})\,
  (k_{12}+x_{15}+x_{35})}
  \Bigg] .
  \end{aligned}
\end{equation}

The complete six-point scalar coefficient is obtained by summing these
representatives over their planar images:
\begin{equation}
  \label{eq:scalar-six-point-reconstructed}
  \begin{aligned}
  \psi^{\trphi}_6
  &=
  \psi^{\trphi}_{6,13,14,15}
  +\mathrm{cyc}_6
  \\
  &\quad+
  \left.\psi^{\trphi}_{6,13,14,15}\right|_{123456\to123654}
  +
  \left.\psi^{\trphi}_{6,13,14,15}\right|_{123456\to321456}
  +\mathrm{cyc}_3
  \\
  &\quad+
  \psi^{\trphi}_{6,13,35,15}
  +
  \left.\psi^{\trphi}_{6,13,35,15}\right|_{123456\to234561}
  .
  \end{aligned}
\end{equation}

At general multiplicity the scalar reconstruction has the similar form.  If
\(\mathcal T_n\) is the set of planar triangulations of the \(n\)-gon,
\(\mathcal E(T)\) is the set of internal chords of a triangulation, and
\(\triangle(T)\) is its set of triangles, then schematically
\begin{equation}
  \label{eq:scalar-n-point-reconstruction}
  \psi_n^{\trphi}
  =
  \sum_{T\in\mathcal T_n}
  (-i)^{|\mathcal E(T)|}
  \int_0^\infty
  \prod_{e\in\mathcal E(T)}
  \frac{dp_e}{2\pi i}\,
  \prod_{e\in\mathcal E(T)}
  \frac{1}{x_e^2+p_e^2}
  \prod_{t\in\triangle(T)}
  \bar C_3(t).
\end{equation}
Here each \(\bar C_3(t)\) is evaluated on the three energies adjacent to the
triangle \(t\), with \(ip_e\) inserted whenever the adjacent edge is an
internal chord.  The \(p\)-integrals can then be evaluated successively by
taking residues in the upper half-plane.
\subsection{Reconstruction strategy}
\label{gluon_reconstruction}

We now turn from the cut structure itself to the reconstruction of full
tree-level Yang--Mills wavefunctions.  The reconstruction is best viewed in two
steps.  First, every contribution with a nontrivial discontinuity in a planar
channel is fixed by gluing lower-point blocks across the corresponding chords.
One then adds the rational terms that are invisible to those cuts.  These extra
terms are not arbitrary: their pole part is fixed by current conservation,
equivalently by the cancellation of the spurious OPE poles introduced by
transverse projectors, while any remaining local ambiguity is fixed by the
flat-space total-energy pole.

We now translate this reconstruction strategy into the polygon language used
throughout the rest of the section.  In this language, the three-point
coefficient is represented by a triangle, which serves as the elementary
building block for higher-point polygons:
\begin{equation}
\label{eq:three-point-polygon}
\psi_3 \quad \longleftrightarrow \quad
\begin{tikzpicture}[baseline={(current bounding box.center)},scale=1.0]
  \coordinate (A) at (-1.20,-0.69);
  \coordinate (B) at ( 0.00, 1.39);
  \coordinate (C) at ( 1.20,-0.69);

  \draw[polygon edge] (A) -- (B) -- (C) -- cycle;

  \node[vertex dot,label={[boundary label]below left:$1$}] at (A) {};
  \node[vertex dot,label={[boundary label]above:$2$}] at (B) {};
  \node[vertex dot,label={[boundary label]below right:$3$}] at (C) {};
\end{tikzpicture}
\quad \longleftrightarrow \quad
\begin{tikzpicture}[baseline={(current bounding box.center)},scale=0.72]
  \coordinate (O) at (0,0);
  \coordinate (A) at (-1.50, 0.87);
  \coordinate (B) at ( 1.50, 0.87);
  \coordinate (C) at ( 0.00,-1.73);

  \draw[gluon] (O) -- (A);
  \draw[gluon] (O) -- (B);
  \draw[gluon] (O) -- (C);
  \fill (O) circle (2pt);

  \node[above left] at (A) {\small \(1\)};
  \node[right] at (B) {\small \(2\)};
  \node[below] at (C) {\small \(3\)};
\end{tikzpicture}
\end{equation}

For Yang--Mills, the three-point wavefunction coefficient takes the form
\begin{equation}
\label{eq:three-point-wavefunction}
\psi_3(\beps_1,\beps_2,\beps_3;\bk_1,\bk_2,\bk_3;k_1,k_2,k_3)
=
V_{123}\,C_3(k_1,k_2,k_3).
\end{equation}
The explicit expression for the three-point Yang--Mills block,
\(\psi_3^{\mathrm{YM}}=V_{123}\,C_3(k_1,k_2,k_3)\), was given in
\eqref{eq:ym-three-point-wavefunction}.
The triangle in Eq.~\eqref{eq:three-point-polygon} should be read as the dual
graph of the purely wavy cubic Feynman diagram shown on the right-hand side:
each side of the triangle is dual to one external wavy leg.

\begin{figure}[H]
\centering
\begin{tikzpicture}[baseline={(current bounding box.center)},scale=0.85]
  \coordinate (v1) at (270:1.55);
  \coordinate (v2) at (210:1.55);
  \coordinate (v3) at (150:1.55);
  \coordinate (v4) at (90:1.55);
  \coordinate (v5) at (30:1.55);
  \coordinate (v6) at (330:1.55);

  \draw[polygon edge] (v1) -- (v2);
  \draw[polygon edge] (v3) -- (v4) -- (v5);
  \draw[polygon edge] (v6) -- (v1);
  \draw[polygon edge] (v1) -- (v4);

  \draw[polygon edge] (v2) -- ($(v2)!0.34!(v3)$);
  \draw[polygon edge] ($(v2)!0.66!(v3)$) -- (v3);
  \foreach \t in {0.42,0.50,0.58}{
    \fill ($(v2)!\t!(v3)$) circle (1.05pt);
  }

  \draw[polygon edge] (v5) -- ($(v5)!0.34!(v6)$);
  \draw[polygon edge] ($(v5)!0.66!(v6)$) -- (v6);
  \foreach \t in {0.42,0.50,0.58}{
    \fill ($(v5)!\t!(v6)$) circle (1.05pt);
  }

  \foreach \i in {1,...,6}{
    \node[vertex dot] at (v\i) {};
  }
\end{tikzpicture}
\(\qquad\Longleftrightarrow\qquad\)
\begin{tikzpicture}[baseline={(current bounding box.center)},scale=0.85]
  \coordinate (L) at (-1.15,0);
  \coordinate (R) at (1.15,0);

  \draw[gluon] (-2.9,-1.15) -- (L);
  \draw[gluon] (-2.9, 1.15) -- (L);
  \draw[gluon] (L) -- (R);
  \draw[gluon] (R) -- (2.9, 1.15);
  \draw[gluon] (R) -- (2.9,-1.15);

  \foreach \y in {-0.18,0,0.18}{
    \fill (-2.9,\y) circle (1.05pt);
    \fill (2.9,\y) circle (1.05pt);
  }

  \fill[white] (L) circle (7pt);
  \fill[white] (R) circle (7pt);
  \path[
    draw=black,
    pattern=north east lines,
    pattern color=black
  ] (L) circle (7pt);
  \path[
    draw=black,
    pattern=north east lines,
    pattern color=black
  ] (R) circle (7pt);
\end{tikzpicture}
\caption{A planar cut of the polygon and the corresponding
diagrammatic dual.  Each planar cut is represented by a wavy diagram.}
\label{fig:polygon-chord-wavy-propagator}
\end{figure}
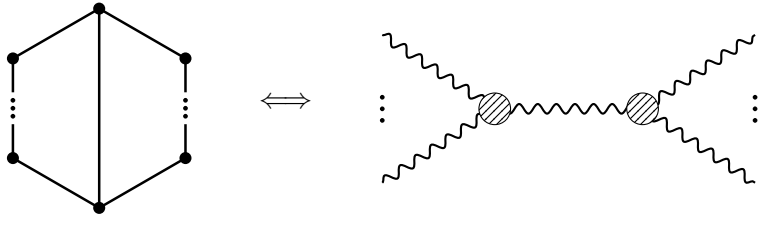

The reconstruction problem can now be phrased recursively.  Suppose that
all lower-point wavefunction blocks have already been determined.  For
each planar chord configuration of the \(n\)-gon, as in
Fig.~\ref{fig:polygon-chord-wavy-propagator}, we first compute the
corresponding discontinuity by cutting the internal chords and gluing the
lower-point blocks with the spectral measure
\[
  \prod_{e\in \mathcal E}
  \frac{dp_e}{2\pi i}\,
  \frac{1}{x_e^2+p_e^2},
\]
together with a transverse polarization sum on every cut gluon line.  This
produces the part of the answer with the prescribed discontinuities in the
channels \(x_e\).  Equivalently, it fixes all terms that can be detected by
iterated cuts.

The glued answer is not yet the full Yang--Mills wavefunction.  The transverse
projectors appearing in the gluing introduce apparent OPE poles \(1/x_e^2\) in
the channel variables \(x_e\).  These poles are the momentum-space imprint of the
longitudinal part of the internal propagator.  Since the currents on the two
sides of the exchanged line must be conserved with respect to the internal
momentum, this longitudinal piece cannot leave a residue in the complete
answer.  Thus, after constructing the known cut part, we expand it near each
channel,
\[
  x_e\to 0,
\]
isolate the singular coefficient of the corresponding OPE pole, and add a
contact-type term with the same pole structure, often written below as a
\(1/x_e^2\) OPE pole, whose singular part cancels it.  This terminology refers
to the OPE singularity in the channel variable \(x_e\), not to a physical
energy discontinuity.  In this sense the contact terms are obtained directly
from the known glued part: they are fixed by the condition
\begin{equation}
\label{eq:ope-residue-cancellation-bootstrap}
  \underset{x_e=0}{\operatorname{Res}}\,
  \psi_n^{\mathrm{YM}}=0
  \qquad
  \text{for every internal channel }e .
\end{equation}
For overlapping channels the same prescription is applied to simultaneous
residues, which is why the five- and six-point contact sectors contain double-
and triple-OPE pole terms.

Accordingly, we write schematically\footnote{At four points there is also a
contact diagram with only a total-energy pole and no OPE pole.  For the
higher-point examples discussed below, the remaining no-cut terms are organized
by the OPE-pole completion.}
\begin{equation}
\label{eq:ym-bootstrap-split}
  \psi_n^{\mathrm{YM}}
  =
  \psi_{n,\mathrm{cut}}^{\mathrm{YM}}
  +
  \psi_{n,\mathrm{OPE}}^{\mathrm{YM}},
\end{equation}
where \(\psi_{n,\mathrm{cut}}^{\mathrm{YM}}\) is the sum over all glued chord
configurations and \(\psi_{n,\mathrm{OPE}}^{\mathrm{YM}}\) is a rational
completion invisible to those cuts.  The pole part of
\(\psi_{n,\mathrm{OPE}}^{\mathrm{YM}}\) is fixed by
\eqref{eq:ope-residue-cancellation-bootstrap}, while any remaining purely
local ambiguity is fixed by the correct flat-space total-energy pole.

The low-point examples below follow the same pattern:
\[
\begin{array}{c|c|c|c}
\text{multiplicity} & \text{cut-detectable data} & \text{OPE completion}
& \text{remaining local input}
\\ \hline
4 & \text{two exchange channels} & b_s,b_t & \text{quartic contact }c\\
5 & \text{two-chord and one-chord pentagons} & B_5 & \text{-}\\
6 & \text{maximal and lower-codimension cuts} & B_6^{(3)},B_6^{(2)}
& \text{-}
\end{array}
\]
Together, the OPE completion and any remaining local input form the
cut-invisible completion.  This table is only a roadmap: the explicit formulas
below show how the singular OPE coefficients are extracted and canceled in each
case.

This reconstruction procedure is a reorganization of the momentum-space
Feynman rules collected in Appendix~\ref{Feynman_rule_for_YM}.  The part
detected by cuts corresponds to transverse bulk-to-bulk propagation, represented
below by wavy internal lines.  The remaining no-cut part is built from the
purely longitudinal sector and from the elementary quartic Yang--Mills contact
vertex, as will be made explicit in the low-point examples.

\subsection{Four-point reconstruction}
\label{subsec:four-point-reconstruction}

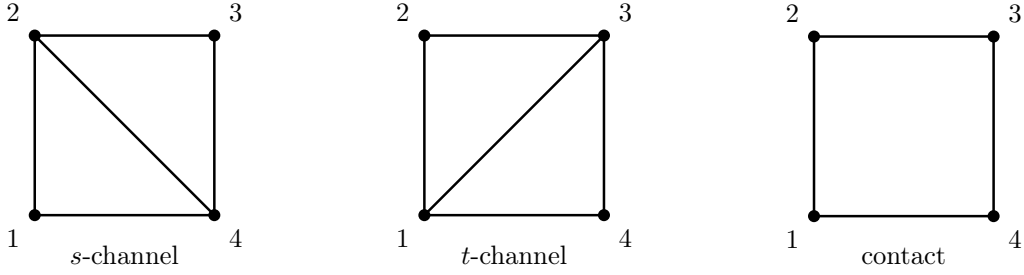
\begin{figure}[h]
\centering
\begin{minipage}[t]{0.31\textwidth}
\centering
\begin{tikzpicture}[scale=0.95]
  \coordinate (A) at (-1.25, 1.25);
  \coordinate (B) at ( 1.25, 1.25);
  \coordinate (C) at ( 1.25,-1.25);
  \coordinate (D) at (-1.25,-1.25);

  \draw[polygon edge] (A) -- (B) -- (C) -- (D) -- cycle;
  \draw[triangulation] (A) -- (C);

  \node[vertex dot,label={[boundary label]135:$2$}] at (A) {};
  \node[vertex dot,label={[boundary label]45:$3$}] at (B) {};
  \node[vertex dot,label={[boundary label]-45:$4$}] at (C) {};
  \node[vertex dot,label={[boundary label]-135:$1$}] at (D) {};

  \node at (0,-1.8) {\small \(s\)-channel};
\end{tikzpicture}
\end{minipage}
\hfill
\begin{minipage}[t]{0.31\textwidth}
\centering
\begin{tikzpicture}[scale=0.95]
  \coordinate (A) at (-1.25, 1.25);
  \coordinate (B) at ( 1.25, 1.25);
  \coordinate (C) at ( 1.25,-1.25);
  \coordinate (D) at (-1.25,-1.25);

  \draw[polygon edge] (A) -- (B) -- (C) -- (D) -- cycle;
  \draw[triangulation] (B) -- (D);

  \node[vertex dot,label={[boundary label]135:$2$}] at (A) {};
  \node[vertex dot,label={[boundary label]45:$3$}] at (B) {};
  \node[vertex dot,label={[boundary label]-45:$4$}] at (C) {};
  \node[vertex dot,label={[boundary label]-135:$1$}] at (D) {};

  \node at (0,-1.8) {\small \(t\)-channel};
\end{tikzpicture}
\end{minipage}
\hfill
\begin{minipage}[t]{0.31\textwidth}
\centering
\begin{tikzpicture}[scale=0.95]
  \coordinate (A) at (-1.25, 1.25);
  \coordinate (B) at ( 1.25, 1.25);
  \coordinate (C) at ( 1.25,-1.25);
  \coordinate (D) at (-1.25,-1.25);

  \draw[polygon edge] (A) -- (B) -- (C) -- (D) -- cycle;

  \node[vertex dot,label={[boundary label]135:$2$}] at (A) {};
  \node[vertex dot,label={[boundary label]45:$3$}] at (B) {};
  \node[vertex dot,label={[boundary label]-45:$4$}] at (C) {};
  \node[vertex dot,label={[boundary label]-135:$1$}] at (D) {};

  \node at (0,-1.8) {\small contact};
\end{tikzpicture}
\end{minipage}
\caption{Polygon description of the planar four-point coefficient.}
\label{fig1:polygonsfor4pt}
\end{figure}

Four points are the first place where the distinction between the part generated
by gluing and the cut-invisible completion becomes visible.  In polygon
language the answer receives three contributions, shown in
Fig.~\ref{fig1:polygonsfor4pt}: the two planar exchange channels and a
zero-chord contact term.  The exchange channels are reconstructed directly from
their discontinuities, while the remaining contact contribution is fixed by
current conservation and the flat-space limit. 

\begin{equation}
\label{eq:four-point-s-channel}
\begin{aligned}
\begin{tikzpicture}[baseline={(current bounding box.center)},scale=0.38]
  \coordinate (A) at (-1.25, 1.25);
  \coordinate (B) at ( 1.25, 1.25);
  \coordinate (C) at ( 1.25,-1.25);
  \coordinate (D) at (-1.25,-1.25);

  \draw[polygon edge] (A) -- (B) -- (C) -- (D) -- cycle;
  \draw[triangulation] (A) -- (C);

  \node[vertex dot,label={[boundary label]135:$2$}] at (A) {};
  \node[vertex dot,label={[boundary label]45:$3$}] at (B) {};
  \node[vertex dot,label={[boundary label]-45:$4$}] at (C) {};
  \node[vertex dot,label={[boundary label]-135:$1$}] at (D) {};
\end{tikzpicture}
&=
\,-i\int_0^\infty
\frac{dp_{13}}{2\pi i\,(x_{13}^2+p_{13}^2)}
\sum_h
\Disc_{p_{13}}\psi_3(\beps_1,\beps_2,\beps_I^{(h)};\bk_1,\bk_2,-\bk_1-\bk_2;k_1,k_2,ip_{13})
\\
&\qquad\qquad\times
    \Disc_{p_{13}}\psi_3(\beps_I^{(-h)},\beps_3,\beps_4;\bk_1+\bk_2,\bk_3,\bk_4;ip_{13},k_3,k_4)
\\
&=
\,-i\int_0^\infty
\frac{dp_{13}}{2\pi i}\,
\frac{
  V_{12\mu}\,\Pi_{\bk_{12}}^{\mu\nu}\,V_{\nu34}}{x_{13}^2+p_{13}^2}\,
\bar C(k_1,k_2,i p_{13})\bar C(i p_{13},k_3,k_4)
\\
&=
V_{12\mu}\,\Pi_{\bk_{12}}^{\mu\nu}\,V_{\nu34}\,
\psi^{\trphi}_{4,13}
\end{aligned}
\end{equation}
Here \(\psi^{\trphi}_{4,13}\) is the scalar four-point \(x_{13}\)-channel
block evaluated in Eq.~\eqref{eq:scalar-four-point-x13-residue}.

This four-point example displays the basic reconstruction mechanism.
The cut integral factorizes into a scalar discontinuity block and a tensor
numerator obtained by gluing lower-point Yang--Mills vertices across the
internal channel.  The helicity sum on the cut line produces the transverse
projector, \(\sum_h \beps_I^{(h)\mu}\beps_I^{(-h)\nu}=\Pi^{\mu\nu}\), so each
planar chord contributes a local numerator dressing together with a purely
spectral scalar factor.
{ This organizational principle---decomposing momentum-space AdS$_4$
vector Witten diagrams into vertex factors, projectors, and scalar graphs---was
discussed previously in Ref.~\cite{Albayrak:2019asr}, and an analogous structure
will reappear in the higher-point reconstructions below.}

The exchange contributions reconstructed from the discontinuity are not yet the
complete four-point Yang--Mills wavefunction.  In the \(s\)-channel, associated
with \(x_{13}\) (equivalently \(x_{24}\)), current conservation requires the
residue of the \(x_{13}^2\) pole to vanish.  This pole is introduced by the
transverse projector \(\Pi_{\bk_{12}}^{\mu\nu}\) when we reconstruct the
four-point exchange from two three-point building blocks.  Its cancellation
requires the additional terms \(b_s\) and \(b_t\), given in
Eq.~\eqref{eq:four-point-b}.
\begin{equation}
\label{eq:four-point-b}
\begin{aligned}
b_s
&=
-\frac{(\beps_1\!\cdot\!\beps_2)(\beps_3\!\cdot\!\beps_4)
    (k_1-k_2)(k_3-k_4)}
{4\,x_{13}^2\,(k_1+k_2+k_3+k_4)},
\\
b_t
&=
\left.b_s\right|_{2\leftrightarrow4}
=
\frac{(\beps_1\!\cdot\!\beps_4)(\beps_2\!\cdot\!\beps_3)
    (k_1-k_4)(k_2-k_3)}
{4\,x_{24}^2\,(k_1+k_2+k_3+k_4)}.
\end{aligned}
\end{equation}
A contact contribution \(c\), given in
Eq.~\eqref{eq:four-point-c}, is fixed by the flat-space limit.
\begin{equation}
\label{eq:four-point-c}
\begin{aligned}
c
&=
\frac{V_4^{1234}}{E_4}
=
\frac{1}{2(k_1+k_2+k_3+k_4)}
\bigl[
  (\beps_1\!\cdot\!\beps_3)(\beps_2\!\cdot\!\beps_4)
  - \frac{1}{2}(\beps_1\!\cdot\!\beps_2)(\beps_3\!\cdot\!\beps_4)
  - \frac{1}{2}(\beps_1\!\cdot\!\beps_4)(\beps_2\!\cdot\!\beps_3)
\bigr].
\end{aligned}
\end{equation}
Together, these terms assemble into the full zero-chord contribution

\begin{equation}
\label{eq:four-point-contact}
\begin{aligned}
&B_4\!\left(
\begin{array}{c}
\beps_1,\beps_2,\beps_3,\beps_4\\
k_1,k_2,k_3,k_4\\
x_{13},x_{24}
\end{array}
\right)
=
\vcenter{\hbox{\begin{tikzpicture}[scale=0.38]
  \coordinate (A) at (-1.25, 1.25);
  \coordinate (B) at ( 1.25, 1.25);
  \coordinate (C) at ( 1.25,-1.25);
  \coordinate (D) at (-1.25,-1.25);

  \draw[polygon edge] (A) -- (B) -- (C) -- (D) -- cycle;

  \node[vertex dot,label={[boundary label]135:$2$}] at (A) {};
  \node[vertex dot,label={[boundary label]45:$3$}] at (B) {};
  \node[vertex dot,label={[boundary label]-45:$4$}] at (C) {};
  \node[vertex dot,label={[boundary label]-135:$1$}] at (D) {};
\end{tikzpicture}}}
=
b_s+b_t+c
\end{aligned}
\end{equation}

Combining the two exchange reconstructions with the contact polygon, the
four-point Yang--Mills wavefunction is written in polygon language as
\begin{equation}
\label{eq:four-point-polygon-language}
\psi_4^{\mathrm{YM}}
=
\begin{tikzpicture}[baseline={(current bounding box.center)},scale=0.30]
  \coordinate (A) at (-1.25, 1.25);
  \coordinate (B) at ( 1.25, 1.25);
  \coordinate (C) at ( 1.25,-1.25);
  \coordinate (D) at (-1.25,-1.25);

  \draw[polygon edge] (A) -- (B) -- (C) -- (D) -- cycle;
  \draw[triangulation] (A) -- (C);

  \node[vertex dot,label={[boundary label]135:$2$}] at (A) {};
  \node[vertex dot,label={[boundary label]45:$3$}] at (B) {};
  \node[vertex dot,label={[boundary label]-45:$4$}] at (C) {};
  \node[vertex dot,label={[boundary label]-135:$1$}] at (D) {};
\end{tikzpicture}
+
\begin{tikzpicture}[baseline={(current bounding box.center)},scale=0.30]
  \coordinate (A) at (-1.25, 1.25);
  \coordinate (B) at ( 1.25, 1.25);
  \coordinate (C) at ( 1.25,-1.25);
  \coordinate (D) at (-1.25,-1.25);

  \draw[polygon edge] (A) -- (B) -- (C) -- (D) -- cycle;
  \draw[triangulation] (B) -- (D);

  \node[vertex dot,label={[boundary label]135:$2$}] at (A) {};
  \node[vertex dot,label={[boundary label]45:$3$}] at (B) {};
  \node[vertex dot,label={[boundary label]-45:$4$}] at (C) {};
  \node[vertex dot,label={[boundary label]-135:$1$}] at (D) {};
\end{tikzpicture}
+
\begin{tikzpicture}[baseline={(current bounding box.center)},scale=0.30]
  \coordinate (A) at (-1.25, 1.25);
  \coordinate (B) at ( 1.25, 1.25);
  \coordinate (C) at ( 1.25,-1.25);
  \coordinate (D) at (-1.25,-1.25);

  \draw[polygon edge] (A) -- (B) -- (C) -- (D) -- cycle;

  \node[vertex dot,label={[boundary label]135:$2$}] at (A) {};
  \node[vertex dot,label={[boundary label]45:$3$}] at (B) {};
  \node[vertex dot,label={[boundary label]-45:$4$}] at (C) {};
  \node[vertex dot,label={[boundary label]-135:$1$}] at (D) {};
\end{tikzpicture}.
\end{equation}

For the color ordering \((1,2,3,4)\), there is no independent planar
\(u\)-channel triangulation of the quadrilateral.
Comparing Eqs.~\eqref{eq:four-point-s-channel}--\eqref{eq:four-point-c} with
the explicit Feynman-rule computation summarized in
Appendix~\ref{Feynman_rule_for_YM}, Eq.~\eqref{eq:four-point-s-channel-tr}, we
find exact agreement: the two exchange polygons reproduce the transverse
exchange diagrams, while the zero-chord quadrilateral, through the
mixed-vertex rule and the shared dashed propagator in
Eqs.~\eqref{eq:vijz-radial-rule} and
\eqref{eq:ym-longitudinal-propagator}, reproduces the longitudinal exchanges.
In addition, there is also a four-point contact diagram.
The resulting four-point answer satisfies the conformal Ward identities and
the expected soft behavior.

\subsection{Five-point reconstruction}
\label{subsec:five-point-reconstruction}

At five points the reconstruction problem becomes richer.  Besides
the fully triangulated polygons reconstructed from maximal cuts, one must also
include one-chord and zero-chord sectors, and the role of current conservation
is correspondingly more intricate.  In polygon language, the answer is organized
by the pentagon together with all its planar chord configurations.

For the color ordering \((1,2,3,4,5)\), we place the boundary labels clockwise
starting from the lower-left vertex.  The relevant configurations are displayed
below.

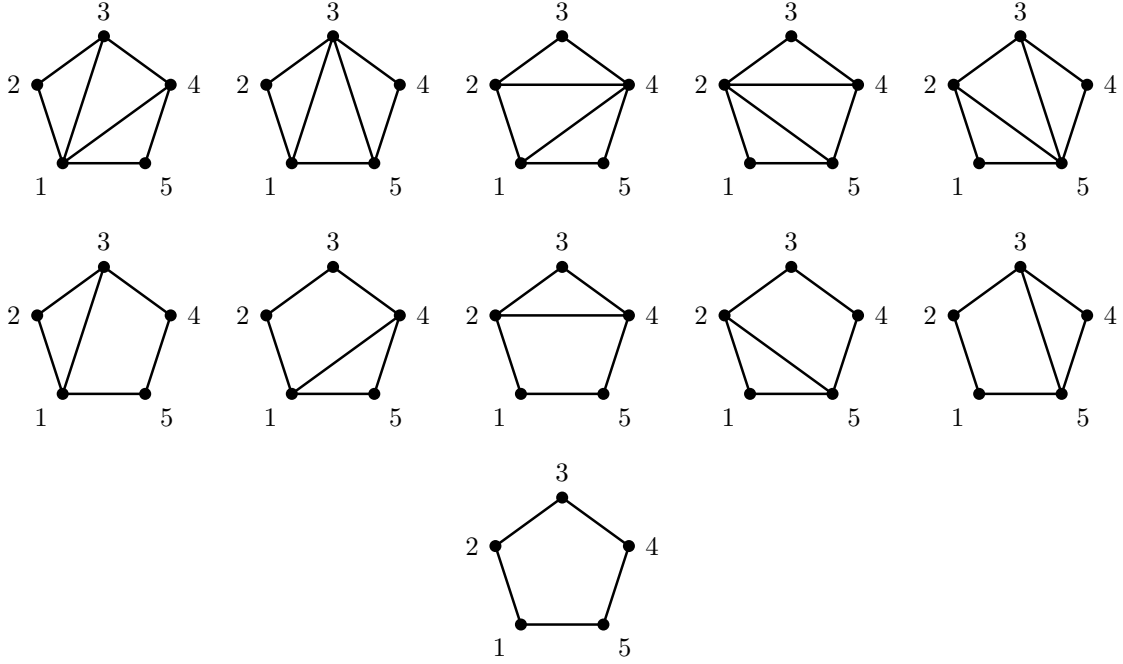
\begin{figure}[H]
\centering
\begin{minipage}[t]{0.19\textwidth}
\centering
\begin{tikzpicture}[scale=0.58]
  \coordinate (A) at (234:1.6);
  \coordinate (B) at (162:1.6);
  \coordinate (C) at (90:1.6);
  \coordinate (D) at (18:1.6);
  \coordinate (E) at (-54:1.6);

  \draw[polygon edge] (A) -- (B) -- (C) -- (D) -- (E) -- cycle;
  \draw[triangulation] (A) -- (C);
  \draw[triangulation] (A) -- (D);

  \node[vertex dot,label={[boundary label]below left:$1$}] at (A) {};
  \node[vertex dot,label={[boundary label]left:$2$}] at (B) {};
  \node[vertex dot,label={[boundary label]above:$3$}] at (C) {};
  \node[vertex dot,label={[boundary label]right:$4$}] at (D) {};
  \node[vertex dot,label={[boundary label]below right:$5$}] at (E) {};
\end{tikzpicture}
\end{minipage}
\hfill
\begin{minipage}[t]{0.19\textwidth}
\centering
\begin{tikzpicture}[scale=0.58]
  \coordinate (A) at (234:1.6);
  \coordinate (B) at (162:1.6);
  \coordinate (C) at (90:1.6);
  \coordinate (D) at (18:1.6);
  \coordinate (E) at (-54:1.6);

  \draw[polygon edge] (A) -- (B) -- (C) -- (D) -- (E) -- cycle;
  \draw[triangulation] (A) -- (C);
  \draw[triangulation] (C) -- (E);

  \node[vertex dot,label={[boundary label]below left:$1$}] at (A) {};
  \node[vertex dot,label={[boundary label]left:$2$}] at (B) {};
  \node[vertex dot,label={[boundary label]above:$3$}] at (C) {};
  \node[vertex dot,label={[boundary label]right:$4$}] at (D) {};
  \node[vertex dot,label={[boundary label]below right:$5$}] at (E) {};
\end{tikzpicture}
\end{minipage}
\hfill
\begin{minipage}[t]{0.19\textwidth}
\centering
\begin{tikzpicture}[scale=0.58]
  \coordinate (A) at (234:1.6);
  \coordinate (B) at (162:1.6);
  \coordinate (C) at (90:1.6);
  \coordinate (D) at (18:1.6);
  \coordinate (E) at (-54:1.6);

  \draw[polygon edge] (A) -- (B) -- (C) -- (D) -- (E) -- cycle;
  \draw[triangulation] (A) -- (D);
  \draw[triangulation] (B) -- (D);

  \node[vertex dot,label={[boundary label]below left:$1$}] at (A) {};
  \node[vertex dot,label={[boundary label]left:$2$}] at (B) {};
  \node[vertex dot,label={[boundary label]above:$3$}] at (C) {};
  \node[vertex dot,label={[boundary label]right:$4$}] at (D) {};
  \node[vertex dot,label={[boundary label]below right:$5$}] at (E) {};
\end{tikzpicture}
\end{minipage}
\hfill
\begin{minipage}[t]{0.19\textwidth}
\centering
\begin{tikzpicture}[scale=0.58]
  \coordinate (A) at (234:1.6);
  \coordinate (B) at (162:1.6);
  \coordinate (C) at (90:1.6);
  \coordinate (D) at (18:1.6);
  \coordinate (E) at (-54:1.6);

  \draw[polygon edge] (A) -- (B) -- (C) -- (D) -- (E) -- cycle;
  \draw[triangulation] (B) -- (D);
  \draw[triangulation] (B) -- (E);

  \node[vertex dot,label={[boundary label]below left:$1$}] at (A) {};
  \node[vertex dot,label={[boundary label]left:$2$}] at (B) {};
  \node[vertex dot,label={[boundary label]above:$3$}] at (C) {};
  \node[vertex dot,label={[boundary label]right:$4$}] at (D) {};
  \node[vertex dot,label={[boundary label]below right:$5$}] at (E) {};
\end{tikzpicture}
\end{minipage}
\hfill
\begin{minipage}[t]{0.19\textwidth}
\centering
\begin{tikzpicture}[scale=0.58]
  \coordinate (A) at (234:1.6);
  \coordinate (B) at (162:1.6);
  \coordinate (C) at (90:1.6);
  \coordinate (D) at (18:1.6);
  \coordinate (E) at (-54:1.6);

  \draw[polygon edge] (A) -- (B) -- (C) -- (D) -- (E) -- cycle;
  \draw[triangulation] (B) -- (E);
  \draw[triangulation] (C) -- (E);

  \node[vertex dot,label={[boundary label]below left:$1$}] at (A) {};
  \node[vertex dot,label={[boundary label]left:$2$}] at (B) {};
  \node[vertex dot,label={[boundary label]above:$3$}] at (C) {};
  \node[vertex dot,label={[boundary label]right:$4$}] at (D) {};
  \node[vertex dot,label={[boundary label]below right:$5$}] at (E) {};
\end{tikzpicture}
\end{minipage}
\hfill
\par\medskip
\begin{minipage}[t]{0.19\textwidth}
\centering
\begin{tikzpicture}[scale=0.58]
  \coordinate (A) at (234:1.6);
  \coordinate (B) at (162:1.6);
  \coordinate (C) at (90:1.6);
  \coordinate (D) at (18:1.6);
  \coordinate (E) at (-54:1.6);

  \draw[polygon edge] (A) -- (B) -- (C) -- (D) -- (E) -- cycle;
  \draw[triangulation] (A) -- (C);

  \node[vertex dot,label={[boundary label]below left:$1$}] at (A) {};
  \node[vertex dot,label={[boundary label]left:$2$}] at (B) {};
  \node[vertex dot,label={[boundary label]above:$3$}] at (C) {};
  \node[vertex dot,label={[boundary label]right:$4$}] at (D) {};
  \node[vertex dot,label={[boundary label]below right:$5$}] at (E) {};
\end{tikzpicture}
\end{minipage}
\hfill
\begin{minipage}[t]{0.19\textwidth}
\centering
\begin{tikzpicture}[scale=0.58]
  \coordinate (A) at (234:1.6);
  \coordinate (B) at (162:1.6);
  \coordinate (C) at (90:1.6);
  \coordinate (D) at (18:1.6);
  \coordinate (E) at (-54:1.6);

  \draw[polygon edge] (A) -- (B) -- (C) -- (D) -- (E) -- cycle;
  \draw[triangulation] (A) -- (D);

  \node[vertex dot,label={[boundary label]below left:$1$}] at (A) {};
  \node[vertex dot,label={[boundary label]left:$2$}] at (B) {};
  \node[vertex dot,label={[boundary label]above:$3$}] at (C) {};
  \node[vertex dot,label={[boundary label]right:$4$}] at (D) {};
  \node[vertex dot,label={[boundary label]below right:$5$}] at (E) {};
\end{tikzpicture}
\end{minipage}
\hfill
\begin{minipage}[t]{0.19\textwidth}
\centering
\begin{tikzpicture}[scale=0.58]
  \coordinate (A) at (234:1.6);
  \coordinate (B) at (162:1.6);
  \coordinate (C) at (90:1.6);
  \coordinate (D) at (18:1.6);
  \coordinate (E) at (-54:1.6);

  \draw[polygon edge] (A) -- (B) -- (C) -- (D) -- (E) -- cycle;
  \draw[triangulation] (B) -- (D);

  \node[vertex dot,label={[boundary label]below left:$1$}] at (A) {};
  \node[vertex dot,label={[boundary label]left:$2$}] at (B) {};
  \node[vertex dot,label={[boundary label]above:$3$}] at (C) {};
  \node[vertex dot,label={[boundary label]right:$4$}] at (D) {};
  \node[vertex dot,label={[boundary label]below right:$5$}] at (E) {};
\end{tikzpicture}
\end{minipage}
\hfill
\begin{minipage}[t]{0.19\textwidth}
\centering
\begin{tikzpicture}[scale=0.58]
  \coordinate (A) at (234:1.6);
  \coordinate (B) at (162:1.6);
  \coordinate (C) at (90:1.6);
  \coordinate (D) at (18:1.6);
  \coordinate (E) at (-54:1.6);

  \draw[polygon edge] (A) -- (B) -- (C) -- (D) -- (E) -- cycle;
  \draw[triangulation] (B) -- (E);

  \node[vertex dot,label={[boundary label]below left:$1$}] at (A) {};
  \node[vertex dot,label={[boundary label]left:$2$}] at (B) {};
  \node[vertex dot,label={[boundary label]above:$3$}] at (C) {};
  \node[vertex dot,label={[boundary label]right:$4$}] at (D) {};
  \node[vertex dot,label={[boundary label]below right:$5$}] at (E) {};
\end{tikzpicture}
\end{minipage}
\hfill
\begin{minipage}[t]{0.19\textwidth}
\centering
\begin{tikzpicture}[scale=0.58]
  \coordinate (A) at (234:1.6);
  \coordinate (B) at (162:1.6);
  \coordinate (C) at (90:1.6);
  \coordinate (D) at (18:1.6);
  \coordinate (E) at (-54:1.6);

  \draw[polygon edge] (A) -- (B) -- (C) -- (D) -- (E) -- cycle;
  \draw[triangulation] (C) -- (E);

  \node[vertex dot,label={[boundary label]below left:$1$}] at (A) {};
  \node[vertex dot,label={[boundary label]left:$2$}] at (B) {};
  \node[vertex dot,label={[boundary label]above:$3$}] at (C) {};
  \node[vertex dot,label={[boundary label]right:$4$}] at (D) {};
  \node[vertex dot,label={[boundary label]below right:$5$}] at (E) {};
\end{tikzpicture}
\end{minipage}
\hfill
\par\medskip
\begin{minipage}[t]{0.19\textwidth}
\centering
\begin{tikzpicture}[scale=0.58]
  \coordinate (A) at (234:1.6);
  \coordinate (B) at (162:1.6);
  \coordinate (C) at (90:1.6);
  \coordinate (D) at (18:1.6);
  \coordinate (E) at (-54:1.6);

  \draw[polygon edge] (A) -- (B) -- (C) -- (D) -- (E) -- cycle;

  \node[vertex dot,label={[boundary label]below left:$1$}] at (A) {};
  \node[vertex dot,label={[boundary label]left:$2$}] at (B) {};
  \node[vertex dot,label={[boundary label]above:$3$}] at (C) {};
  \node[vertex dot,label={[boundary label]right:$4$}] at (D) {};
  \node[vertex dot,label={[boundary label]below right:$5$}] at (E) {};
\end{tikzpicture}
\end{minipage}
\caption{The five-point pentagon, its five one-chord diagrams, and its five planar triangulations.}
\end{figure}

We begin with a representative two-chord contribution, using
\(\bk_{12}:=\bk_1+\bk_2\) and \(\bk_{123}:=\bk_1+\bk_2+\bk_3\):
\begin{equation}
\label{eq:five-point-first-diagram}
\begin{aligned}
\begin{tikzpicture}[baseline={(current bounding box.center)},scale=0.38]
  \coordinate (A) at (234:1.6);
  \coordinate (B) at (162:1.6);
  \coordinate (C) at (90:1.6);
  \coordinate (D) at (18:1.6);
  \coordinate (E) at (-54:1.6);

  \draw[polygon edge] (A) -- (B) -- (C) -- (D) -- (E) -- cycle;
  \draw[triangulation] (A) -- (C);
  \draw[triangulation] (A) -- (D);

  \node[vertex dot,label={[boundary label]below left:$1$}] at (A) {};
  \node[vertex dot,label={[boundary label]left:$2$}] at (B) {};
  \node[vertex dot,label={[boundary label]above:$3$}] at (C) {};
  \node[vertex dot,label={[boundary label]right:$4$}] at (D) {};
  \node[vertex dot,label={[boundary label]below right:$5$}] at (E) {};
\end{tikzpicture}
&=
-\int_0^\infty \frac{dp_{13}}{2\pi i}
\int_0^\infty \frac{dp_{14}}{2\pi i}\,
\frac{
V_{12\mu}\,\Pi_{\bk_{12}}^{\mu\nu}\,V_{\nu3\rho}\,\Pi_{\bk_{123}}^{\rho\sigma}\,V_{\sigma45}
}
{(x_{13}^2+p_{13}^2)(x_{14}^2+p_{14}^2)}
\\
&\qquad\times
\bar C(k_1,k_2,i p_{13})\,
\bar C(i p_{13},k_3,i p_{14})\,
\bar C(i p_{14},k_4,k_5).
\\
&=
V_{12\mu}\,\Pi_{\bk_{12}}^{\mu\nu}\,V_{\nu3\rho}\,\Pi_{\bk_{123}}^{\rho\sigma}\,V_{\sigma45}\,
\psi^{\trphi}_{5,13,14}.
\end{aligned}
\end{equation}
As at four points, the result factorizes into a tensor numerator and the
stripped scalar wavefunction \(\psi^{\trphi}_{5,13,14}\), given in
\eqref{eq:scalar-five-point-raylike-result}.

Next consider a representative one-chord contribution.  With
\(\bk_{12}:=\bk_1+\bk_2\),
\begin{equation}
\label{eq:five-point-first-single-chord}
\begin{gathered}
\begin{tikzpicture}[baseline={(current bounding box.center)},scale=0.34]
  \coordinate (A) at (234:1.6);
  \coordinate (B) at (162:1.6);
  \coordinate (C) at (90:1.6);
  \coordinate (D) at (18:1.6);
  \coordinate (E) at (-54:1.6);

  \draw[polygon edge] (A) -- (B) -- (C) -- (D) -- (E) -- cycle;
  \draw[triangulation] (A) -- (C);

  \node[vertex dot,label={[boundary label]below left:$1$}] at (A) {};
  \node[vertex dot,label={[boundary label]left:$2$}] at (B) {};
  \node[vertex dot,label={[boundary label]above:$3$}] at (C) {};
  \node[vertex dot,label={[boundary label]right:$4$}] at (D) {};
  \node[vertex dot,label={[boundary label]below right:$5$}] at (E) {};
\end{tikzpicture}
\\[-0.3em]
\begin{aligned}
&=
-i\int_0^\infty
\frac{dp_{13}}{2\pi i\,(x_{13}^2+p_{13}^2)}
\sum_h
\Disc_{p_{13}}\psi_3\!\left(
\begin{array}{c}
\beps_1,\beps_2,\beps_I^{(h)}\\
\bk_1,\bk_2,-\bk_{12}\\
k_1,k_2,ip_{13}
\end{array}
\right)
\Disc_{p_{13}}B_4\!\left(
\begin{array}{c}
\beps_I^{(-h)},\beps_3,\beps_4,\beps_5\\
ip_{13},k_3,k_4,k_5\\
x_{14},x_{35}
\end{array}
\right).
\\
&=
\left(V_{12\mu}\,\Pi_{\bk_{12}}^{\mu\nu}\,\beps_{3\nu}\right)
(\beps_{4}\!\cdot\!\beps_{5})
\left[\frac{(k_4-k_5)(2k_3+k_4+k_5)}{4\,\bk_{45}^2}\right]
\psi_{5,\,12|345}^{\mathrm{scalar}}+\,(3\leftrightarrow 5)
\\
&\quad+
V_{12\mu}\,\Pi_{\bk_{12}}^{\mu\nu}\,V_{\nu345}\,
\psi_{5,\,12|345}^{\mathrm{scalar}}
\end{aligned}
\end{gathered}
\end{equation}
Here \(\psi_{5,\,12|345}^{\mathrm{scalar}}\) denotes the scalar one-chord
factor given explicitly in \eqref{eq:five-point-one-chord-scalar}.  The first
line is the longitudinal contribution from the quadrilateral, together with the
reflected channel obtained by \(3\leftrightarrow 5\).
The one-chord sector exhibits the first structural feature absent at four
points: once a three-point block is glued to a lower-point contact block, the
result is no longer captured by a single purely transverse tensor structure.
It instead splits into a transverse piece and a pair of mixed corrections
related by reflection.

By cyclically summing the double-chord contribution in
\eqref{eq:five-point-first-diagram} and the single-chord contribution in
\eqref{eq:five-point-first-single-chord}, we obtain the part of the
five-point answer generated purely by gluing lower-point polygon blocks:
\begin{equation}
\label{eq:five-point-temporary-gluing-sum}
\begin{aligned}
\psi_{5,\mathrm{cut}}^{\mathrm{YM}}
&=
\begin{tikzpicture}[baseline={(current bounding box.center)},scale=0.28]
  \coordinate (A) at (234:1.6);
  \coordinate (B) at (162:1.6);
  \coordinate (C) at (90:1.6);
  \coordinate (D) at (18:1.6);
  \coordinate (E) at (-54:1.6);

  \draw[polygon edge] (A) -- (B) -- (C) -- (D) -- (E) -- cycle;
  \draw[triangulation] (A) -- (C);
  \draw[triangulation] (A) -- (D);

  \node[vertex dot,label={[boundary label]below left:$1$}] at (A) {};
  \node[vertex dot,label={[boundary label]left:$2$}] at (B) {};
  \node[vertex dot,label={[boundary label]above:$3$}] at (C) {};
  \node[vertex dot,label={[boundary label]right:$4$}] at (D) {};
  \node[vertex dot,label={[boundary label]below right:$5$}] at (E) {};
\end{tikzpicture}
+
\begin{tikzpicture}[baseline={(current bounding box.center)},scale=0.28]
  \coordinate (A) at (234:1.6);
  \coordinate (B) at (162:1.6);
  \coordinate (C) at (90:1.6);
  \coordinate (D) at (18:1.6);
  \coordinate (E) at (-54:1.6);

  \draw[polygon edge] (A) -- (B) -- (C) -- (D) -- (E) -- cycle;
  \draw[triangulation] (A) -- (C);

  \node[vertex dot,label={[boundary label]below left:$1$}] at (A) {};
  \node[vertex dot,label={[boundary label]left:$2$}] at (B) {};
  \node[vertex dot,label={[boundary label]above:$3$}] at (C) {};
  \node[vertex dot,label={[boundary label]right:$4$}] at (D) {};
  \node[vertex dot,label={[boundary label]below right:$5$}] at (E) {};
\end{tikzpicture}
+\mathrm{cyclic}.
\end{aligned}
\end{equation}
This gluing-generated sum is the cut-detectable five-point part, but it is not
yet the full answer.  Some single-OPE residues remain, so current conservation
is not yet manifest.  The missing completion therefore belongs to the
zero-chord sector.
Equivalently, the cut-detectable part can be summarized as
\begin{equation}
\label{eq:five-point-cut-bookkeeping}
\psi_{5,\mathrm{cut}}^{\mathrm{YM}}
=
\sum_{\mathrm{cyc}}
\left(
\psi_{5}^{(2\mathrm{chord})}
+\psi_{5}^{(1\mathrm{chord})}
\right),
\end{equation}
where representative two-chord and one-chord terms are given in
\eqref{eq:five-point-first-diagram} and
\eqref{eq:five-point-first-single-chord}.

\paragraph{Zero-chord completion.}

To impose current conservation at five points, we require the residue of every
single-OPE pole associated with an internal chord to vanish.  For the pentagon
these channels are \(x_{13}\), \(x_{14}\), \(x_{24}\), \(x_{25}\), and
\(x_{35}\); equivalently, the coefficient of each corresponding
\(1/x_{ij}^2\) pole must cancel in the full sum over polygon contributions.
In practice, the zero-chord sector can be fixed by isolating the simultaneous
residue of two single-OPE poles.  For example, cancellation of the
\(x_{13}^{-2}x_{14}^{-2}\) double pole fixes
\begin{equation}
\label{eq:five-point-b5}
\begin{aligned}
B_5\!\left(
\begin{array}{c}
\beps_1,\beps_2,\beps_3,\beps_4,\beps_5\\
\bk_1,\bk_2,\bk_3,\bk_4,\bk_5\\
k_1,k_2,k_3,k_4,k_5\\
x_{13},x_{14}
\end{array}
\right)
&=
-\frac{
  (k_1-k_2)(k_4-k_5)
  (\beps_1\!\cdot\!\beps_2)
  \bigl[\beps_3\!\cdot\!(\bk_4+\bk_5)\bigr]
  (\beps_4\!\cdot\!\beps_5)
}{
  4\,(k_1+k_2+k_3+k_4+k_5)\,x_{13}^2x_{14}^2
}
\;+\;\mathrm{cyclic}.
\end{aligned}
\end{equation}
The representative zero-chord polygon is therefore
\begin{equation}
\label{eq:five-point-zero-chord-current-conservation}
\begin{tikzpicture}[baseline={(current bounding box.center)},scale=0.28]
  \coordinate (A) at (234:1.6);
  \coordinate (B) at (162:1.6);
  \coordinate (C) at (90:1.6);
  \coordinate (D) at (18:1.6);
  \coordinate (E) at (-54:1.6);

  \draw[polygon edge] (A) -- (B) -- (C) -- (D) -- (E) -- cycle;

  \node[vertex dot,label={[boundary label]below left:$1$}] at (A) {};
  \node[vertex dot,label={[boundary label]left:$2$}] at (B) {};
  \node[vertex dot,label={[boundary label]above:$3$}] at (C) {};
  \node[vertex dot,label={[boundary label]right:$4$}] at (D) {};
  \node[vertex dot,label={[boundary label]below right:$5$}] at (E) {};
\end{tikzpicture}
=
B_5\!\left(
\begin{array}{c}
\beps_1,\beps_2,\beps_3,\beps_4,\beps_5\\
\bk_1,\bk_2,\bk_3,\bk_4,\bk_5\\
k_1,k_2,k_3,k_4,k_5\\
x_{13},x_{14}
\end{array}
\right).
\end{equation}
After adding this cyclic zero-chord contribution, the single-OPE singular
parts cancel in the full cyclic sum.  This is the five-point implementation of
the current-conservation condition
\eqref{eq:ope-residue-cancellation-bootstrap}.

Thus the full five-point Yang--Mills wavefunction takes the polygon form
\begin{equation}
\label{eq:five-point-full-polygon-sum}
\begin{aligned}
\psi_5^{\mathrm{YM}}
&=
\left(
\begin{tikzpicture}[baseline={(current bounding box.center)},scale=0.28]
  \coordinate (A) at (234:1.6);
  \coordinate (B) at (162:1.6);
  \coordinate (C) at (90:1.6);
  \coordinate (D) at (18:1.6);
  \coordinate (E) at (-54:1.6);

  \draw[polygon edge] (A) -- (B) -- (C) -- (D) -- (E) -- cycle;
  \draw[triangulation] (A) -- (C);
  \draw[triangulation] (A) -- (D);

  \node[vertex dot,label={[boundary label]below left:$1$}] at (A) {};
  \node[vertex dot,label={[boundary label]left:$2$}] at (B) {};
  \node[vertex dot,label={[boundary label]above:$3$}] at (C) {};
  \node[vertex dot,label={[boundary label]right:$4$}] at (D) {};
  \node[vertex dot,label={[boundary label]below right:$5$}] at (E) {};
\end{tikzpicture}
+
\begin{tikzpicture}[baseline={(current bounding box.center)},scale=0.28]
  \coordinate (A) at (234:1.6);
  \coordinate (B) at (162:1.6);
  \coordinate (C) at (90:1.6);
  \coordinate (D) at (18:1.6);
  \coordinate (E) at (-54:1.6);

  \draw[polygon edge] (A) -- (B) -- (C) -- (D) -- (E) -- cycle;
  \draw[triangulation] (A) -- (C);

  \node[vertex dot,label={[boundary label]below left:$1$}] at (A) {};
  \node[vertex dot,label={[boundary label]left:$2$}] at (B) {};
  \node[vertex dot,label={[boundary label]above:$3$}] at (C) {};
  \node[vertex dot,label={[boundary label]right:$4$}] at (D) {};
  \node[vertex dot,label={[boundary label]below right:$5$}] at (E) {};
\end{tikzpicture}
+\mathrm{cyclic}
\right)
+
\begin{tikzpicture}[baseline={(current bounding box.center)},scale=0.28]
  \coordinate (A) at (234:1.6);
  \coordinate (B) at (162:1.6);
  \coordinate (C) at (90:1.6);
  \coordinate (D) at (18:1.6);
  \coordinate (E) at (-54:1.6);

  \draw[polygon edge] (A) -- (B) -- (C) -- (D) -- (E) -- cycle;

  \node[vertex dot,label={[boundary label]below left:$1$}] at (A) {};
  \node[vertex dot,label={[boundary label]left:$2$}] at (B) {};
  \node[vertex dot,label={[boundary label]above:$3$}] at (C) {};
  \node[vertex dot,label={[boundary label]right:$4$}] at (D) {};
  \node[vertex dot,label={[boundary label]below right:$5$}] at (E) {};
\end{tikzpicture}.
\end{aligned}
\end{equation}
Equivalently,
\begin{equation}
\label{eq:five-point-full-bookkeeping}
\psi_5^{\mathrm{YM}}
=
\sum_{\mathrm{cyc}}
\left(
\psi_{5}^{(2\mathrm{chord})}
+\psi_{5}^{(1\mathrm{chord})}
\right)
+B_5 .
\end{equation}
The full data are provided in the ancillary file
\texttt{5ptYM and 6pt YM data.zip}.  The final five-point Yang--Mills
wavefunction satisfies the expected
\((A)dS\) soft theorem and reproduces the correct flat-space limit.
Comparing Eqs.~\eqref{eq:five-point-first-diagram}--\eqref{eq:five-point-b5}
with the explicit Feynman-rule organization in
Appendix~\ref{Feynman_rule_for_YM},
Eqs.~\eqref{eq:five-point-double-tr}--\eqref{eq:five-point-double-zz}, we find
that the reconstruction matches the direct Feynman-diagram computation: the
maximally cut and one-chord polygon sectors reproduce the transverse and mixed
exchange topologies, while the remaining zero-chord pentagon supplies the
contact and longitudinal completion.  { For comparison,
Ref.~\cite{Albayrak:2018tam} obtained the four- and five-point momentum-space
vector Witten diagrams in AdS$_4$ by direct bulk perturbation theory and
outlined a strategy for the six-point calculation.}  { We have also checked that our
five-point result agrees with the Berends--Giele recursive construction
presented in Ref.~\cite{Gomez:2026yno}.}  The one-chord contribution in
Eq.~\eqref{eq:five-point-first-single-chord} contains the factor
\((k_4-k_5)(2k_3+k_4+k_5)\), which has a natural Feynman-diagrammatic
interpretation in terms of the local propagator operations reviewed in
Appendix~\ref{appendixA1}.

\subsection{Six-point reconstruction}
\label{subsec:six-point-reconstruction}
\newcommand{\SixPointCutFigure}[3]{%
\begin{minipage}[t]{0.42\textwidth}
\centering
\begin{tikzpicture}[scale=0.58]
  \coordinate (A) at (240:1.6);
  \coordinate (B) at (180:1.6);
  \coordinate (C) at (120:1.6);
  \coordinate (D) at (60:1.6);
  \coordinate (E) at (0:1.6);
  \coordinate (F) at (300:1.6);
  \coordinate (v1) at (A);
  \coordinate (v2) at (B);
  \coordinate (v3) at (C);
  \coordinate (v4) at (D);
  \coordinate (v5) at (E);
  \coordinate (v6) at (F);

  \draw[polygon edge] (A) -- (B) -- (C) -- (D) -- (E) -- (F) -- cycle;
  \foreach \a/\b in {#1}{
    \draw[triangulation] (v\a) -- (v\b);
  }

  \node[vertex dot,label={[boundary label]below left:$1$}] at (A) {};
  \node[vertex dot,label={[boundary label]left:$2$}] at (B) {};
  \node[vertex dot,label={[boundary label]above left:$3$}] at (C) {};
  \node[vertex dot,label={[boundary label]above right:$4$}] at (D) {};
  \node[vertex dot,label={[boundary label]right:$5$}] at (E) {};
  \node[vertex dot,label={[boundary label]below right:$6$}] at (F) {};
\end{tikzpicture}

\small \textbf{(#2)} \(#3\)
\end{minipage}%
}

At six points the space of chord configurations is large enough that it is most
efficient to organize the reconstruction by topology.  We begin with the
ray-like maximal cuts, then turn to the non-ray-like maximal cuts, followed by
the lower-codimension chord sectors, and finally determine the zero-chord part
by cancelling the remaining higher-order OPE poles.  Since the number of
individual building blocks grows quickly at this multiplicity, we emphasize
below the representative topologies and the logic by which they combine; more
exhaustive diagrammatic summaries are collected in the appendix.

The representatives used below are:
\begin{center}
\small
\begin{tabular}{p{0.21\textwidth}|p{0.31\textwidth}|p{0.13\textwidth}|p{0.23\textwidth}}
sector & representatives & orbit sizes & building blocks\\ \hline
maximal cuts &
\begin{tabular}[t]{@{}l@{}}
\((13|14|15),(13|14|46)\)\\
\((24|14|15),(13|15|35)\)
\end{tabular} &
\(6,3,3,2\) &
\(\psi_3^4\)\\
two-chord sectors &
\begin{tabular}[t]{@{}l@{}}
\((13|46),(24|14)\)\\
\((13|15),(13|14)\)
\end{tabular} &
\(3,6,6,6\) &
\(\psi_3 B_4\psi_3\) or \(\psi_3^2B_4\)\\
one-chord sectors &
\((14),(13)\) &
\(3,6\) &
\(B_4B_4\) or \(\psi_3B_5\)\\
zero-chord sector &
\(B_6^{(3)},B_6^{(2)}\) &
\(6+3+3,\;6+3\) &
OPE subtractions
\end{tabular}
\end{center}
Here \((13|14|15)\) denotes the set of chords in the representative polygon,
and the orbit size counts the number of distinct cyclic or reflected images
included in the corresponding sum.

\begin{figure}[H]
\centering
\begin{minipage}[t]{0.42\textwidth}
\centering
\begin{tikzpicture}[scale=0.58]
  \coordinate (A) at (240:1.6);
  \coordinate (B) at (180:1.6);
  \coordinate (C) at (120:1.6);
  \coordinate (D) at (60:1.6);
  \coordinate (E) at (0:1.6);
  \coordinate (F) at (300:1.6);
  \coordinate (v1) at (A);
  \coordinate (v2) at (B);
  \coordinate (v3) at (C);
  \coordinate (v4) at (D);
  \coordinate (v5) at (E);
  \coordinate (v6) at (F);
  \draw[polygon edge] (A) -- (B) -- (C) -- (D) -- (E) -- (F) -- cycle;
  \foreach \a/\b in {1/3,1/4,1/5}{
    \draw[triangulation] (v\a) -- (v\b);
  }
  \node[vertex dot,label={[boundary label]below left:$1$}] at (A) {};
  \node[vertex dot,label={[boundary label]left:$2$}] at (B) {};
  \node[vertex dot,label={[boundary label]above left:$3$}] at (C) {};
  \node[vertex dot,label={[boundary label]above right:$4$}] at (D) {};
  \node[vertex dot,label={[boundary label]right:$5$}] at (E) {};
  \node[vertex dot,label={[boundary label]below right:$6$}] at (F) {};
\end{tikzpicture}

\scriptsize
\(\mathbf{(13),(14),(15)}\)\\[-1pt]
orbit size \(= 6\);\\[-1pt]
\(\psi_3\times\psi_3\times\psi_3\times\psi_3\)
\end{minipage}%
\hfill
\begin{minipage}[t]{0.42\textwidth}
\centering
\begin{tikzpicture}[scale=0.58]
  \coordinate (A) at (240:1.6);
  \coordinate (B) at (180:1.6);
  \coordinate (C) at (120:1.6);
  \coordinate (D) at (60:1.6);
  \coordinate (E) at (0:1.6);
  \coordinate (F) at (300:1.6);
  \coordinate (v1) at (A);
  \coordinate (v2) at (B);
  \coordinate (v3) at (C);
  \coordinate (v4) at (D);
  \coordinate (v5) at (E);
  \coordinate (v6) at (F);
  \draw[polygon edge] (A) -- (B) -- (C) -- (D) -- (E) -- (F) -- cycle;
  \foreach \a/\b in {1/3,1/4,4/6}{
    \draw[triangulation] (v\a) -- (v\b);
  }
  \node[vertex dot,label={[boundary label]below left:$1$}] at (A) {};
  \node[vertex dot,label={[boundary label]left:$2$}] at (B) {};
  \node[vertex dot,label={[boundary label]above left:$3$}] at (C) {};
  \node[vertex dot,label={[boundary label]above right:$4$}] at (D) {};
  \node[vertex dot,label={[boundary label]right:$5$}] at (E) {};
  \node[vertex dot,label={[boundary label]below right:$6$}] at (F) {};
\end{tikzpicture}

\scriptsize
\(\mathbf{(13),(14),(46)}\)\\[-1pt]
orbit size \(= 3\);\\[-1pt]
\(\psi_3\times\psi_3\times\psi_3\times\psi_3\)
\end{minipage}%

\vspace{0.45em}

\begin{minipage}[t]{0.42\textwidth}
\centering
\begin{tikzpicture}[scale=0.58]
  \coordinate (A) at (240:1.6);
  \coordinate (B) at (180:1.6);
  \coordinate (C) at (120:1.6);
  \coordinate (D) at (60:1.6);
  \coordinate (E) at (0:1.6);
  \coordinate (F) at (300:1.6);
  \coordinate (v1) at (A);
  \coordinate (v2) at (B);
  \coordinate (v3) at (C);
  \coordinate (v4) at (D);
  \coordinate (v5) at (E);
  \coordinate (v6) at (F);
  \draw[polygon edge] (A) -- (B) -- (C) -- (D) -- (E) -- (F) -- cycle;
  \foreach \a/\b in {2/4,1/4,1/5}{
    \draw[triangulation] (v\a) -- (v\b);
  }
  \node[vertex dot,label={[boundary label]below left:$1$}] at (A) {};
  \node[vertex dot,label={[boundary label]left:$2$}] at (B) {};
  \node[vertex dot,label={[boundary label]above left:$3$}] at (C) {};
  \node[vertex dot,label={[boundary label]above right:$4$}] at (D) {};
  \node[vertex dot,label={[boundary label]right:$5$}] at (E) {};
  \node[vertex dot,label={[boundary label]below right:$6$}] at (F) {};
\end{tikzpicture}

\scriptsize
\(\mathbf{(24),(14),(15)}\)\\[-1pt]
orbit size \(= 3\);\\[-1pt]
\(\psi_3\times\psi_3\times\psi_3\times\psi_3\)
\end{minipage}%
\hfill
\begin{minipage}[t]{0.42\textwidth}
\centering
\begin{tikzpicture}[scale=0.58]
  \coordinate (A) at (240:1.6);
  \coordinate (B) at (180:1.6);
  \coordinate (C) at (120:1.6);
  \coordinate (D) at (60:1.6);
  \coordinate (E) at (0:1.6);
  \coordinate (F) at (300:1.6);
  \coordinate (v1) at (A);
  \coordinate (v2) at (B);
  \coordinate (v3) at (C);
  \coordinate (v4) at (D);
  \coordinate (v5) at (E);
  \coordinate (v6) at (F);
  \draw[polygon edge] (A) -- (B) -- (C) -- (D) -- (E) -- (F) -- cycle;
  \foreach \a/\b in {1/3,1/5,3/5}{
    \draw[triangulation] (v\a) -- (v\b);
  }
  \node[vertex dot,label={[boundary label]below left:$1$}] at (A) {};
  \node[vertex dot,label={[boundary label]left:$2$}] at (B) {};
  \node[vertex dot,label={[boundary label]above left:$3$}] at (C) {};
  \node[vertex dot,label={[boundary label]above right:$4$}] at (D) {};
  \node[vertex dot,label={[boundary label]right:$5$}] at (E) {};
  \node[vertex dot,label={[boundary label]below right:$6$}] at (F) {};
\end{tikzpicture}

\scriptsize
\(\mathbf{(13),(15),(35)}\)\\[-1pt]
orbit size \(= 2\);\\[-1pt]
\(\psi_3\times\psi_3\times\psi_3\times\psi_3\)
\end{minipage}%
\caption{Representative maximal six-point chord configurations: the ray-like
class and its first non-ray-like variants.}
\end{figure}
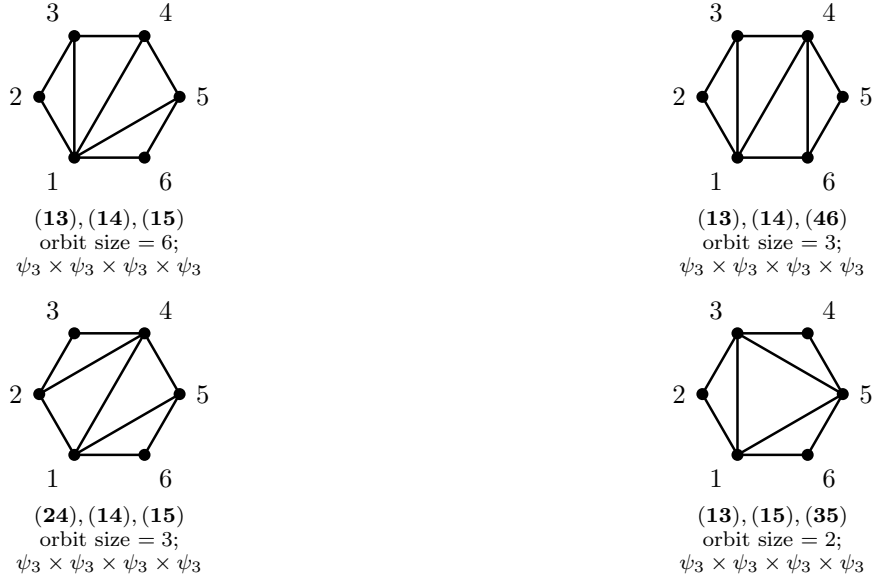
A representative gluing expression is
\begin{equation}
\label{eq:six-point-raylike-131415-gluing}
\begin{aligned}
\psi_{6,\,13|14|15}^{\mathrm{YM}}(1,2,3,4,5,6)
&=
i\int_0^\infty
\frac{dp_{13}}{2\pi i}
\frac{dp_{14}}{2\pi i}
\frac{dp_{15}}{2\pi i}\,
\frac{
V_{12i}\,
\Pi_{\bk_{12}}^{ij}\,
V_{j3l}\,
\Pi_{\bk_{123}}^{lm}\,
V_{m4n}\,
\Pi_{\bk_{56}}^{nr}\,
V_{r56}
}{
(x_{13}^2+p_{13}^2)
(x_{14}^2+p_{14}^2)
(x_{15}^2+p_{15}^2)
}
\\
&\quad\times
\bar C(k_1,k_2,ip_{13})\,
\bar C(ip_{13},k_3,ip_{14})\,
\bar C(ip_{14},k_4,ip_{15})\,
\bar C(ip_{15},k_5,k_6),\\
&=
\Bigl(
V_{12i}\,\Pi_{\bk_{12}}^{ij}\,
V_{j3l}\,\Pi_{\bk_{123}}^{lm}\,
V_{m4n}\,\Pi_{\bk_{56}}^{nr}\,
V_{r56}
\Bigr)\psi^{\trphi}_{6,13,14,15}.
\end{aligned}
\end{equation}

For a permutation \(i j k l m n\) of \(123456\), define
\begin{equation}
\label{eq:six-point-rho-permutation}
\rho_{i j k l m n}:\quad 123456\longmapsto i j k l m n .
\end{equation}
This relabelling acts on all labels, momenta, polarizations, and channels in
the expression on its right.  The two remaining maximally cut ray-like
representatives are then generated from
\eqref{eq:six-point-raylike-131415-gluing} by
\begin{subequations}
\begin{align}
\psi_{6,\,13|14|46}^{\mathrm{YM}}
&=
\rho_{123654}\,\psi_{6,\,13|14|15}^{\mathrm{YM}},
\label{eq:six-point-131446-gluing}\\
\psi_{6,\,24|14|15}^{\mathrm{YM}}
&=
\rho_{321456}\,\psi_{6,\,13|14|15}^{\mathrm{YM}}.
\label{eq:six-point-241415-gluing}
\end{align}
\end{subequations}

\begin{equation}
\label{eq:six-point-131535-gluing}
\begin{aligned}
\psi_{6,\,13|15|35}^{\mathrm{YM}}(1,2,3,4,5,6)
&=
i\int_0^\infty
\frac{dp_{13}}{2\pi i}
\frac{dp_{15}}{2\pi i}
\frac{dp_{35}}{2\pi i}\,
\frac{
V_{12i}\,
\Pi_{\bk_{12}}^{ij}\,
V_{jln}\,
\Pi_{\bk_{34}}^{lm}\,
V_{m34}\,
\Pi_{\bk_{56}}^{nr}\,
V_{r56}
}{
(x_{13}^2+p_{13}^2)
(x_{15}^2+p_{15}^2)
(x_{35}^2+p_{35}^2)
}
\\
&\quad\times
\bar C(k_1,k_2,ip_{13})\,
\bar C(ip_{13},ip_{35},ip_{15})\,
\bar C(ip_{35},k_3,k_4)\,
\bar C(ip_{15},k_5,k_6),\\
&=
\Bigl(
V_{12i}\,\Pi_{\bk_{12}}^{ij}\,
V_{jln}\,\Pi_{\bk_{34}}^{lm}\,
V_{m34}\,\Pi_{\bk_{56}}^{nr}\,
V_{r56}
\Bigr)
\psi_{6,\,12|34|56}^{\mathrm{scalar}}.
\end{aligned}
\end{equation}
Together, \eqref{eq:six-point-raylike-131415-gluing}--
\eqref{eq:six-point-131535-gluing} provide one representative for each maximal
six-point topology class; the remaining maximal cuts are generated by cyclic
images in \eqref{eq:six-point-minimal-cut-sum}.

\paragraph{Two-chord sectors.}

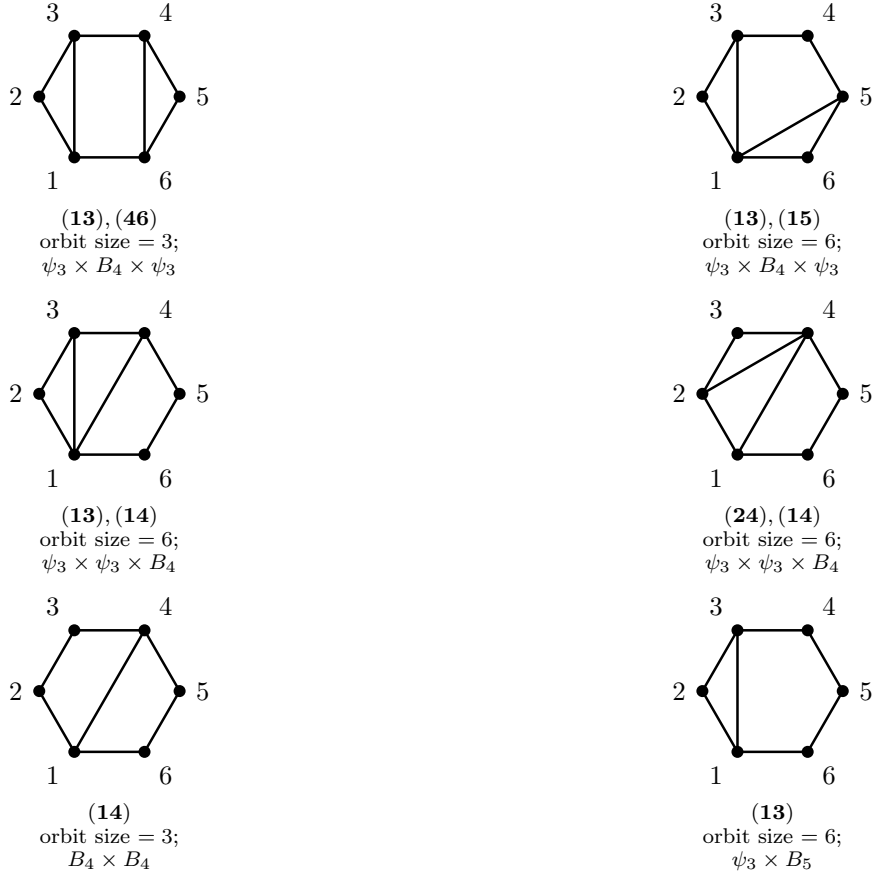
\begin{figure}[H]
\centering
\begin{minipage}[t]{0.42\textwidth}
\centering
\begin{tikzpicture}[scale=0.58]
  \coordinate (A) at (240:1.6);
  \coordinate (B) at (180:1.6);
  \coordinate (C) at (120:1.6);
  \coordinate (D) at (60:1.6);
  \coordinate (E) at (0:1.6);
  \coordinate (F) at (300:1.6);
  \coordinate (v1) at (A);
  \coordinate (v2) at (B);
  \coordinate (v3) at (C);
  \coordinate (v4) at (D);
  \coordinate (v5) at (E);
  \coordinate (v6) at (F);
  \draw[polygon edge] (A) -- (B) -- (C) -- (D) -- (E) -- (F) -- cycle;
  \foreach \a/\b in {1/3,4/6}{
    \draw[triangulation] (v\a) -- (v\b);
  }
  \node[vertex dot,label={[boundary label]below left:$1$}] at (A) {};
  \node[vertex dot,label={[boundary label]left:$2$}] at (B) {};
  \node[vertex dot,label={[boundary label]above left:$3$}] at (C) {};
  \node[vertex dot,label={[boundary label]above right:$4$}] at (D) {};
  \node[vertex dot,label={[boundary label]right:$5$}] at (E) {};
  \node[vertex dot,label={[boundary label]below right:$6$}] at (F) {};
\end{tikzpicture}

\scriptsize
\(\mathbf{(13),(46)}\)\\[-1pt]
orbit size \(= 3\);\\[-1pt]
\(\psi_3\times B_4\times\psi_3\)
\end{minipage}%
\hfill
\begin{minipage}[t]{0.42\textwidth}
\centering
\begin{tikzpicture}[scale=0.58]
  \coordinate (A) at (240:1.6);
  \coordinate (B) at (180:1.6);
  \coordinate (C) at (120:1.6);
  \coordinate (D) at (60:1.6);
  \coordinate (E) at (0:1.6);
  \coordinate (F) at (300:1.6);
  \coordinate (v1) at (A);
  \coordinate (v2) at (B);
  \coordinate (v3) at (C);
  \coordinate (v4) at (D);
  \coordinate (v5) at (E);
  \coordinate (v6) at (F);
  \draw[polygon edge] (A) -- (B) -- (C) -- (D) -- (E) -- (F) -- cycle;
  \foreach \a/\b in {1/3,1/5}{
    \draw[triangulation] (v\a) -- (v\b);
  }
  \node[vertex dot,label={[boundary label]below left:$1$}] at (A) {};
  \node[vertex dot,label={[boundary label]left:$2$}] at (B) {};
  \node[vertex dot,label={[boundary label]above left:$3$}] at (C) {};
  \node[vertex dot,label={[boundary label]above right:$4$}] at (D) {};
  \node[vertex dot,label={[boundary label]right:$5$}] at (E) {};
  \node[vertex dot,label={[boundary label]below right:$6$}] at (F) {};
\end{tikzpicture}

\scriptsize
\(\mathbf{(13),(15)}\)\\[-1pt]
orbit size \(= 6\);\\[-1pt]
\(\psi_3\times B_4\times\psi_3\)
\end{minipage}%

\vspace{0.45em}

\begin{minipage}[t]{0.42\textwidth}
\centering
\begin{tikzpicture}[scale=0.58]
  \coordinate (A) at (240:1.6);
  \coordinate (B) at (180:1.6);
  \coordinate (C) at (120:1.6);
  \coordinate (D) at (60:1.6);
  \coordinate (E) at (0:1.6);
  \coordinate (F) at (300:1.6);
  \coordinate (v1) at (A);
  \coordinate (v2) at (B);
  \coordinate (v3) at (C);
  \coordinate (v4) at (D);
  \coordinate (v5) at (E);
  \coordinate (v6) at (F);
  \draw[polygon edge] (A) -- (B) -- (C) -- (D) -- (E) -- (F) -- cycle;
  \foreach \a/\b in {1/3,1/4}{
    \draw[triangulation] (v\a) -- (v\b);
  }
  \node[vertex dot,label={[boundary label]below left:$1$}] at (A) {};
  \node[vertex dot,label={[boundary label]left:$2$}] at (B) {};
  \node[vertex dot,label={[boundary label]above left:$3$}] at (C) {};
  \node[vertex dot,label={[boundary label]above right:$4$}] at (D) {};
  \node[vertex dot,label={[boundary label]right:$5$}] at (E) {};
  \node[vertex dot,label={[boundary label]below right:$6$}] at (F) {};
\end{tikzpicture}

\scriptsize
\(\mathbf{(13),(14)}\)\\[-1pt]
orbit size \(= 6\);\\[-1pt]
\(\psi_3\times\psi_3\times B_4\)
\end{minipage}%
\hfill
\begin{minipage}[t]{0.42\textwidth}
\centering
\begin{tikzpicture}[scale=0.58]
  \coordinate (A) at (240:1.6);
  \coordinate (B) at (180:1.6);
  \coordinate (C) at (120:1.6);
  \coordinate (D) at (60:1.6);
  \coordinate (E) at (0:1.6);
  \coordinate (F) at (300:1.6);
  \coordinate (v1) at (A);
  \coordinate (v2) at (B);
  \coordinate (v3) at (C);
  \coordinate (v4) at (D);
  \coordinate (v5) at (E);
  \coordinate (v6) at (F);
  \draw[polygon edge] (A) -- (B) -- (C) -- (D) -- (E) -- (F) -- cycle;
  \foreach \a/\b in {2/4,1/4}{
    \draw[triangulation] (v\a) -- (v\b);
  }
  \node[vertex dot,label={[boundary label]below left:$1$}] at (A) {};
  \node[vertex dot,label={[boundary label]left:$2$}] at (B) {};
  \node[vertex dot,label={[boundary label]above left:$3$}] at (C) {};
  \node[vertex dot,label={[boundary label]above right:$4$}] at (D) {};
  \node[vertex dot,label={[boundary label]right:$5$}] at (E) {};
  \node[vertex dot,label={[boundary label]below right:$6$}] at (F) {};
\end{tikzpicture}

\scriptsize
\(\mathbf{(24),(14)}\)\\[-1pt]
orbit size \(= 6\);\\[-1pt]
\(\psi_3\times\psi_3\times B_4\)
\end{minipage}%

\vspace{0.45em}

\begin{minipage}[t]{0.42\textwidth}
\centering
\begin{tikzpicture}[scale=0.58]
  \coordinate (A) at (240:1.6);
  \coordinate (B) at (180:1.6);
  \coordinate (C) at (120:1.6);
  \coordinate (D) at (60:1.6);
  \coordinate (E) at (0:1.6);
  \coordinate (F) at (300:1.6);
  \coordinate (v1) at (A);
  \coordinate (v2) at (B);
  \coordinate (v3) at (C);
  \coordinate (v4) at (D);
  \coordinate (v5) at (E);
  \coordinate (v6) at (F);
  \draw[polygon edge] (A) -- (B) -- (C) -- (D) -- (E) -- (F) -- cycle;
  \foreach \a/\b in {1/4}{
    \draw[triangulation] (v\a) -- (v\b);
  }
  \node[vertex dot,label={[boundary label]below left:$1$}] at (A) {};
  \node[vertex dot,label={[boundary label]left:$2$}] at (B) {};
  \node[vertex dot,label={[boundary label]above left:$3$}] at (C) {};
  \node[vertex dot,label={[boundary label]above right:$4$}] at (D) {};
  \node[vertex dot,label={[boundary label]right:$5$}] at (E) {};
  \node[vertex dot,label={[boundary label]below right:$6$}] at (F) {};
\end{tikzpicture}

\scriptsize
\(\mathbf{(14)}\)\\[-1pt]
orbit size \(= 3\);\\[-1pt]
\(B_4\times B_4\)
\end{minipage}%
\hfill
\begin{minipage}[t]{0.42\textwidth}
\centering
\begin{tikzpicture}[scale=0.58]
  \coordinate (A) at (240:1.6);
  \coordinate (B) at (180:1.6);
  \coordinate (C) at (120:1.6);
  \coordinate (D) at (60:1.6);
  \coordinate (E) at (0:1.6);
  \coordinate (F) at (300:1.6);
  \coordinate (v1) at (A);
  \coordinate (v2) at (B);
  \coordinate (v3) at (C);
  \coordinate (v4) at (D);
  \coordinate (v5) at (E);
  \coordinate (v6) at (F);
  \draw[polygon edge] (A) -- (B) -- (C) -- (D) -- (E) -- (F) -- cycle;
  \foreach \a/\b in {1/3}{
    \draw[triangulation] (v\a) -- (v\b);
  }
  \node[vertex dot,label={[boundary label]below left:$1$}] at (A) {};
  \node[vertex dot,label={[boundary label]left:$2$}] at (B) {};
  \node[vertex dot,label={[boundary label]above left:$3$}] at (C) {};
  \node[vertex dot,label={[boundary label]above right:$4$}] at (D) {};
  \node[vertex dot,label={[boundary label]right:$5$}] at (E) {};
  \node[vertex dot,label={[boundary label]below right:$6$}] at (F) {};
\end{tikzpicture}

\scriptsize
\(\mathbf{(13)}\)\\[-1pt]
orbit size \(= 6\);\\[-1pt]
\(\psi_3\times B_5\)
\end{minipage}%
\caption{Lower-codimension six-point chord sectors used in the reconstruction.}
\end{figure}

\begin{equation}
\label{eq:six-point-1346-gluing}
\begin{aligned}
\psi_{6,\,13|46}^{\mathrm{YM}}(1,2,3,4,5,6)
&=
-\int_0^\infty
\frac{dp_{13}}{2\pi i}
\frac{dp_{46}}{2\pi i}\,
\frac{
V_{12i}\,
\Pi_{\bk_{12}}^{ij}\,
B_{4,j3l6}\,
\Pi_{\bk_{45}}^{lm}\,
V_{m45}
}{
(x_{13}^2+p_{13}^2)
(x_{46}^2+p_{46}^2)
}
\\
&\quad\times
\bar C(k_1,k_2,ip_{13})\,
\bar C(ip_{46},k_4,k_5).
\\
&=
\psi_{6,\,12|36|45}^{(\mu\mu)}
+
\left(
  \rho_{123654}
  +
   \rho_{216345}
\right)
\psi_{6,\,12|3|4|56}^{(\mu z\mu)}.
\end{aligned}
\end{equation}
where
\[
B_{4,j3l6}
:=
\frac{\partial}{\partial \beps_I^{j}}
\frac{\partial}{\partial \beps_J^{l}}\,
\Disc_{p_{13}}\Disc_{p_{46}}\,
B_4\!\left(
\begin{array}{c}
\beps_I,\beps_3,\beps_J,\beps_6\\
ip_{13},k_3,ip_{46},k_6\\
x_{14},x_{36}
\end{array}
\right).
\]

The Feynman-rule blocks used in the last line,
\(\psi_{6,\,12|36|45}^{(\mu\mu)}\) and
\(\psi_{6,\,12|3|4|56}^{(\mu z\mu)}\), are given in
\eqref{eq:six-point-chain-tr-mm-123645} and the indicated relabellings of
\eqref{eq:six-point-chain-mzm}. 

\begin{equation}
\label{eq:six-point-1315-gluing}
\begin{aligned}
\psi_{6,\,13|15}^{\mathrm{YM}}(1,2,3,4,5,6)
&=
-\int_0^\infty
\frac{dp_{13}}{2\pi i}
\frac{dp_{15}}{2\pi i}\,
\frac{
V_{12i}\,
\Pi_{\bk_{12}}^{ij}\,
B_{4,j34l}\,
\Pi_{\bk_{56}}^{lm}\,
V_{m56}
}{
(x_{13}^2+p_{13}^2)
(x_{15}^2+p_{15}^2)
}
\\
&\quad\times
\bar C(k_1,k_2,ip_{13})\,
\bar C(ip_{15},k_5,k_6).
\\
&=
\psi_{6,\,12|34|56}^{(\mu\mu)}
+
\psi_{6,\,12|3|4|56}^{(\mu z\mu)}
+
\rho_{345612}\,
\psi_{6,\,12|34|56}^{(\mu\mu z)}.
\end{aligned}
\end{equation}
where
\[
B_{4,j34l}
:=
\frac{\partial}{\partial \beps_I^{j}}
\frac{\partial}{\partial \beps_J^{l}}\,
\Disc_{p_{13}}\Disc_{p_{15}}\,
B_4\!\left(
\begin{array}{c}
\beps_I,\beps_3,\beps_4,\beps_J\\
ip_{13},k_3,k_4,ip_{15}\\
x_{14},x_{35}
\end{array}
\right).
\]

The three terms in the last line,
\(\psi_{6,\,12|34|56}^{(\mu\mu)}\),
\(\psi_{6,\,12|3|4|56}^{(\mu z\mu)}\), and
\(\psi_{6,\,12|34|56}^{(\mu\mu z)}\), are displayed in
\eqref{eq:six-point-chain-tr-mm-123456-pairs},
\eqref{eq:six-point-chain-mzm}, and
\eqref{eq:six-point-star-mmz}, respectively.

\begin{equation}
\label{eq:six-point-1314-gluing}
\begin{aligned}
\psi_{6,\,13|14}^{\mathrm{YM}}(1,2,3,4,5,6)
&=
-\int_0^\infty
\frac{dp_{13}}{2\pi i}
\frac{dp_{14}}{2\pi i}\,
\frac{
V_{12i}\,
\Pi_{\bk_{12}}^{ij}\,
V_{j3l}\,
\Pi_{\bk_{123}}^{lm}\,
B_{4,m456}
}{
(x_{13}^2+p_{13}^2)
(x_{14}^2+p_{14}^2)
}
\\
&\quad\times
\bar C(k_1,k_2,ip_{13})\,
\bar C(ip_{13},k_3,ip_{14}).
\\
&=
\psi_{6,\,12|3|456}^{(\mu\mu)}
\;+\;
\left(1+\rho_{123654}\right)
\psi_{6,\,12|3|4|56}^{(\mu\mu z)}.
\end{aligned}
\end{equation}
The two block types in the last line,
\(\psi_{6,\,12|3|456}^{(\mu\mu)}\) and
\(\psi_{6,\,12|3|4|56}^{(\mu\mu z)}\), are given in
\eqref{eq:six-point-chain-tr-mm-123456} and
\eqref{eq:six-point-chain-mmz-main}.  The reflected partner is
\begin{equation}
\label{eq:six-point-2414-gluing}
\psi_{6,\,24|14}^{\mathrm{YM}}
=
\rho_{321456}\,\psi_{6,\,13|14}^{\mathrm{YM}}.
\end{equation}

\paragraph{One-chord sectors.}

\begin{equation}
\label{eq:six-point-14-gluing}
\begin{aligned}
\psi_{6,\,14}^{\mathrm{YM}}(1,2,3,4,5,6)
&=
-i\int_0^\infty
\frac{dp_{14}}{2\pi i}\,
\frac{
B_{4,123i}\,
\Pi_{\bk_{123}}^{ij}\,
B_{4,j456}
}{
x_{14}^2+p_{14}^2
}\\
&=
\psi_{6,\,123|456}^{(\mu)}
+\left(
  1+\rho_{123654}+\rho_{321456}+\rho_{321654}
\right)
\psi_{6,\,12|3|4|56}^{(z\mu z)}
\\
&\quad
+\left(
  1+\rho_{321456}+\rho_{654321}
  +\rho_{456321}
\right)
\psi_{6,\,12|3|456}^{(z\mu)}.
\end{aligned}
\end{equation}
where
\[
\begin{aligned}
B_{4,123i}
&:=
\frac{\partial}{\partial \beps_I^{i}}\,
\Disc_{p_{14}}\,
B_4\!\left(
\begin{array}{c}
\beps_1,\beps_2,\beps_3,\beps_I\\
k_1,k_2,k_3,ip_{14}\\
x_{13},x_{24}
\end{array}
\right),\\
B_{4,j456}
&:=
\frac{\partial}{\partial \beps_J^{j}}\,
\Disc_{p_{14}}\,
B_4\!\left(
\begin{array}{c}
\beps_J,\beps_4,\beps_5,\beps_6\\
ip_{14},k_4,k_5,k_6\\
x_{15},x_{46}
\end{array}
\right).
\end{aligned}
\]

The three block types in the last two lines,
\(\psi_{6,\,123|456}^{(\mu)}\),
\(\psi_{6,\,12|3|4|56}^{(z\mu z)}\), and
\(\psi_{6,\,12|3|456}^{(z\mu)}\), are given in
\eqref{eq:six-point-chain-tr-mu-123456},
\eqref{eq:six-point-chain-zmuz-main}, and
\eqref{eq:six-point-chain-zmu-main}, with the displayed relabellings.

\begin{equation}
\begin{aligned}
\psi_{6,\,13}^{\mathrm{YM}}(1,2,3,4,5,6)
&=
-i\int_0^\infty
\frac{dp_{13}}{2\pi i}\,
\frac{
V_{12i}\,
\Pi_{\bk_{12}}^{ij}\,
B_{5,j3456}
}{
x_{13}^2+p_{13}^2
}
\bar C(k_1,k_2,ip_{13}).\\
&=
\left(
  1+\rho_{123654}
  +\rho_{216345}
  +\rho_{216543}
\right)
\psi_{6,\,12|3|4|56}^{(\mu zz)}
+\psi_{6,\,12|34|56}^{(\mu zz)}.
\end{aligned}
\label{eq:six-point-13-gluing}
\end{equation}
where
\[
B_{5,j3456}
:=
\frac{\partial}{\partial \beps_I^{j}}\,
\Disc_{p_{13}}\,
B_5\!\left(
\begin{array}{c}
\beps_I,\beps_3,\beps_4,\beps_5,\beps_6\\
\bk_{12},\bk_3,\bk_4,\bk_5,\bk_6\\
ip_{13},k_3,k_4,k_5,k_6\\
x_{14},x_{15},x_{35},x_{36},x_{46}
\end{array}
\right).
\]

The two block types in the last line,
\(\psi_{6,\,12|3|4|56}^{(\mu zz)}\) and
\(\psi_{6,\,12|34|56}^{(\mu zz)}\), are given in
\eqref{eq:six-point-chain-mzz-main} and
\eqref{eq:six-point-chain-muzz-pairs}, together with the displayed
relabellings.

Let
\[
E_6:=k_1+k_2+k_3+k_4+k_5+k_6,
\]
and let \(\sigma\) denote the one-step cyclic relabelling
\[
\sigma:\quad i\mapsto i+1\quad (\mathrm{mod}\ 6),
\qquad
x_{ij}\mapsto x_{i+1,j+1}.
\]
In the sums below, \(\sum_{\sigma\in\mathbb Z_m}\sigma[\cdots]\) means the sum
over the \(m\) distinct cyclic images generated by this relabelling.

Summing the maximally cut and lower-codimension sectors constructed above, we
obtain the cut-detectable six-point part
\begin{equation}
\label{eq:six-point-minimal-cut-sum}
\begin{aligned}
\psi_{6,\mathrm{cut}}^{\mathrm{YM}}
:={}&
\sum_{\text{cut sectors}}\psi_{6,\text{sector}}^{\mathrm{YM}}
\\
={}&
\sum_{\sigma\in\mathbb Z_6}\sigma\!\left[
  \psi_{6,\,13|14|15}^{\mathrm{YM}}\right]
+
\sum_{\sigma\in\mathbb Z_3}\sigma\!\left[
  \psi_{6,\,13|14|46}^{\mathrm{YM}}\right]
+
\sum_{\sigma\in\mathbb Z_3}\sigma\!\left[
  \psi_{6,\,24|14|15}^{\mathrm{YM}}\right]
+
\sum_{\sigma\in\mathbb Z_2}\sigma\!\left[
  \psi_{6,\,13|15|35}^{\mathrm{YM}}\right]
\\
&+
\sum_{\sigma\in\mathbb Z_3}\sigma\!\left[
  \psi_{6,\,13|46}^{\mathrm{YM}}\right]
+
\sum_{\sigma\in\mathbb Z_6}\sigma\!\left[
  \psi_{6,\,24|14}^{\mathrm{YM}}\right]
+
\sum_{\sigma\in\mathbb Z_6}\sigma\!\left[
  \psi_{6,\,13|15}^{\mathrm{YM}}\right]
+
\sum_{\sigma\in\mathbb Z_6}\sigma\!\left[
  \psi_{6,\,13|14}^{\mathrm{YM}}\right]
\\
&+
\sum_{\sigma\in\mathbb Z_3}\sigma\!\left[
  \psi_{6,\,14}^{\mathrm{YM}}\right]
+
\sum_{\sigma\in\mathbb Z_6}\sigma\!\left[
  \psi_{6,\,13}^{\mathrm{YM}}\right] .
\end{aligned}
\end{equation}

\paragraph{Zero-chord completion.}

After summing the cut polygons, the remaining zero-chord six-point sector is
determined by the requirement that the spurious higher-order OPE poles cancel
in the full answer.  In practice, this means subtracting the triple-OPE and
double-OPE pieces that remain in the minimal-basis expression obtained from the
cut contributions.

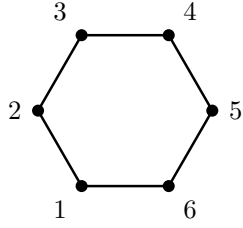
\begin{figure}[H]
\centering
\begin{tikzpicture}[scale=0.72]
  \coordinate (A) at (240:1.6);
  \coordinate (B) at (180:1.6);
  \coordinate (C) at (120:1.6);
  \coordinate (D) at (60:1.6);
  \coordinate (E) at (0:1.6);
  \coordinate (F) at (300:1.6);

  \draw[polygon edge] (A) -- (B) -- (C) -- (D) -- (E) -- (F) -- cycle;

  \node[vertex dot,label={[boundary label]below left:$1$}] at (A) {};
  \node[vertex dot,label={[boundary label]left:$2$}] at (B) {};
  \node[vertex dot,label={[boundary label]above left:$3$}] at (C) {};
  \node[vertex dot,label={[boundary label]above right:$4$}] at (D) {};
  \node[vertex dot,label={[boundary label]right:$5$}] at (E) {};
  \node[vertex dot,label={[boundary label]below right:$6$}] at (F) {};
\end{tikzpicture}
\caption{The zero-chord six-point polygon.  Its cyclic orbit has size \(6\).}
\end{figure}

First, the triple-OPE sector is
\[
B_6^{(3)}
=
\sum_{\sigma\in\mathbb Z_6}\sigma\!\left[B_{6,\,13|14|15}^{(3)}\right]
+
\sum_{\sigma\in\mathbb Z_3}\sigma\!\left[B_{6,\,13|14|46}^{(3)}\right]
+
\sum_{\sigma\in\mathbb Z_3}\sigma\!\left[B_{6,\,13|36|46}^{(3)}\right],
\]
with representatives
\[
B_{6,\,13|14|15}^{(3)}
=
\frac{
(k_1-k_2)(k_5-k_6)
(\beps_1\!\cdot\!\beps_2)
\bigl[\beps_3\!\cdot\!(\bk_1+\bk_2)\bigr]
\bigl[\beps_4\!\cdot\!(\bk_1+\bk_2+\bk_3)\bigr]
(\beps_5\!\cdot\!\beps_6)
}{
4E_6 x_{13}^2x_{14}^2x_{15}^2
},
\]
\[
B_{6,\,13|14|46}^{(3)}
=
\frac{
(k_1-k_2)(k_4-k_5)
(\beps_1\!\cdot\!\beps_2)
\bigl[\beps_3\!\cdot\!(\bk_1+\bk_2)\bigr]
(\beps_4\!\cdot\!\beps_5)
\bigl[\beps_6\!\cdot\!(\bk_4+\bk_5)\bigr]
}{
4E_6 x_{13}^2x_{14}^2x_{46}^2
},
\]
and
\[
B_{6,\,13|36|46}^{(3)}
=
-\frac{
(k_1-k_2)(k_4-k_5)
(\beps_1\!\cdot\!\beps_2)
\bigl[\beps_3\!\cdot\!(\bk_4+\bk_5)\bigr]
(\beps_4\!\cdot\!\beps_5)
\bigl[\beps_6\!\cdot\!(\bk_3+\bk_4+\bk_5)\bigr]
}{
4E_6 x_{13}^2x_{36}^2x_{46}^2
}.
\]

Second, the double-OPE sector is
\[
B_6^{(2)}
=
\sum_{\sigma\in\mathbb Z_6}\sigma\!\left[B_{6,\,13|15}^{(2)}\right]
+
\sum_{\sigma\in\mathbb Z_3}\sigma\!\left[B_{6,\,13|46}^{(2)}\right],
\]
with representatives
\[
B_{6,\,13|15}^{(2)}
=
-\frac{
(k_1-k_2)(k_5-k_6)
(\beps_1\!\cdot\!\beps_2)
(\beps_3\!\cdot\!\beps_4)
(\beps_5\!\cdot\!\beps_6)
}{
16E_6 x_{13}^2x_{15}^2
},
\]
and
\[
B_{6,\,13|46}^{(2)}
=
\frac{
(k_1-k_2)(k_4-k_5)
(\beps_1\!\cdot\!\beps_2)
(\beps_3\!\cdot\!\beps_6)
(\beps_4\!\cdot\!\beps_5)
}{
8E_6 x_{13}^2x_{46}^2
}.
\]

The zero-chord contact contribution is therefore
\[
\psi_{6,\mathrm{OPE}}^{\mathrm{YM}}
=
-B_6^{(3)}
-
B_6^{(2)}.
\]
This agrees with the Feynman-rule representative in
Appendix~\ref{Feynman_rule_for_YM}: Eq.~\eqref{eq:six-point-chain-mzz} gives the
double-OPE contact subtractions, while Eq.~\eqref{eq:six-point-chain-zzz}
gives the triple-OPE completion.

Thus the full six-point Yang--Mills wavefunction is
\[
\psi_6^{\mathrm{YM}}
=
\psi_{6,\mathrm{cut}}^{\mathrm{YM}}
+
\psi_{6,\mathrm{OPE}}^{\mathrm{YM}}.
\]
The full data are provided in the ancillary file
\texttt{5ptYM and 6pt YM data.zip}.  One can check that this full answer
satisfies the single-OPE constraints:
the OPE singular coefficient in every single-channel limit vanishes after the
cyclic sums in \(B_6^{(3)}\) and \(B_6^{(2)}\) are included, and the result also
reproduces the expected soft behavior and correct flat-space limit.

\subsection{All-\texorpdfstring{$n$}{n} structure suggested by low points}
\label{subsec:all-n-structure}

The low-point results suggest that the useful scalar reference theory is not
pure \(\phi^3\), but the planar scalar theory with cubic and quartic vertices.
At fixed cyclic ordering, the scalar sectors are dissections of an \(n\)-gon
into triangles and quadrilaterals.  For the first three nontrivial
multiplicities this gives
\[
\begin{array}{c|c|c|c|c}
n & \text{all cubic} & \text{one quartic vertex}
& \text{two quartic vertices} & \text{total}\\ \hline
4 & 2 & 1 & 0 & 3\\
5 & 5 & 5 & 0 & 10\\
6 & 14 & 21 & 3 & 38
\end{array}
\]
Thus the no-dashed Yang--Mills terms follow the same counting as the
color-ordered \(\phi^3+\phi^4\) scalar wavefunction.  Each such scalar tubing
carries the total-energy and partial-energy poles, while Yang--Mills supplies a
local numerator built from cubic or quartic vertices and transverse projectors.

The longitudinal sector acts on a smaller subset of these scalar tubings.  In a
candidate replacement, one or more
transverse internal propagators are replaced by dashed propagators.  Most
candidate replacements are absent or vanish, because the Yang--Mills three- and
four-point vertices have special longitudinal contractions.  When the
replacement is nonzero and the dashed line touches an external rooted block, the
effect is still simple: the dashed propagator collapses the original tubing to a
more contact-like scalar wavefunction, and the numerator is reduced by the local
collapse factors reviewed in Appendix~\ref{appendixA1}.  At five points this is
already visible in the mixed \(3\ast3\ast3\) graphs, which reuse the scalar
wavefunction of the \(3\ast4\) sector, and in the fully longitudinal graph,
which is proportional to the contact scalar factor \(1/E_5\).

Six points make the same rule more explicit.  The ordinary no-dashed sectors are
the \(38\) planar \(\phi^3+\phi^4\) scalar tubings listed above.  The nonzero
dashed replacements include single- and double-dashed variants of the chain and
star representatives.  The terms in Eqs.~\eqref{eq:six-point-chain-mmz} and
\eqref{eq:six-point-chain-mzz}, for example, still have the expected form:
a simplified Yang--Mills numerator multiplying an ordinary scalar tubing or a
collapsed contact-type scalar tubing.  Purely longitudinal chains, such as
Eq.~\eqref{eq:six-point-pure-zz}, reduce further to zero-chord contact
contributions.

The first correction to this proportionality appears when the dashed propagator
is purely internal.  In that case the local collapse is no longer simply a
Yang--Mills numerator times the scalar wavefunction associated with the original
tubing.  The six-point examples in Eqs.~\eqref{eq:six-point-chain-mzm} and
\eqref{eq:six-point-star-mmz} show this directly: the answer can still be
written as a local operation on scalar tubings, but it contains an additional
internal-collapse correction.  This correction is not an arbitrary new pole
structure.  It is localized, tied to the dashed internal line, and constrained
by current conservation and by the cancellation of spurious OPE singularities.

The resulting all-\(n\) pattern is therefore concrete.  One starts with the
planar \(\phi^3+\phi^4\) scalar tubings and dresses each no-dashed tubing by its
Yang--Mills numerator.  One then applies the allowed dashed-propagator
replacements: many vanish, while the nonzero replacements either collapse to
more contact-like scalar wavefunctions or, from six points onward, generate
localized internal-collapse corrections when the dashed line is purely internal.
Schematically,
\[
\psi_n^{\mathrm{YM}}
=
\sum_{\Gamma\in{\cal T}^{\phi^3+\phi^4}_n}
N_\Gamma^{\mathrm{YM}}\,
\psi_\Gamma^{\phi^3+\phi^4}
+
\sum_{\Gamma\in{\cal D}_n}
\widetilde N_\Gamma^{\mathrm{YM}}\,
\psi_{\mathrm{coll}(\Gamma)}^{\phi^3+\phi^4}
+
\Delta_n^{\mathrm{int}} .
\]
Here \({\cal T}^{\phi^3+\phi^4}_n\) denotes ordinary planar scalar tubings,
\({\cal D}_n\) denotes the smaller set of nonzero dashed replacements, and
\(\Delta_n^{\mathrm{int}}\) denotes the purely internal dashed-line corrections.
The pole structure is therefore largely inherited from scalar wavefunctions.
The remaining all-\(n\) problem is to organize the Yang--Mills numerators and the
sparse internal-collapse corrections, perhaps through graph relations analogous
to BCJ Jacobi identities {\cite{Bern:2008qj,Armstrong:2020woi,
Albayrak:2020fyp,Diwakar:2021juk,Herderschee:2022ntr,
Alday:2021odx,Zhou:2021gnu}} or
through a compact set of local moves on scalar tubings.

\section{Conclusions and outlook}

In this paper we studied tree-level Yang--Mills de Sitter wavefunction
coefficients in momentum space from the viewpoint of discontinuities.
Our main goal was to turn the general cutting formalism into an explicit
reconstruction framework for spinning observables.  Concretely, we formulated
the gluing rule for gluon discontinuities, worked out representative maximal
and non-maximal cuts, reconstructed the four-, five-, and six-point
wavefunction coefficients, and compared the resulting expressions with direct
momentum-space Feynman-rule computations.  Along the way, the examples revealed
an organizing principle that appears to persist beyond low multiplicity: a
cut-detectable part controlled by lower-point gluing data together with a
cut-invisible completion fixed by current conservation, spurious-pole
cancellation, local longitudinal-sector corrections, and the flat-space limit.
The explicit results through six points should therefore be viewed both as a
concrete validation of the discontinuity method and as low-point data from which
a more systematic all-\(n\) organization can be sought.

The low-point answers also point to a possible geometric interpretation.  In
our representation, many terms look like cosmological-polytope-type scalar
objects dressed by Yang--Mills numerators.  This suggests asking whether the
dressed graphs can be combined into a single geometric or combinatorial
object, rather than organized diagram by diagram.  Such a structure would have
to encode the gluing/completion split directly in momentum space, for example
through its boundaries, canonical form, or stratification.  Even a partial
answer would clarify how far the scalar polytope picture extends once tensor
numerators are included
\cite{Arkani-Hamed:2017fdk,Arkani-Hamed:2018bjr,Benincasa:2024leu,Arkani-Hamed:2024jbp}.

One natural next step is to pass from wavefunctions to correlators.  Recent work
has developed several tools for this purpose, including dressing rules,
dispersive reconstruction, physical cut bases, and correlator discontinuities
\cite{Stefanyszyn:2024msm,Stefanyszyn:2023qov,Tong:2021wai,Ema:2024hkj,
Das:2025qsh,Colipi-Marchant:2025oin,Chowdhury:2026upp,Das:2026vfv,
Ansari:2026xkm,De:2024zic,Jain:2025maa,Chowdhury:2023arc,
Chowdhury:2025ohm,Liu:2024xyi,Xianyu:2022jwk}.  The explicit wavefunction data
constructed here provide a momentum-space starting point for an analogous
reconstruction of spinning correlators.

Gravity is another immediate target.  The ingredients used here---factorized
discontinuities, gluing of lower-point building blocks, and a cut-invisible
completion sector---should have gravitational analogues, although the tensor
algebra is more restrictive and the size of the completion is not yet clear.
Existing results on spinning \((A)dS\) observables,
on-shell bootstrap constructions for gluons and gravitons, and
amplitude-inspired routes from \(AdS\) data to cosmological observables give
several possible starting points
\cite{Baumann:2020dch,Baumann:2024ttn,Armstrong:2022mfr,Armstrong:2023phb,
Chowdhury:2024wwe,Mei:2024abu,Mei:2024sqz,Albayrak:2018tam,Mei:2023jkb}. A concrete question is whether any double-copy-like organization survives at the
level of de Sitter wavefunctions.

There are also kinematical and representation-theoretic refinements to pursue.
One is a spinor-helicity description of the results obtained here.  Since
low-point answers often simplify in that language, special helicity sectors,
especially all-plus configurations, may admit a more compact reconstruction
from discontinuity data.  Another is the possible relation to
Grassmannian- or on-shell-diagram-like structures.  In flat-space amplitudes,
such frameworks reorganize factorization data more efficiently than
Feynman-diagram expansions; the open question is whether an analogue remains
for cosmological observables in momentum space
\cite{Arundine:2026fbr,Huang:2026tsh,Bala:2026bdx,Bala:2026hdm}.

Finally, the relation to positive geometry remains open.  Some aspects of the
organization are more transparent in Mellin space than in momentum space, but
the momentum-space formulas here make the gluing data and completion terms
explicit enough to test proposed geometric descriptions directly
\cite{Arkani-Hamed:2017fdk,Arkani-Hamed:2018bjr,Arkani-Hamed:2024jbp,Cao:2025lzv}.
At present, the precise relation to scaffolding-type constructions in momentum
space remains unclear and deserves a more focused future discussion
\cite{Arkani-Hamed:2023jry,He:2026dql}.

The main lesson is that discontinuity data for spinning de Sitter observables
are not only consistency checks.  They are explicit enough to reconstruct
momentum-space wavefunctions and structured enough to expose patterns that are
hard to see in a direct Feynman-diagram expansion.

\acknowledgments
We would like to thank Qu Cao for collaboration at an early stage of this
project.  S.H. and Y.M. are supported by the National Natural Science Foundation
of China under Grant Nos. 12225510 and 12447101, and by the New Cornerstone
Science Foundation.  J.M. is supported by the European Union (ERC, UNIVERSE
PLUS, 101118787).
Views and opinions expressed, however, are those of the authors only and do not
necessarily reflect those of the European Union or the European Research Council
Executive Agency.  Neither the European Union nor the granting authority can be
held responsible for them.

\appendix
\section{Momentum-space Feynman rules for Yang--Mills theory}
\label{Feynman_rule_for_YM}
For convenience, we collect here the conventions and elementary
momentum-space Feynman rules used throughout the main text.  We work in the
radial coordinate \(z\), with \(k=|\bk|\) denoting the boundary energy
associated with a spatial momentum \(\bk\).  The external bulk-to-boundary
propagator is
\begin{equation}
K^i(z)=\beps^i e^{-zk},
\end{equation}
while the transverse bulk-to-bulk propagator is written in spectral form as
\begin{equation}
G^\perp_{ij}(z_1,z_2;k_s)
:=
\Pi_{ij}^{(k_s)}
\int_{-\infty}^{\infty}\frac{dp}{\pi}\,
\frac{\sin(pz_1)\sin(pz_2)}
{(k_s^2+p^2)},
\qquad
\Pi_{ij}^{(k_s)}
=
\delta_{ij}
-\frac{k_s^i k_s^j}{k_s^2}.
\end{equation}
The longitudinal propagator is local in the radial coordinate:
\begin{equation}
G^\parallel_{zz}(z_1,z_2;k_I)
=
\frac{\delta(z_1-z_2)}{k_I^2}.
\label{eq:ym-longitudinal-propagator}
\end{equation}
Each bulk vertex is accompanied by an integration,
\begin{equation}
\int_0^\infty dz .
\end{equation}
\paragraph{Three-point vertices.}

Latin indices denote spatial components, while a \(z\) index denotes the radial
direction.  With all momenta taken outgoing, the cubic vertices that enter our
computations are
\begin{align}
V_{ij\ell}(\bk_1,\bk_2,\bk_3)
&=
\frac{1}{2}\Bigl[
\delta_{ij}(\bk_1^\ell-\bk_2^\ell)
+\delta_{j\ell}(\bk_2^i-\bk_3^i)
+\delta_{\ell i}(\bk_3^j-\bk_1^j)
\Bigr], 
\label{eq:vertex-vijk}
\\
V_{ijz}&=
-\frac{i}{2}\,\delta_{ij}\,
\bigl(\partial_{z_i}-\partial_{z_j}\bigr),
\label{eq:vertex-vijz-radial}
\\
V_{izz}(\bk_2,\bk_3)
&=
\frac{1}{2}\bigl(\bk_2^i-\bk_3^i\bigr).
\end{align}
\begin{figure}[H]
\centering
\begin{tikzpicture}[scale=0.95]
  \coordinate (V) at (0,0);
  \draw[scalar] (V) -- (0,1.35);
  \draw[gluon] (-2.2,0) -- (V);
  \draw[gluon] (V) -- (2.2,0);
  \fill (V) circle (1.5pt);
  \node[above] at (0,1.35) {\small \(z\)};
  \node[below] at (-1.75,-0.10) {\small \(j\)};
  \node[below] at (1.75,-0.10) {\small \(i\)};
  \node[above] at (-0.58,0.28) {\small \(\partial_{z_j}\)};
  \node[above] at (0.58,0.28) {\small \(\partial_{z_i}\)};
\end{tikzpicture}
\caption{Radial-derivative assignment for the mixed vertex
\eqref{eq:vertex-vijz-radial}.  When the \(i\) and \(j\) legs are external, the
derivatives act directly on the corresponding bulk-to-boundary propagators, giving
\eqref{eq:vijz-on-external-btb}.  When one or both of these legs is internal, the
derivative is understood to act from the vertex endpoint on the adjacent
transverse bulk-to-bulk propagator.}
\end{figure}
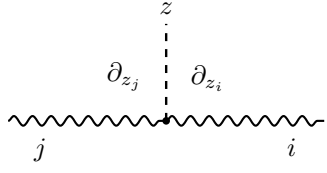

\paragraph{Four-point vertex.}

The quartic Yang--Mills vertex is
\begin{equation}
V_{\mu\nu\rho\sigma}
=
\frac{1}{2}\eta_{\mu\rho}\eta_{\nu\sigma}
-\frac{1}{4}\bigl(\eta_{\mu\nu}\eta_{\rho\sigma}+\eta_{\mu\sigma}\eta_{\nu\rho}\bigr).
\end{equation}

Here Greek indices run over both spatial and radial directions,
\(\mu=i,z\), while Latin indices \(i,j,k,\ell,m,\ldots\) are spatial.
The metric is the Minkowski metric, and we use \(\eta_{zz}=1\) and
\(\eta_{ij}=\delta_{ij}\).

Since \(\eta_{zi}=0\), the only nonvanishing mixed \(z/i\) components of
\(V_{\mu\nu\rho\sigma}\), apart from the purely spatial \(V_{ijkl}\), are
\begin{equation}
V_{zzij}=V_{zijz}=V_{izzj}=V_{ijzz}
=
-\frac{1}{4}\eta_{zz}\eta_{ij},
\qquad
V_{zizj}=V_{izjz}
=
\frac{1}{2}\eta_{zz}\eta_{ij}.
\end{equation}

All remaining mixed components vanish; in particular, any component with an odd number of
$z$ indices is zero, and $V_{zzzz}=0$.

The propagator factors above should be multiplied by the appropriate external
waves and integrated over the radial coordinates of all vertices.

\subsection{Special cases of internal longitudinal propagator}
\label{appendixA1}

\noindent\textbf{Mixed-vertex radial rule and collapse identity.}

We now spell out the elementary identity responsible for the collapse terms
appearing in the main text.  In radial space the mixed cubic vertex with one
radial index is represented by the differential operator
\begin{equation}
  \label{eq:vijz-radial-rule}
  V_{ijz}
  =
  -\frac{i}{2}\,\delta_{ij}\,
  \bigl(\partial_{z_i}-\partial_{z_j}\bigr).
\end{equation}
When both legs adjacent to the derivative are external, this operator acts on
the two bulk-to-boundary propagators as
\begin{equation}
  \label{eq:vijz-on-external-btb}
  -\frac{i}{2}
  \bigl(\partial_{z_i}-\partial_{z_j}\bigr)
  \Bigl(e^{-k_i z}e^{-k_j z}\Bigr)
  =
  \frac{i}{2}(k_i-k_j)\,e^{-k_i z}e^{-k_j z},
\end{equation}
which reproduces the corresponding momentum-space mixed-vertex factor with
$q_i=k_i$ and $q_j=k_j$.

The same derivative can instead hit an adjacent internal transverse line.  For
this purpose it is enough to use the scalar part of the spectral kernel,
\begin{equation}
  G(z_1,z_2;k_s)
  :=
  \int_{-\infty}^{\infty}\frac{dp}{\pi}\,
  \frac{\sin(p z_1)\sin(p z_2)}{2i\,(k_s^2+p^2)}.
\end{equation}
Acting with the sum of derivatives at the two endpoints gives
\begin{equation}
  \label{eq:btb-derivative-sum}
  \begin{aligned}
    (\partial_{z_1}+\partial_{z_2})\,G(z_1,z_2;k_s)
    &=
    \int_{-\infty}^{\infty}\frac{dp}{\pi}\,
    \frac{
      p\cos(p z_1)\sin(p z_2)
      +
      p\sin(p z_1)\cos(p z_2)
    }{2i\,(k_s^2+p^2)}\\
    &=
    \int_{-\infty}^{\infty}\frac{dp}{\pi}\,
    \frac{p\,\sin\!\bigl(p(z_1+z_2)\bigr)}{2i\,(k_s^2+p^2)}\\
    &=
    -\frac{i}{2}
    \int_{-\infty}^{\infty}\frac{dp}{\pi}\,
    \frac{p\,\sin\!\bigl(p(z_1+z_2)\bigr)}{k_s^2+p^2}.
  \end{aligned}
\end{equation}
For \(L>0\), the standard Fourier integral is
\begin{equation}
  \int_{-\infty}^{\infty}\frac{dp}{\pi}\,
  \frac{\cos(pL)}{k_s^2+p^2}
  =
  \frac{e^{-k_s L}}{k_s},
\end{equation}
and differentiating with respect to \(L\) gives
\begin{equation}
  \int_{-\infty}^{\infty}\frac{dp}{\pi}\,
  \frac{p\,\sin(pL)}{k_s^2+p^2}
  =
  e^{-k_s L}.
\end{equation}
Setting \(L=z_1+z_2\) in \eqref{eq:btb-derivative-sum}, we therefore obtain
\begin{equation}
  \label{eq:btb-collapse-source}
  (\partial_{z_1}+\partial_{z_2})\,G(z_1,z_2;k_s)
  =
  -\frac{i}{2}\,e^{-k_s(z_1+z_2)}.
\end{equation}
Equivalently, the same propagator can be written as
\begin{equation}
  G(z_1,z_2;k_s)
  =
  -\frac{i}{4k_s}
  \Bigl(
    e^{-k_s|z_1-z_2|}
    -
    e^{-k_s(z_1+z_2)}
  \Bigr),
\end{equation}
so \((\partial_{z_1}+\partial_{z_2})\) annihilates the direct piece
\(e^{-k_s|z_1-z_2|}\) and leaves only the image term
\(e^{-k_s(z_1+z_2)}\).  This surviving exponential is the
propagator-collapse contribution packaged below into the
operators $\widehat{\mathcal C}_I$.
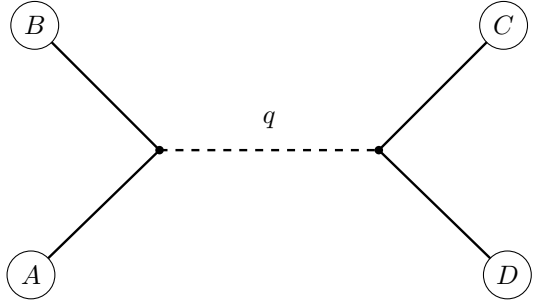
\DiagramFigure{\LocalOperatorDoubleSketch}{Schematic rooted-branch configuration for the local operator rule with two mixed vertices sharing a dashed propagator.}{fig:local-operator-double-schematic}

After the radial integration, it is useful to package a mixed cubic vertex $(V_{ijz})$ with
one radial index as a local operator acting on a rooted scalar skeleton.  For
two rooted branches \(I,J\) meeting at such a vertex, as in
Fig.~\ref{fig:local-operator-double-schematic}, define
\begin{equation}
  \label{eq:local-operator-single}
  \widehat{\mathbb V}_{I,J}
  :=
  \frac{1}{4}
  \Big[
    (k_I-\widehat{\mathcal C}_I)
    -
    (k_J-\widehat{\mathcal C}_J)
  \Big],
\end{equation}
Here \(I,J\) stand for either the pair \(A,B\) or the pair \(C,D\) in
Fig.~\ref{fig:local-operator-double-schematic}, and \(k_I,k_J\) denote the sums
of the energies carried by the outermost external legs of those rooted
branches.  The operators
$\widehat{\mathcal C}_I$ and $\widehat{\mathcal C}_J$ are the
propagator-collapse operators associated with the two rooted branches, namely
the extra terms generated when one moves $\partial_z$ outward so that it acts
on the outermost external leg. Each time the derivative crosses an internal
transverse propagator line, one picks up the corresponding product of
bulk-to-boundary propagators.
Let \(\psi_{0}^{\mathrm{cs}}\) denote the rooted scalar skeleton obtained
after shrinking the dashed line.

For a pair of mixed vertices sharing the same dashed propagator $1/q^2$, with
four rooted branches $A,B,C,D$, this becomes
\begin{equation}
  \label{eq:local-operator-double}
  \psi_{(A,B)\,||\,(C,D)}^{\mathrm{sc}}
  =
  \frac{1}{4\,q^2}\,
  \widehat{\mathbb V}_{A,B}\,
  \widehat{\mathbb V}_{C,D}\,
  \psi_{0}^{\mathrm{cs}}.
\end{equation}
Expanding the two operators gives
\begin{equation}
  \label{eq:local-operator-double-expanded}
  \begin{aligned}
    \psi_{(A,B)\,||\,(C,D)}^{\mathrm{sc}}
    &=
    \frac{1}{4\,q^2}
    \Big[
      \bigl((k_A-k_B)-\widehat{\mathcal C}_A+\widehat{\mathcal C}_B\bigr)
      \bigl((k_C-k_D)-\widehat{\mathcal C}_C+\widehat{\mathcal C}_D\bigr)
    \Big]
    \psi_{0}^{\mathrm{cs}}\\
    &=
    \frac{1}{4\,q^2}
    \Big[
      (k_A-k_B)(k_C-k_D)\,\psi_{0}^{\mathrm{cs}}
      -(k_A-k_B)(\widehat{\mathcal C}_C-\widehat{\mathcal C}_D)\,\psi_{0}^{\mathrm{cs}}\\
    &\hspace{3.7em}
      -(k_C-k_D)(\widehat{\mathcal C}_A-\widehat{\mathcal C}_B)\,\psi_{0}^{\mathrm{cs}}
      +(\widehat{\mathcal C}_A-\widehat{\mathcal C}_B)
       (\widehat{\mathcal C}_C-\widehat{\mathcal C}_D)\,\psi_{0}^{\mathrm{cs}}
    \Big].
  \end{aligned}
\end{equation}
We have checked this relation against \eqref{eq:six-point-chain-mzm}.

Diagrammatically this reads:

\DiagramFigure{\LocalOperatorExpansionSketch}
{Diagrammatic expansion of the double local-operator rule in
eq.~\eqref{eq:local-operator-double-expanded}, after shrinking the shared dashed
propagator to a point. A slash on branch $x$ denotes the action of the collapse
operator $\widehat{\mathcal C}_x$.}
{fig:local-operator-double-expanded-sketch}

\subsection{Summary of \texorpdfstring{$n=4,5,6$}{n=4,5,6} Feynman diagrams}
This subsection records the direct Feynman-rule organization of the
$n=4,5,6$ gluon wavefunctions reconstructed in
Sections~\ref{subsec:four-point-reconstruction}--\ref{subsec:six-point-reconstruction}.
Wavy internal lines
denote transverse propagation, while dashed internal lines denote the
longitudinal sector.  The comparison shows how the polygonal reconstruction
repackages ordinary Feynman diagrams: cut polygons reproduce the transverse and
mixed exchange sectors, whereas the zero-chord terms collect the longitudinal
and contact completions needed for current conservation.  We keep the detailed
diagram-by-diagram formulas here so that the main text can focus on the
reconstruction algorithm and the emerging all-\(n\) pattern.

In this appendix we use \(E_n:=\sum_{a=1}^n k_a\) for the total energy, and we
reserve rooted symbols such as \(\uk{12}\) and \(\uk{123}\) for scalar channel
magnitudes only. Whenever a quantity carries a spatial index, such as a
projector or momentum contraction, we write the corresponding vector
explicitly as \(\bk_{12}\), \(\bk_{123}\), and so on.

\noindent\textbf{Four-point diagrams}

At four points, the transverse exchange contribution is obtained by gluing two
triangles.  The remaining longitudinal-exchange and contact diagrams combine
into the zero-chord boundary term, collected by the quadrilateral.

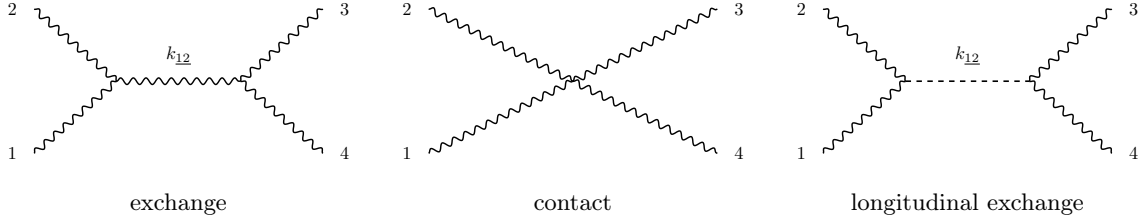
\begin{center}
  \centering
  \CompactDiagramCell{\FourPointExchange}{exchange}
  \hfill
  \CompactDiagramCell{\FourPointContact}{contact}
  \hfill
  \CompactDiagramCell{\FourPointDashedExchange}{longitudinal exchange}
\end{center}

{ We define the four-point tubing}
\[
{
\mathcal T^{(4)}(K_L,K_R;u)
:=
\frac{1}{(K_L+K_R)(K_L+u)(K_R+u)} .
}
\]
Then 
\begin{equation}
  \label{eq:four-point-s-channel-tr}
  \begin{aligned}
    \psi_{4,s}^{\mathrm{YM}}
    &=
   {   V_{12\mu}\,\Pi_{\bk_{12}}^{\mu\nu}\,V_{\nu34}\,
    \mathcal T^{(4)}(k_1+k_2,k_3+k_4;\uk{12})
    =
    \frac{  V_{12\mu}\,\Pi_{\bk_{12}}^{\mu\nu}\,V_{\nu34}}{E_4\,(\uk{12}+k_1+k_2)\,(\uk{12}+k_3+k_4)}},\\
    \psi_4^{\mathrm{c}}
    &=
   \frac{1}{2E_4}
    \left[
      (\beps_1\!\cdot\!\beps_3)(\beps_2\!\cdot\!\beps_4)
      - \frac{1}{2}(\beps_1\!\cdot\!\beps_2)(\beps_3\!\cdot\!\beps_4)
      - \frac{1}{2}(\beps_1\!\cdot\!\beps_4)(\beps_2\!\cdot\!\beps_3)
    \right],\\
    \psi_{4,s}^{(z)}
    &
    =
    -\frac{(\beps_1\!\cdot\!\beps_2)(\beps_3\!\cdot\!\beps_4)
    (k_1-k_2)(k_3-k_4)}{4\,\uk{12}^2\,E_4},
  \end{aligned}
\end{equation}
\noindent Thus the four-point answer already exhibits the basic split between an
exchange piece carrying the nontrivial channel dependence and a completion
piece proportional to the contact scalar factor $1/E_4$.

\noindent\textbf{Five-point diagrams}

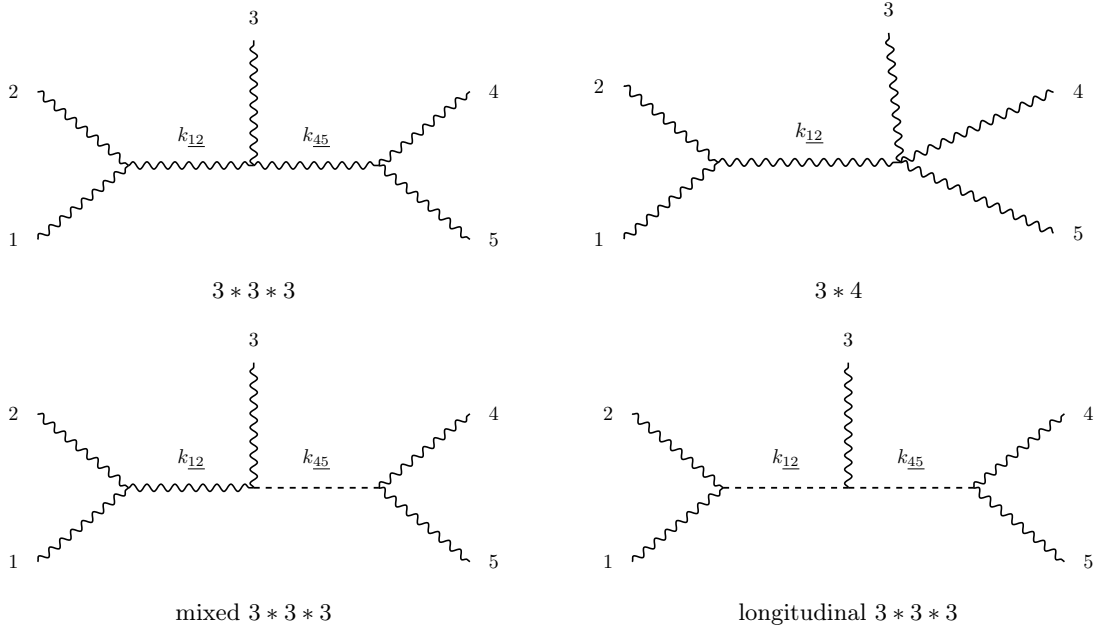
\begin{center}
  \centering
  \CompactDiagramCellWide{\FivePointDoubleExchange{gluon}{gluon}}{$3\ast3\ast3$}
  \hfill
  \CompactDiagramCellWide{\FivePointSingleExchange}{$3\ast4$}

  \vspace{0.45em}

  \CompactDiagramCellWide{\FivePointDoubleExchange{gluon}{scalar}}{mixed $3\ast3\ast3$}
  \hfill
  \CompactDiagramCellWide{\FivePointDoubleExchange{scalar}{scalar}}{longitudinal $3\ast3\ast3$}
\end{center}

At five points it is useful to package the two $3\ast3\ast3$ tubings once and
for all. They are the canonical scalar factors underlying the maximally cut
ray-like sector, and they also reappear in the mixed contributions. For three
rooted blocks with external-energy sums $K_L,K_M,K_R$, two transverse channels
$u,v$, and total energy
\[
  E_5:=K_L+K_M+K_R,
\]
define
\begin{equation}
  \label{eq:five-point-tubing-def}
  \begin{aligned}
    \mathcal T_1^{(5)}(K_L,K_M,K_R;u,v)
    &:=
    \frac{1}
    {E_5\,(K_L+u)\,(K_M+K_R+u)\,(K_R+v)\,(K_M+u+v)},\\
    \mathcal T_2^{(5)}(K_L,K_M,K_R;u,v)
    &:=
    \frac{1}
    {E_5\,(K_L+u)\,(K_L+K_M+v)\,(K_R+v)\,(K_M+u+v)}.
  \end{aligned}
\end{equation}
\begin{equation}
      { \psi_{5,\,12|3|45}^{\mathrm{scalar}}}
  :=
  \mathcal T_1^{(5)}(k_1+k_2,k_3,k_4+k_5;\uk{12},\uk{45})
  +
  \mathcal T_2^{(5)}(k_1+k_2,k_3,k_4+k_5;\uk{12},\uk{45}),
\end{equation}
\begin{equation}
     { \psi_{5,\,12|345}^{\mathrm{scalar}}}
  :=
  { \mathcal T^{(4)}(k_1+k_2,k_3+k_4+k_5;\uk{12})}.
  \label{eq:five-point-one-chord-scalar}
\end{equation}
\begin{equation}
  \label{eq:five-point-double-tr}
  \begin{aligned}
    { \psi_{5,\,12|3|45}^{\mathrm{YM}}}
    &=
    V_{12\mu}\,\Pi_{\bk_{12}}^{\mu\nu}\,V_{\nu 3 \rho}\,\Pi_{\bk_{45}}^{\rho\sigma}\,V_{\sigma45}\,
    { \psi_{5,\,12|3|45}^{\mathrm{scalar}}},\\
    { \psi_{5,\,12|345}^{\mathrm{YM}}}
    &=
    V_{12\mu}\,\Pi_{\bk_{12}}^{\mu\nu}\,V_{\nu345}\,
    { \psi_{5,\,12|345}^{\mathrm{scalar}}},
  \end{aligned}
\end{equation}
so the purely transverse sector consists of the $3\ast3\ast3$ topology and the
$3\ast4$ topology. This is the first multiplicity at which two distinct
gluing sectors coexist: a fully triangulated sector built from maximal-cut
data and a one-chord sector whose scalar factor is the four-point tubing
\(\mathcal T^{(4)}\) evaluated on the enlarged right block
\(k_3+k_4+k_5\), as in \eqref{eq:five-point-one-chord-scalar}.
\begin{equation}
  \label{eq:five-point-double-muz}
  \begin{aligned}
    { \psi_{5,\,12|3|45}^{(\mu z)}}
    &=
    \left(V_{12\mu}\,\Pi_{\bk_{12}}^{\mu\nu}\,\beps_{3\nu}\right)(\beps_4\!\cdot\!\beps_5)
    \left[\frac{(k_4-k_5)(2k_3+k_4+k_5)}{4\,\uk{45}^2}\right]
    { \psi_{5,\,12|345}^{\mathrm{scalar}}},\\
    { \psi_{5,\,12|3|45}^{(z\mu)}}
    &=
    \left.{ \psi_{5,\,12|3|45}^{(\mu z)}}\right|_{(1,2)\leftrightarrow(5,4),\,\uk{12}\leftrightarrow\uk{45}},
  \end{aligned}
\end{equation}
Thus the mixed $3\ast3\ast3$ graphs do not introduce a new scalar denominator
structure; they reuse the same scalar wavefunction as the $3\ast4$ graph, but
with a different tensor numerator.
\begin{equation}
  \label{eq:five-point-double-zz}
 { \psi_{5,\,12|3|45}^{(zz)}}
  =
  \frac{V_{12z}\,V_{z3z}\,V_{z45}}{\uk{12}^2\,\uk{45}^2\,E_5}
  =
  -\frac{(\beps_1\!\cdot\!\beps_2)\,
  (\beps_3\!\cdot\!(\bk_{45}-\bk_{12}))\,
  (\beps_4\!\cdot\!\beps_5)\,
  (k_1-k_2)(k_4-k_5)}
  {8\,\uk{12}^2\,\uk{45}^2\,E_5},
\end{equation}
which is proportional to the five-point contact scalar factor $1/E_5$ and
therefore belongs naturally to the completion sector.  In this way, the
five-point answer already displays the qualitative pattern that persists at
higher points: the cut-detectable part is generated by a small collection of gluing
topologies, while the completion part is supported on comparatively sparse
longitudinal or contact-like completions.

\noindent\textbf{Six-point diagrams}

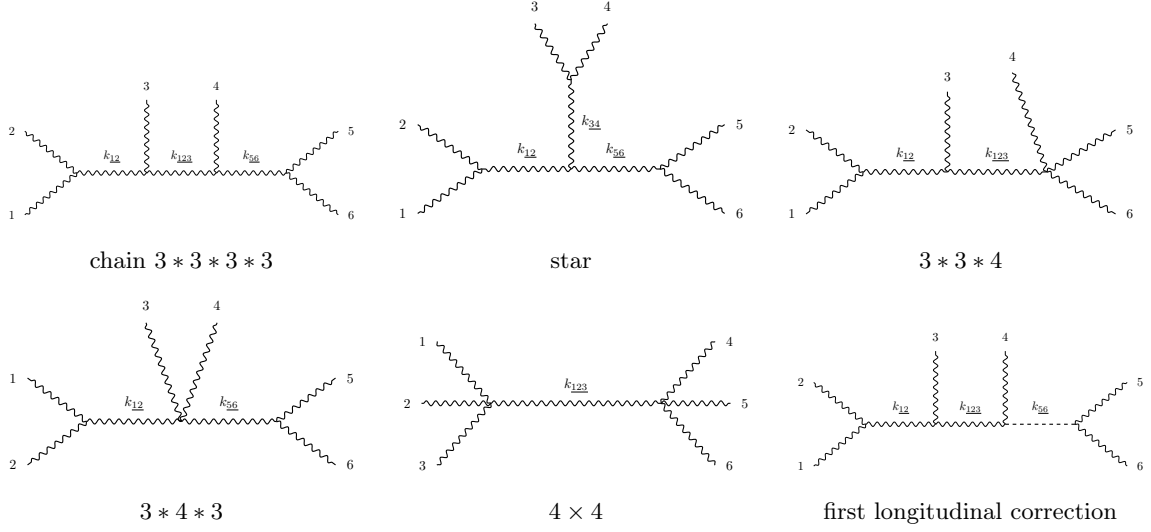
\begin{center}
  \centering
  \CompactDiagramCell{\SixPointChain{gluon}{gluon}{gluon}}{chain $3\ast3\ast3\ast3$}
  \hfill
  \CompactDiagramCell{\SixPointStar{gluon}{gluon}{gluon}}{star}
  \hfill
  \CompactDiagramCell{\SixPointQuarticEnd{gluon}{gluon}}{$3\ast3\ast4$}

  \vspace{0.45em}

  \CompactDiagramCell{\SixPointQuarticMid{gluon}{gluon}}{$3\ast4\ast3$}
  \hfill
  \CompactDiagramCell{\SixPointQuarticPair{gluon}}{$4\times4$}
  \hfill
  \CompactDiagramCell{\SixPointChain{gluon}{gluon}{scalar}}{first longitudinal correction}
\end{center}

We first list the all-transverse topologies, then add longitudinal propagators
one by one, and leave the two corrected cases for the end. This makes it
explicit which six-point structures are inherited directly from gluing and
which first appear as local completion terms. Compared with five
points, the novelty is not a different principle but a richer taxonomy of
topologies: ray-like chains, star-like couplings of three lower-point blocks,
and quartic-end or quartic-middle structures that interpolate between them.
This is precisely why six points are useful: they are the first place where
the low-point reconstruction begins to reveal a genuine space of all-\(n\)
building blocks rather than isolated examples. To keep the appendix readable,
we display one representative formula for each topology class; the remaining
cyclic or reflected images are generated by the relabellings already used in
Section~\ref{subsec:six-point-reconstruction}
and summarized in
\eqref{eq:six-point-rho-permutation}.
For orientation, the appendix representatives match the main-text sectors as
follows:
\begin{center}
\small
\begin{tabular}{>{\raggedright\arraybackslash}m{0.27\textwidth}|>{\raggedright\arraybackslash}m{0.38\textwidth}|>{\raggedright\arraybackslash}m{0.25\textwidth}}
main-text sector & appendix representatives & role\\ \hline
maximal transverse cuts &
\eqref{eq:six-point-chain-tr-raylike},
\eqref{eq:six-point-chain-tr-star} &
purely cut-detectable\\
\arrayrulecolor{black!25}\hline\arrayrulecolor{black}
lower-codimension transverse cuts &
\eqref{eq:six-point-chain-tr-mm-123456},
\eqref{eq:six-point-chain-tr-mm-123456-pairs},
{\eqref{eq:six-point-chain-tr-mm-123645},}
\eqref{eq:six-point-chain-tr-mu-123456} &
gluing with lower-point contact blocks\\
\arrayrulecolor{black!25}\hline\arrayrulecolor{black}
{transverse-cut-replaced longitudinal propagations} &
\eqref{eq:six-point-chain-mmz},
\eqref{eq:six-point-chain-mzz} &
{OPE data with overall normal scalar tubings}\\
\arrayrulecolor{black!25}\hline\arrayrulecolor{black}
{pure longitudinal propagations} &
\eqref{eq:six-point-pure-zz} &
{zero cut/chord completion}\\
\arrayrulecolor{black!25}\hline\arrayrulecolor{black}
{transverse-cut-replaced longitudinal propagations} &
\eqref{eq:six-point-chain-mzm}, \eqref{eq:six-point-star-mmz} &
{modified scalar tubings from local propagator operations}
\end{tabular}
\arrayrulecolor{black}
\end{center}

We first give the list of scalar functions used to present the six-point
diagrammatic results. Apart from the two maximally cut \(\tr\phi^3\) scalar functions
\(\psi^{\trphi}_{6,13,14,15}\) and \(\psi^{\trphi}_{6,13,35,15}\), the scalar
factors appearing below should be read as lower-point tubing types evaluated
on six-point composite rooted blocks, together with the six-point contact
scalar.  The superscripts \((4)\) and \((5)\) label tubing shapes rather than the number
of elementary external legs; in the six-point applications below, one substitutes
\(K_L+K_M+K_R=E_6\).  In particular,
\begin{equation}
  \begin{aligned}
    \psi_{6,\,12|34|56}^{\mathrm{scalar}}
    &:=
    \mathcal T_1^{(5)}(k_1+k_2,k_3+k_4,k_5+k_6;\uk{12},\uk{56})
    \\
    &\quad
    +
    \mathcal T_2^{(5)}(k_1+k_2,k_3+k_4,k_5+k_6;\uk{12},\uk{56}),\\
    \psi_{6,\,12|36|45}^{\mathrm{scalar}}
    &:=
     \psi_{6,\,12|34|56}^{\mathrm{scalar}}|_{4 \leftrightarrow 6},\\
    \psi_{6,\,123|456}^{\mathrm{scalar}}
    &:=
    \mathcal T^{(4)}(k_1+k_2+k_3,k_4+k_5+k_6;\uk{123}),\\
    \psi_{6,\,12|3456}^{\mathrm{scalar}}
    &:=
    \mathcal T^{(4)}(k_1+k_2,k_3+k_4+k_5+k_6;\uk{12}),\\
    \psi_{6,\,12|3|456}^{\mathrm{scalar}}
    &:=
    \mathcal T_1^{(5)}(k_1+k_2,k_3,k_4+k_5+k_6;\uk{12},\uk{456})
    \\
    &\quad
    +
    \mathcal T_2^{(5)}(k_1+k_2,k_3,k_4+k_5+k_6;\uk{12},\uk{456}).
  \end{aligned}
\end{equation}
\begin{subequations}
  \label{eq:six-point-chain-tr}
  \begin{align}
    { \psi_{6,\,12|3|4|56}^{\mathrm{YM}}}
    &=
    V_{12\mu_1}\,\Pi_{\bk_{12}}^{\mu_1\nu_1}\,
    V_{\nu_1 3 \mu_2}\,\Pi_{\bk_{123}}^{\mu_2\nu_2}\,
    V_{\nu_2 4 \mu_3}\,\Pi_{\bk_{56}}^{\mu_3\nu_3}\,
    V_{\nu_3 56}\,
    { \psi^{\trphi}_{6,13,14,15}},
    \label{eq:six-point-chain-tr-raylike}\\
    { \psi_{6,\,12|34|56}^{(\mu\mu\mu)}}
    &=
    V_{12\mu_1}\,\Pi_{\bk_{12}}^{\mu_1\nu_1}\,
    V_{34\mu_2}\,\Pi_{\bk_{34}}^{\mu_2\nu_2}\,
    V_{56\mu_3}\,\Pi_{\bk_{56}}^{\mu_3\nu_3}\,
    V_{\nu_1\nu_2\nu_3}\,
    { \psi^{\trphi}_{6,13,35,15}},
    \label{eq:six-point-chain-tr-star}\\
    { \psi_{6,\,12|3|456}^{(\mu\mu)}}
    &=
    V_{12\mu_1}\,\Pi_{\bk_{12}}^{\mu_1\nu_1}\,
    V_{\nu_1 3 \mu_2}\,\Pi_{\bk_{123}}^{\mu_2\nu_2}\,
    V_{\nu_2 456}\,
    { \psi_{6,\,12|3|456}^{\mathrm{scalar}}},
    \label{eq:six-point-chain-tr-mm-123456}\\
    { \psi_{6,\,12|34|56}^{(\mu\mu)}}
    &=
    V_{12\mu_1}\,\Pi_{\bk_{12}}^{\mu_1\nu_1}\,
    V_{\nu_1 34 \mu_2}\,\Pi_{\bk_{56}}^{\mu_2\nu_2}\,
    V_{\nu_2 56}\,
    { \psi_{6,\,12|34|56}^{\mathrm{scalar}}},
    \label{eq:six-point-chain-tr-mm-123456-pairs}\\
    { \psi_{6,\,12|36|45}^{(\mu\mu)}}
    &=
    V_{12\mu_1}\,\Pi_{\bk_{12}}^{\mu_1\nu_1}\,
    V_{\nu_1\,3\,\mu_2\,6}\,
    \Pi_{\bk_{45}}^{\mu_2\nu_2}\,
    V_{\nu_2\,45}\,
    { \psi_{6,\,12|36|45}^{\mathrm{scalar}}},
    \label{eq:six-point-chain-tr-mm-123645}\\
    { \psi_{6,\,123|456}^{(\mu)}}
    &=
    V_{123\mu}\,\Pi_{\bk_{123}}^{\mu\nu}\,V_{\nu456}\,
    { \psi_{6,\,123|456}^{\mathrm{scalar}}}.
    \label{eq:six-point-chain-tr-mu-123456}
  \end{align}
\end{subequations}

\noindent\textbf{One longitudinal line}

\begin{subequations}
  \label{eq:six-point-chain-mmz}
  \begin{align}
    { \psi_{6,\,12|3|4|56}^{(\mu\mu z)}}
    &=
    \left(
      V_{12\mu_1}\,\Pi_{\bk_{12}}^{\mu_1\nu_1}\,
      V_{\nu_1 3 \mu_2}\,\Pi_{\bk_{123}}^{\mu_2\nu_2}\,
      \beps_{4\nu_2}\,(\beps_5\!\cdot\!\beps_6)
    \right)
    \left[
      \frac{(k_5-k_6)(2k_4+k_5+k_6)}{4\,\uk{56}^2}
    \right]
    { \psi_{6,\,12|3|456}^{\mathrm{scalar}}},
    \label{eq:six-point-chain-mmz-main}\\
    { \psi_{6,\,12|3|456}^{(z\mu)}}
    &=
    \left[
      -\frac{
        (\beps_1\!\cdot\!\beps_2)\,
        (k_1-k_2)\,
        (k_1+k_2+2k_3)
      }{4\,\uk{12}^2}
    \right]
    \Bigl(
      \beps_{3\mu}\,\Pi_{\bk_{123}}^{\mu\nu}\,
      V_{\nu456}
    \Bigr)
    { \psi_{6,\,123|456}^{\mathrm{scalar}}}.
    \label{eq:six-point-chain-zmu-main}
  \end{align}
\end{subequations}
The reflected chain class $(z\mu\mu)$ is obtained from $(\mu\mu z)$ by
$\uk{12}\leftrightarrow\uk{56}$ and $(1,2)\leftrightarrow(6,5)$.
{ There are also two additional cases,
\({\psi}_{6,\,12|3|4|56}^{(\mu z \mu)}\) and
\({\psi}_{6,\,12|34|56}^{(\mu\mu z)}\), given in the later discussion in
Eqs.~\eqref{eq:six-point-chain-mzm} and \eqref{eq:six-point-star-mmz}.}

\noindent\textbf{Two longitudinal lines}

\begin{subequations}
  \label{eq:six-point-chain-mzz}
  \begin{align}
    \psi_{6,\,12|3|4|56}^{(\mu zz)}
    &=
    \left(
      V_{12\mu_1}\,\Pi_{\bk_{12}}^{\mu_1\nu_1}\,
      \beps_{3\nu_1}
    \right)
    \left[
      \frac{
        (\beps_4\!\cdot\!(\bk_{56}-\bk_{123}))\,
        (\beps_5\!\cdot\!\beps_6)\,
        (2k_3+k_4+k_5+k_6)\,
        (k_5-k_6)
      }{8\,\uk{123}^2\,\uk{56}^2}
    \right]
    \notag\\[-0.1em]
    &\quad\times
    \psi_{6,\,12|3456}^{\mathrm{scalar}},
    \label{eq:six-point-chain-mzz-main}\\
    \psi_{6,\,12|3|4|56}^{(z\mu z)}
    &=
    \left[
      \frac{(k_1-k_2)(k_1+k_2+2k_3)}{4\,\uk{12}^2}
    \right]
    \left(
      (\beps_1\!\cdot\!\beps_2)\,
      \beps_{3\mu}\,\Pi_{\bk_{123}}^{\mu\nu}\,\beps_{4\nu}\,
      (\beps_5\!\cdot\!\beps_6)
    \right)
    \notag\\[-0.1em]
    &\quad\times
    \left[
      -\frac{(k_5-k_6)(2k_4+k_5+k_6)}{4\,\uk{56}^2}
    \right]
    \psi_{6,\,123|456}^{\mathrm{scalar}},
    \label{eq:six-point-chain-zmuz-main}\\
    \psi_{6,\,12|34|56}^{(\mu zz)}
    &=
    \left(
      V_{12\mu}\,\Pi_{\bk_{12}}^{\mu\nu}\,
      (\bk_{34}-\bk_{56})_\nu\,
      (\beps_3\!\cdot\!\beps_4)\,
      (\beps_5\!\cdot\!\beps_6)
    \right)
    \left[
      -\frac{(k_3-k_4)(k_5-k_6)}{8\,\uk{34}^2\,\uk{56}^2}
    \right]
    \notag\\[-0.1em]
    &\quad\times
    \psi_{6,\,12|34|56}^{\mathrm{scalar}},
    \label{eq:six-point-chain-muzz-pairs}
  \end{align}
\end{subequations}
Here the reflected class $[zz\mu]$ is obtained from $[\mu zz]$ in the same way.

\begin{subequations}
  \label{eq:six-point-pure-zz}
  \begin{align}
    {\psi}_{6,\,12|34|56}^{(zz)}
    &=
      \frac{
        (\beps_1\!\cdot\!\beps_2)\,
        (\beps_3\!\cdot\!\beps_4)\,
        (\beps_5\!\cdot\!\beps_6)\,
        (k_1-k_2)\,(k_5-k_6)
      }{16\,\uk{12}^2\,\uk{56}^2E_6}
    \label{eq:six-point-chain-zz-pairs}\\
   {\psi}_{6,\,12|36|45}^{(zz)}
    &=
     - \frac{
        (\beps_1\!\cdot\!\beps_2)\,
        (\beps_3\!\cdot\!\beps_6)\,
        (\beps_4\!\cdot\!\beps_5)\,
        (k_1-k_2)\,(k_4-k_5)
      }{8\,\uk{12}^2\,\uk{45}^2E_6}
    \label{eq:six-point-chain-zz-pairs2c}
  \end{align}
\end{subequations}

\noindent\textbf{Three longitudinal lines}

\begin{equation}
  \label{eq:six-point-chain-zzz}
  \psi_{6,\,12|3|4|56}^{(zzz)}
  =
  -\frac{
    (\beps_1\!\cdot\!\beps_2)\,
    (\beps_3\!\cdot\!(\bk_{456}-\bk_{12}))\,
    (\beps_4\!\cdot\!(\bk_{56}-\bk_{123}))\,
    (\beps_5\!\cdot\!\beps_6)\,
    (k_1-k_2)(k_5-k_6)
  }{16\,\uk{12}^2\,\uk{123}^2\,\uk{56}^2\,E_6},
\end{equation}
so the fully longitudinal chain is proportional to the six-point contact scalar
factor $1/E_6$.

\noindent\textbf{Propagator-collapse effects and one longitudinal line}

The two remaining six-point graphs are the only ones in this subsection
that are not simply a tensor numerator times overall ordinary scalar tubings.
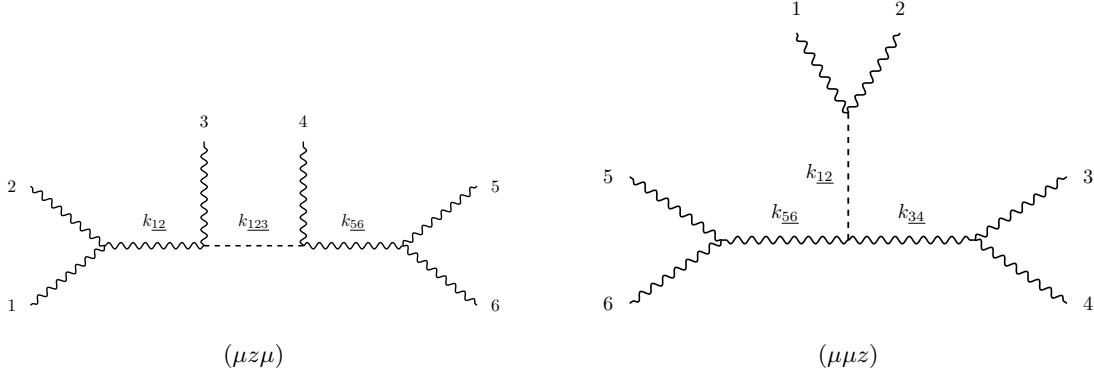
\begin{center}
  \centering
  \CompactDiagramCellWide{\SixPointChain{gluon}{scalar}{gluon}}{$(\mu z\mu)$}
  \hfill
  \CompactDiagramCellWide{\SixPointStarFigureSixteen}{$(\mu\mu z)$}
\end{center}
We first consider the \((\mu z\mu)\) case.
\begin{equation}
  \label{eq:six-point-chain-mzm}
  \psi_{6,\,12|3|4|56}^{(\mu z \mu)}
  =
  \left(
    V_{12\mu_1}\,\Pi_{\bk_{12}}^{\mu_1\nu_1}\,
    \beps_{3\nu_1}
  \right)
  \left(
    \beps_{4\mu_2}\,\Pi_{\bk_{56}}^{\mu_2\nu_2}\,
    V_{\nu_2 56}
  \right)
  \psi_{6,\,12|3|4|56}^{\mathrm{sc},\,mzm}.
\end{equation}
Here the middle longitudinal propagator contributes only the local factor
$1/\uk{123}^2$, so the two mixed cubic vertices share the same bulk radial
coordinate. Defining
\begin{equation*}
  \begin{aligned}
    A&:=k_1+k_2,\qquad
    B:=k_3+k_4,\qquad
    C:=k_5+k_6,\\
    u&:=\uk{12},\qquad
    v:=\uk{56},\qquad
    E_t:=A+B+C=E_6.
  \end{aligned}
\end{equation*}
The scalar factor can be written as
\begin{equation}
  \label{eq:six-point-chain-mzm-scalar}
  \psi_{6,\,12|3|4|56}^{\mathrm{sc},\,mzm}
  =
  \frac{1}{4\,\uk{123}^2}
  \Big[
    (k_3-A)(k_4-C)\,\psi_{6,\,12|34|56}^{\mathrm{scalar}}
    + E_t\big((k_4-C)T_1+(k_3-A)T_2\big)
    + C_0
  \Big],
\end{equation}
where
\begin{equation*}
  \begin{aligned}
    \psi_{6,\,12|34|56}^{\mathrm{scalar}}
    &=
    \mathcal T_1^{(5)}(A,B,C;u,v)
    +
    \mathcal T_2^{(5)}(A,B,C;u,v),\\
    C_0
    &=
    \frac{1}{(A+u)(B+u+v)(C+v)}.
  \end{aligned}
\end{equation*}
Equivalently, \eqref{eq:six-point-chain-mzm-scalar} can be packaged as a local
operator rule acting on the rooted scalar skeleton
\[
  \psi_{0}^{\mathrm{cs}}
  :=
  \psi_{6,\,12|34|56}^{\mathrm{scalar}}
  =
  \mathcal T_1^{(5)}(A,B,C;u,v)
  +
  \mathcal T_2^{(5)}(A,B,C;u,v).
\]
Let $\widehat{\mathcal C}_L$ and $\widehat{\mathcal C}_R$ denote the left and
right collapse operators associated with the two transverse kernels adjacent to
the shared dashed propagator, defined by
\[
  \widehat{\mathcal C}_L\,\psi_{0}^{\mathrm{cs}}
  =
  E_t\,\mathcal T_1^{(5)}(A,B,C;u,v),
  \qquad
  \widehat{\mathcal C}_R\,\psi_{0}^{\mathrm{cs}}
  =
  E_t\,\mathcal T_2^{(5)}(A,B,C;u,v),
  \qquad
  \widehat{\mathcal C}_L\widehat{\mathcal C}_R\,\psi_{0}^{\mathrm{cs}}=C_0.
\]
Then the scalar factor becomes
\begin{equation}
  \label{eq:six-point-chain-mzm-operator}
  \psi_{6,\,12|3|4|56}^{\mathrm{sc},\,mzm}
  =
  \frac{1}{4\,\uk{123}^2}
  \Big[
    (k_3-A)+\widehat{\mathcal C}_L
  \Big]
  \Big[
    (k_4-C)+\widehat{\mathcal C}_R
  \Big]
  \psi_{0}^{\mathrm{cs}}.
\end{equation}
Expanding \eqref{eq:six-point-chain-mzm-operator} reproduces the four expected
pieces: the main term, two single-collapse terms, and the double-collapse
contact-like term.
The contour prescription follows the transverse bulk-to-bulk measure
$dp/\pi$, so each spectral integral is implemented by $2i$ times the sum of
upper-half-plane residues.  In \eqref{eq:six-point-chain-mzm-scalar}, the first
term comes from moving both radial derivatives onto external exponentials, the
two $E_t$ terms come from collapsing one transverse kernel, and $C_0$ is the
double-collapse contact-like remainder.

{ We now move to the \((\mu\mu z)\) case.}
\begin{equation}
  \label{eq:six-point-star-mmz}
  { \psi}_{6,\,12|34|56}^{(\mu\mu z)}
  =
  -\frac{(k_1-k_2)(\beps_1\!\cdot\!\beps_2)}{\uk{12}^2}
  \left(
    { V_{34 i}\,\Pi_{\bk_{34}}^{ij}\,
    \Pi_{\bk_{56}\,j}^{\ \ k}\,V_{k56}}
  \right)
  { \psi}_{6,\,12|34|56}^{\mathrm{sc},\,mmz}.
\end{equation}
The two transverse branches are contracted through the
product of projectors $\Pi_{\bk_{34}}\Pi_{\bk_{56}}$, while the nontrivial scalar
piece is carried by the shared-radial-coordinate derivatives acting on the two
transverse kernels. With
\begin{equation*}
  \begin{aligned}
    A&:=k_1+k_2,\qquad
    B:=k_3+k_4,\qquad
    C:=k_5+k_6,\\
    u&:=\uk{34},\qquad
    v:=\uk{56},\qquad
    E_t:=A+B+C=E_6,
  \end{aligned}
\end{equation*}
the corrected scalar factor is
\begin{equation}
  \label{eq:six-point-star-mmz-scalar}
  \psi_{6,\,12|34|56}^{\mathrm{sc},\,mmz}
  =
  \frac{A(B-C+u-v)+2Bu-2Cv}
  {4\,E_t\,(B+u)\,(A+C+u)\,(A+B+v)\,(C+v)\,(A+u+v)}.
\end{equation}
Equivalently, using the generic five-point tubings from
\eqref{eq:five-point-tubing-def} with the identification
\[
  (K_L,K_M,K_R;u,v)=(B,A,C;u,v),
\]
one has
\begin{equation}
  \label{eq:six-point-star-mmz-decompose}
  \begin{aligned}
    \psi_{6,\,12|34|56}^{\mathrm{sc},\,mmz}
    &=
    \frac{1}{4}\Big[
      (B-C)\,\psi_{6,\,12|34|56}^{\mathrm{scalar}}
      + E_t\Big(
        \mathcal T_2^{(5)}(B,A,C;u,v)
        -
        \mathcal T_1^{(5)}(B,A,C;u,v)
      \Big)
    \Big] \\
    &=
    \frac{1}{4}\Big[
      (u-C)\,\mathcal T_1^{(5)}(B,A,C;u,v)
      +
      (B-v)\,\mathcal T_2^{(5)}(B,A,C;u,v)
    \Big].
  \end{aligned}
\end{equation}
In \eqref{eq:six-point-star-mmz-decompose}, the first term is the expected
effective five-point tubing combination, while the second term is the
propagator-collapse correction generated when the radial derivative is
moved through one of the two transverse kernels.

Taken together, these formulas provide the direct Feynman-rule checks of the
reconstructed answers in Sections~\ref{subsec:four-point-reconstruction}--\ref{subsec:six-point-reconstruction}.
They also make visible, diagram by diagram, the
gluing/completion organization discussed there: transverse and mixed exchange
topologies account for the cut-detectable part, while longitudinal and contact
topologies supply the sparse completion required by current conservation.
\bibliographystyle{JHEP}\bibliography{refs_v3,cosmological_discontinuity_refs_v3}

\end{document}